\documentclass[preprint,12pt]{elsarticle}

\usepackage{amssymb}
\usepackage{amsmath}

\usepackage{lineno}

\biboptions{compress}

\usepackage[table]{xcolor}
\usepackage{multirow}
\usepackage[scriptsize,tight,figbotcap]{subfigure}
\usepackage[cm]{fullpage}

\usepackage{bm}   
\usepackage{doi}
\usepackage{hyperref}
\hypersetup{
    colorlinks,
    linkcolor={red!50!black},
    citecolor={blue!50!black},
    urlcolor={blue!80!black}
}

\journal{Journal of Non-Newtonian Fluid Mechanics}

\begin{document}
\newcommand{\vf}[1]{\bm{\mathrm{#1}}}
\newcommand{\pd}[2]{\frac{\partial #1}{\partial #2}}

\begin{frontmatter}



\title{Performance of the finite volume method in solving regularised Bingham flows: inertia effects in the lid-driven 
cavity flow}


\author[dms,oc]{Alexandros Syrakos\corref{cor1}}
\ead{syrakos.alexandros@ucy.ac.cy}

\author[dms,oc]{Georgios C. Georgiou}
\ead{georgios@ucy.ac.cy}

\author[dmme]{Andreas N. Alexandrou}
\ead{andalexa@ucy.ac.cy}

\cortext[cor1]{Corresponding author}

\address[dms]{Department of Mathematics and Statistics, University of Cyprus, PO Box 20537, 1678 Nicosia, Cyprus}
\address[oc]{Oceanography Centre, University of Cyprus, PO Box 20537, 1678 Nicosia, Cyprus}
\address[dmme]{Department of Mechanical and Manufacturing Engineering, University of Cyprus, PO Box 20537, 1678 Nicosia,
Cyprus}

\begin{abstract}
We extend our recent work on the creeping flow of a Bingham fluid in a lid-driven cavity, to the study of inertial 
effects, using a finite volume method and the Papanastasiou regularisation of the Bingham constitutive model  [J. 
Rheology 31 (1987) 385-404]. The finite volume method used belongs to a very popular class of methods for solving 
Newtonian flow problems, which use the SIMPLE algorithm to solve the discretised set of equations, and have matured 
over the years. By regularising the Bingham constitutive equation it is easy to extend such a solver to Bingham flows 
since all that this requires is to modify the viscosity function. This is a tempting approach, since it requires 
minimum programming effort and makes available all the existing features of the mature finite volume solver. On the 
other hand, regularisation introduces a parameter which controls the error in addition to the grid spacing, and makes it 
difficult to locate the yield surfaces. Furthermore, the equations become stiffer and more difficult to solve, while 
the discontinuity at the yield surfaces causes large truncation errors. The present work attempts to investigate the 
strengths and weaknesses of such a method by applying it to the lid-driven cavity problem for a range of Bingham and 
Reynolds numbers (up to 100 and 5000 respectively). By employing techniques such as multigrid, local grid refinement, 
and an extrapolation procedure to reduce the effect of the regularisation parameter on the calculation of the yield 
surfaces (Liu et al. J. Non-Newtonian Fluid Mech. 102 (2002) 179-191), satisfactory results are obtained, although the 
weaknesses of the method become more noticeable as the Bingham number is increased.
\end{abstract}


\end{frontmatter}

This is the accepted version of the article published in: Journal of Non-Newtonian Fluid Mechanics 208--209 (2014) 88--107, 
\doi{10.1016/j.jnnfm.2014.03.004}

\textcopyright 2016. This manuscript version is made available under the CC-BY-NC-ND 4.0 license 
\url{http://creativecommons.org/licenses/by-nc-nd/4.0/}



\section{Introduction}
\label{sec: introduction}

Finite volume methods (see \cite{Ferziger_02} for a detailed description) are popular for the solution of fluid flow 
problems. They divide the computational domain into a number of small volumes, and integrate the governing differential 
equations over each volume. The resulting integrals, which consist of the fluxes and source terms of the relevant 
physical quantities on each volume, are approximated by algebraic expressions involving the values of the dependent 
variables at selected points of the domain, such as the volume centres. This results in a system of non-linear algebraic 
equations, the solution of which gives approximate values of the dependent variables at the selected points. In the 
literature one can find that many different algebraic solvers have been used to solve this system, but the most popular 
appear to be SIMPLE \cite{Patankar_72} and its variants. These are iterative methods which consecutively solve a series 
of linear systems in each iteration: one linear system for each momentum equation, and a linear system for pressure 
which tries to enforce satisfaction of the continuity equation by relating velocity corrections to pressure corrections. 
Although SIMPLE is an old algorithm, it is still very popular and it has been used successfully in numerous studies to 
solve complex flow phenomena, \cite{Ferras_2013} being a recent example involving non-Newtonian flow. SIMPLE is a slowly 
converging algorithm, but its performance can be greatly enhanced by using it in a multigrid context 
\cite{Sivaloganathan_88, Hortmann_90}. The popularity of SIMPLE and its variants is due in part to the fact that they 
allow easy extension of the solver to account for additional physical phenomena: for each additional differential 
equation to be solved, the solution of a corresponding linear system is added to each SIMPLE iteration. Over the years, 
codes which implement this method have been developed and expanded to include many features, such as meshing 
capabilities, discretisation schemes, boundary condition choices, graphical user interfaces, turbulence models, and 
choices of models for different physical phenomena (heat transfer, combustion, chemical reactions, phase change, flow 
with a free surface etc.). In the present work we investigate the ability of such a solver to solve viscoplastic flows. 
This would be desirable, as it would greatly reduce the programming effort and it would allow one to also use other 
features of the solver, contrary to developing a specialised solver from scratch.

Viscoplastic materials behave as solids at low stress levels, but flow when the stress exceeds a critical value, the 
\textit{yield stress}, $\tau_y$. Suspensions of particles or macromolecules, such as pastes, gels, foams, drilling 
fluids, food products, and nanocomposites, are typical viscoplastic fluids. The simplest model for a viscoplastic 
material is the Bingham model, which exhibits a linear stress to rate-of-strain relationship during flow. The constant 
of proportionality of this linear relationship is called the \textit{plastic viscosity}, $\mu$. Thus, the Bingham 
constitutive equation is written as follows:

\begin{equation} \label{eq: Bingham_constitutive}
 \left\{ \begin{array}{ll} 
   \vf{\dot{\gamma}} \;=\; \vf{0} \;, \qquad  &  \tau \leq \tau_y
\\
   \vf{\tau} \;=\; \left( \dfrac{\tau_y}{\dot{\gamma}} + \mu \right)\vf{\dot{\gamma}} \;, & \tau > \tau_y
 \end{array} \right.
\end{equation}
where $\vf{\tau}$ is the stress tensor and $\vf{\dot{\gamma}}$ is the rate-of-strain tensor, $\vf{\dot{\gamma}} \equiv
\nabla\vf{u} + (\nabla\vf{u})^{\mathrm{T}}$, $\vf{u}$ being the velocity vector. The magnitudes of these two tensors are
$\tau \equiv ( \frac{1}{2} \vf{\tau} : \vf{\tau}) ^{1/2}$ and $\dot{\gamma} \equiv ( \frac{1}{2} \vf{\dot{\gamma}} : 
\vf{\dot{\gamma}}) ^{1/2}$.

Thus the Bingham constitutive equation has two branches, with different physical laws applying to yielded and unyielded 
areas. The solution of Bingham flows is difficult because the location of the interface between the yielded and 
unyielded regions is a priori unknown. The most popular approaches to tackle this problem fall into two categories. One 
category includes methods which approximate equation (\ref{eq: Bingham_constitutive}) by a regularised constitutive 
equation, which treats the whole material as a fluid of variable viscosity and is applicable throughout. The unyielded 
regions are approximated by locally assigning a very high value to the viscosity. This category includes methods such 
as that proposed by Bercovier and Engelman \cite{Bercovier_80}, and that by Papanastasiou \cite{Papanastasiou_87}. A 
related method (although not strictly a regularisation method since it does not remove the discontinuity) is the 
bi-viscosity method of O'Donovan and Tanner \cite{Donovan_84}, which treats the unyielded material as a separate, very 
viscous fluid. Regularisation methods make use of a parameter which, depending on the model, should be given a very 
large or a very small value, in order for the results to be a good approximation to the actual Bingham flow solution. It 
can be shown (e.g.\ \cite{Frigaard_05, Dean_2007}) that in the limit when this parameter approaches asymptotically to 
infinity or zero, the velocity field obtained with the regularisation method converges to that of the actual Bingham 
flow. However, in practice, the range of usable values for the regularisation parameter is limited by the fact that 
using extreme values causes numerical problems. A disadvantage of the regularisation approach is that since the whole 
material is a fluid, yield surfaces are not clearly defined. Usually the yield surfaces are approximated by the surfaces 
where the magnitude of the stress equals the yield stress. On the other hand, to implement such a method in an existing 
solver one only has to define appropriately the function which calculates the viscosity. This makes regularisation 
methods very suitable for the use examined in this work, that is for enabling existing solvers to tackle viscoplastic 
flows. In fact many commercial solvers have built-in or allow user-defined viscosity functions, and there exist a 
number of published works where commercial finite volume (e.g. in \cite{Naccache_2007, Mossaz_2010, Turan_2010, 
Filali_2013}) or finite element packages (e.g.\ in \cite{Tokpavi_2008, Nirmalkar_2012, Zamankhan_2012, Filali_2013}) were 
used in this way, applying either a regularisation method or the bi-viscosity method to solve viscoplastic flows.

The other category includes methods which start by deriving a variational inequality whose solution is equivalent to the 
solution of the original problem (variational inequality formulation of the problem). Their solution relies on the use 
of multiplier functions (projection methods and augmented Lagrangian methods) \cite{Fortin_83, Glowinski_84, Dean_2007, 
Glowinski_2011}. These methods avoid the use of regularisation and thus solve the actual problem directly. They are 
mathematically more involved and the concepts may be difficult to follow for people who do not have the relevant 
mathematical background. Unfortunately, contrary to regularisation methods, it is not straightforward to incorporate 
them into existing general purpose solvers, and thus are not suitable for the present investigation. They are 
implemented usually in combination with a finite element discretisation scheme, but a finite volume scheme can also be 
used \cite{Vinay_2005}. The availability of these techniques has led some researchers to support that regularisation 
methods should be avoided. However, apart from the aforementioned practical advantages of regularisation methods, one 
should also keep in mind that the Bingham and other viscoplastic models only approximate the behaviour of real 
materials. There is some ambiguity concerning the manifestation of yield stress by actual materials (see 
\cite{Barnes_1999} or \cite{Balmforth_2014} for discussions and further references) and some researchers have suggested 
that regularised constitutive equations may in fact be better representations of the physical reality 
\cite{Mendes_2004}. In any case, there is a large literature with successful application of regularisation methods to a 
wide range of problems - see \cite{Glowinski_2011} for some references. The results produced by regularisation methods 
should be interpreted with some caution though as pointed out in \cite{Frigaard_05}, because it is not easy to know when 
the regularisation parameter is large enough. A direct comparison between Papanastasiou regularisation and the augmented 
Lagrangian method is made in a recent publication \cite{Dimakopoulos_2013}.

The present work aims to explore the potential and the limitations of a finite volume method that uses the SIMPLE 
algebraic solver, together with a regularised constitutive equation, for solving viscoplastic flows. This was partly 
explored also in our previous publication \cite{Syrakos_13}, but the present work differs in the following: a) non-zero 
Reynolds number flows are also treated; b) the detrimental effect of the discontinuity exhibited at the yield surfaces 
on the accuracy of the finite volume solution is studied in more depth, by calculating the truncation error generated, 
and local grid refinement is proposed as a treatment; and c) more attention is given to the accurate determination of 
the yield surfaces, and ways to increase this accuracy are investigated.

Adaptive mesh refinement for viscoplastic flows was applied successfully in a Finite Element context already by Roquet 
and Saramito \cite{Roquet_2003} and Zhang \cite{Zhang_10}, with an augmented Lagrangian method in both cases, with the 
aim of improving the accuracy of the calculated yield surfaces. The present work employs a Finite Volume variant of 
adaptive mesh refinement, proposed in \cite{Syrakos_06b, Syrakos_12}. The focus is not restricted to the yield surfaces, 
but the advantages of local grid refinement are assessed with respect to the overall flow field accuracy as well.

We employ the Papanastasiou \cite{Papanastasiou_87} regularisation of equation (\ref{eq: Bingham_constitutive}). It 
introduces an exponential term to replace the discontinuous constitutive equation (\ref{eq: Bingham_constitutive}) by a 
single equation, applicable throughout the material:

\begin{equation} \label{eq: papanastasiou, dimensional}
 \vf{\tau} \;=\; \left[ \frac{\tau_y}{\dot{\gamma}} \{ 1-\exp(-m\dot{\gamma})\} \,+\, \mu \right] \vf{\dot{\gamma}}
\end{equation}
where $m$ is a stress growth parameter, which is required to be ``sufficiently'' large so that the ideal Bingham 
behaviour is approximated with satisfactory accuracy.

The method is tested on the square lid-driven cavity problem, which is probably the most popular problem for testing 
numerical methods in computational fluid dynamics. Consider a square cavity of side $L$, filled with a fluid which is 
set to motion by the lid of the cavity which moves with a tangential velocity $U$. In the case of Bingham flow, the 
flow field is characterised by two dimensionless numbers: the Reynolds number, which is defined in terms of the plastic 
viscosity $\mu$,

\begin{equation} \label{eq: Re}
 Re \;\equiv \; \frac{\rho U L}{\mu}
\end{equation}
and the Bingham number $Bn$, defined as

\begin{equation} \label{eq: Bn}
 Bn \;\equiv\; \frac{\tau_y L}{\mu U}
\end{equation}

The two-dimensional Newtonian lid-driven cavity problem has been studied numerically in great detail - see e.g.\ the 
review papers \cite{Shankar_00, Erturk_09}. Accurate numerical results can be found in the literature for Reynolds 
numbers up to 10,000 \cite{Ghia_82, Botella_98, Syrakos_12}. The few available experimental studies  \cite{Pan_67, 
Koseff_84, Rhee_84} have shown that, even at moderate Reynolds numbers, the flow field is in fact three-dimensional. 
Three-dimensional computational studies have confirmed this and provided a value of $Re \approx 785$ as a critical value 
beyond which 3-dimensional features appear \cite{Albensoeder_01, Albensoeder_05}. Nevertheless, the two-dimensional 
lid-driven cavity problem, although unrealistic beyond a critical Reynolds number, is very attractive for testing 
numerical methods, as it is easy to set up due to the simple geometry and boundary conditions, and the flow field is 
complex enough to provide a good test for the numerical method.

Flow of viscoplastic fluids in a lid-driven cavity has also been studied numerically, although in the majority of cases 
the problem was only used as a validation test case and limited results were obtained. Most studies only considered 
creeping flow or flow at very low Reynolds numbers, such as the works of Bercovier and Engelman \cite{Bercovier_80} and 
of Mitsoulis and Zisis \cite{Mitsoulis_01}, who used regularisation methods, and the works of Sanchez \cite{Sanchez_98}, 
Yu and Wachs \cite{Yu_07}, Muravleva and Olshanskii \cite{Muravleva_08}, Zhang \cite{Zhang_10} and Glowinski and Wachs 
\cite{Glowinski_2011}, who used the augmented Lagrangian method. The results of all these studies agree qualitatively, 
in that two unyielded zones appear in general: one at the bottom of the cavity which is motionless, and one near the 
vortex centre which moves with solid body rotation. However, it is rather surprising that there is not an agreement on 
the precise shape of these zones, even among the augmented Lagrangian studies. A possible explanation is that, as noted 
by Glowinski and Wachs \cite{Glowinski_2011}, the solution of the augmented Lagrangian method only becomes equivalent to 
the original Bingham problem when iterations have fully converged, but in any numerical procedure convergence is assumed 
when a certain non-zero tolerance has been reached. So, the accuracy depends on the choice of this tolerance, in a 
similar fashion that the accuracy of regularisation methods depends on the choice of regularisation parameter. To this 
one has to add the errors generated by the discretisation procedure. Our previous results \cite{Syrakos_13} are closer 
to those of Mitsoulis and Zisis \cite{Mitsoulis_01}, and Glowinski and Wachs \cite{Glowinski_2011}.

There are also some studies containing results for moderate or high Reynolds numbers. We note the works of Neophytou 
\cite{Neofytou_05}, Elias et al. \cite{Elias_06}, Frey et al. \cite{Frey_10}, Prashant and Derksen \cite{Prashant_11} 
and dos Santos et al. \cite{dos_Santos_11} who used regularisation methods or the bi-viscosity method, and the work of 
Vola et al. \cite{Vola_03} who used an augmented Lagrangian method. It is noteworthy that whereas in the creeping flow 
case the multipliers methods are dominant, in the inertia case the situation is reversed.

Concerning the discretisation methods, it appears that the finite element method is the usual choice. The finite volume 
method has been used in only two of the aforementioned studies, those of Neophytou \cite{Neofytou_05} and Glowinski and 
Wachs \cite{Glowinski_2011}. The method of Neophytou \cite{Neofytou_05} is a finite volume method which employs the 
SIMPLE algebraic solver and uses the Papanastasiou regularisation, and thus is of the category of methods examined here. 
It does not use more advanced techniques such as multigrid and local grid refinement, and unfortunately the results are 
limited to very low Bingham numbers, $Bn \leq 1$. The finite volume method of Glowinski and Wachs \cite{Glowinski_2011} 
on the other hand is adapted to the needs of the augmented Lagrangian method, and deals only with creeping flow.

The aim of the present work is to investigate the capabilities and limitations of the popular finite volume / SIMPLE 
method coupled with the Papanastasiou regularisation, by applying it to the simulation of Bingham flow in a lid driven 
cavity. Where limitations are detected, an effort is made to overcome them using techniques that are relatively easy to 
apply in the framework of such a method. We provide results for a wide range of Bingham and Reynolds numbers, up to 100 
and 5000 respectively, and study systematically the effects of these dimensionless numbers on the flow. The rest of the 
paper is organised as follows: The associated governing equations are presented in Section \ref{sec: governing 
equations}. The numerical method for solving these equations is briefly described in Section \ref{sec: numerical 
method}, while the results of the numerical experiments are given and discussed in Section \ref{sec: results}. Finally, 
Section \ref{sec: conclusions} summarises the conclusions of this research.

\section{Governing equations}
\label{sec: governing equations}

The flow is assumed to be steady-state, two-dimensional, incompressible, and isothermal. By scaling the fluid velocity 
by the lid velocity $U$, and the pressure and stress by $\mu U/L$, the continuity and momentum equations can be written 
in the following dimensionless forms:

\begin{equation} \label{eq: continuity}
 \nabla \cdot \vf{u} \;=\; 0
\end{equation}

\begin{equation} \label{eq: momentum}
 Re \; \vf{u} \cdot \nabla \vf{u} \;=\; -\nabla p \;+\; \nabla \cdot \vf{\tau}
\end{equation}
where $\vf{u}$ denotes here the dimensionless velocity of the fluid, $p$ is the dimensionless pressure, and $\vf{\tau}$ 
is the dimensionless stress tensor (for the sake of simplicity, we kept the same symbols for the dedimensionalised 
variables). The dimensionless form of the Papanastasiou constitutive equation becomes

\begin{equation} \label{eq: Papanastasiou, dimensionless}
 \vf{\tau} \;=\; \left[ \frac{Bn}{\dot{\gamma}} \{ 1-\exp(-M\dot{\gamma})\} \,+\, 1 \right] \vf{\dot{\gamma}}
\end{equation}
where $M$ is the dimensionless stress growth parameter, defined as:

\begin{equation} \label{eq: M}
 M \;\equiv\; \frac{m U}{L}
\end{equation}
and $\vf{\dot{\gamma}}$ is now the dimensionless rate of strain tensor. The Papanastasiou regularisation (\ref{eq: 
Papanastasiou, dimensionless}) corresponds to a dimensionless apparent viscosity of

\begin{equation} \label{eq: viscosity}
 \eta \;=\; \frac{Bn}{\dot{\gamma}} \{ 1-\exp(-M\dot{\gamma})\} \,+\, 1
\end{equation}
The higher the value of $M$, the better Eq.\ (\ref{eq: Papanastasiou, dimensionless}) approximates the actual Bingham 
constitutive equation, $\vf{\tau} = [Bn/\dot{\gamma} + 1]\vf{\dot{\gamma}}$, in the yielded regions of the flow field 
($\tau>Bn$), and the higher the apparent viscosity is in the unyielded regions, making them behave approximately as 
solid bodies. For practical reasons though, $M$ must not be so high as to cause convergence problems to the numerical 
methods used to solve the above equations.

Equations (\ref{eq: continuity}) -- (\ref{eq: Papanastasiou, dimensionless}) together with the no-slip wall boundary 
conditions fully determine the flow problem which is solved numerically. In the present study, the direction of motion 
of the lid is towards the right.

\section{Numerical method}
\label{sec: numerical method}

\subsection{Solution without local grid refinement}
\label{ssec: method without refinement}

Most of the results presented in the following section were computed on Cartesian grids without local grid refinement, 
consisting of $512 \times 512$ square control volumes. Coarser grids are constructed by removing every second grid line 
from the immediately finer grid. The finite volume method used to solve the governing equations on these grids is 
described in detail in Syrakos et al. \cite{Syrakos_13}. According to the finite volume methodology, the continuity and 
momentum equations are integrated over each control volume and the integrals are approximated by algebraic expressions 
involving the values of the flow variables at discrete points. In the present work, all variables (velocity components, 
pressure, and viscosity) are stored at control volume centres. Both the convective and viscous fluxes are discretised 
using 2$^{\text{nd}}$-order accurate central differences. The mass fluxes are discretised using momentum interpolation 
as described in \cite{Syrakos_06a}, to suppress spurious pressure oscillations between control volume centres.

The resulting algebraic system is solved using the SIMPLE algorithm, with the only modification being that at the start 
of every SIMPLE iteration the viscosity is updated according to Eq.\ (\ref{eq: viscosity}), using the current estimate of 
the velocity field. To accelerate convergence, SIMPLE is used in a geometric multigrid framework. Due to the high 
degree of nonlinearity of the problem, the standard multigrid algorithm fails to converge except at small Bingham 
numbers, $Bn < 0.5$, as the results in \cite{Syrakos_13} show. To overcome this problem, the modification suggested by 
Ferziger and Peric \cite{Ferziger_02} has been applied; on coarse grids the viscosity is not updated according to Eq.\ 
(\ref{eq: viscosity}), but it is interpolated (restricted) from the immediately finer grid and held constant within the 
multigrid cycle. Therefore the viscosity is updated only on the finest grid, which means that the procedure is not 
purely multigrid, but it has single-grid features. This technique was observed to slow down the multigrid convergence, 
but it makes the algorithm more robust and capable of achieving convergence up to high Bingham numbers (depending also 
on the value of $M$). Other measures that were found necessary in order to achieve convergence are the following: a 
large number of pre- and post-smoothing steps should be used (4 or more, depending on the value of $Bn$); a number of 
additional SIMPLE iterations (e.g.\ 5 -- 20) may have to be performed on the finest grid between multigrid cycles; very 
small values of the underrelaxation factor for pressure, denoted here by $a_p$, should be used in the SIMPLE smoother 
(e.g.\ 0.01), and in the case of high Reynolds numbers ($Re \geq 2000$) also relatively small values of the 
underrelaxation factor for velocity, $a_u$, (e.g.\ 0.3 -- 0.4) should be used; and the coarse grid corrections may have 
to be underrelaxed by a constant $\alpha_{MG} < 1$ prior to prolongation to the fine grid (usually $\alpha_{MG} \approx 
0.9$ suffices). We use W cycles, denoted by W($\nu_1$,$\nu_2$)-$\nu_3$, where $\nu_1$ SIMPLE iterations are performed 
prior to restriction, $\nu_2$ SIMPLE iterations are performed after prolongation, and $\nu_3$ extra SIMPLE iterations 
are applied only on the finest grid at the end of each cycle. More details can be found in \cite{Syrakos_13}. For more 
information on multigrid in general, the reader is referred to \cite{Brandt_77} or \cite{Trottenberg_01}.

As noted in \cite{Syrakos_13}, the SIMPLE/multigrid procedure becomes less efficient as either $Bn$ or $M$ increase, 
although the multigrid efficiency is always much higher than that of SIMPLE as a single grid solver. It has been 
observed that it is useful, or sometimes necessary, to use a good initial guess. This could be, for example, a solution 
on a coarser grid, or a solution obtained with a smaller value of $M$, both of which are more easily computable. In the 
present work we used mostly the former choice, but the latter choice was also useful in some ``difficult'' cases, and 
also led to the following idea: Instead of using a fully-converged lower $M$ solution as the initial guess, start with a 
very low value of $M$, say $M=1$, and progressively increase the value of $M$ every $n_M$, say, multigrid cycles, $n_M$ 
being a small constant of the order 1--4, until $M$ obtains its maximum value, beyond which point its value is held 
fixed until the multigrid cycles converge. This technique can increase the efficiency in some cases, as will be shown in 
the results section.

\subsection{Calculation of the truncation error}
\label{ssec: method truncation error}

The truncation error is the natural measure of the discrepancy between the integrals of the differential equations to 
be solved and their finite volume approximations. It consists of all the terms of the Taylor series expansions that 
were truncated in order to obtain the discrete finite volume approximations of the differential equations, and which 
have the form of products of powers of the grid spacing times higher order derivatives of the flow variables. Grid 
refinement reduces the truncation error, and therefore an efficient grid refinement strategy is to refine the grid 
locally where the truncation error is large, instead of applying uniform grid refinement throughout the domain. The 
truncation error is unknown, but can be approximated using various techniques. In the present study the method described 
in \cite{Syrakos_06a, Syrakos_12} is utilised, which originates from multigrid theory \cite{Brandt_77}.

Suppose $P$ is a finite volume of a grid with characteristic spacing $h$, $\phi$ is the unknown function (there could be 
more than one, for example in our case we have the two velocity components and pressure), and $N_P(\phi)~=~0$ is the 
equation obtained by integrating the differential equation over $P$ and dividing by its volume, while $N_{h,P}$ is the 
finite volume discrete approximation to $N_h$. The truncation error, $\zeta_P$, is defined by

\begin{equation} \label{eq: truncation error}
 \zeta_P \;=\; N_P(\phi) \;-\; N_{h,P}(\phi_h) 
\end{equation}
where $\phi_h$ is the vector of the values of $\phi$ at the centres of the finite volumes of grid $h$. The equation to 
be solved, $N_P(\phi)~=~0$, is equivalent to $N_{h,P}(\phi_h) + \zeta_P~=~0$ but since $\zeta_P$ is unknown the finite 
volume procedure solves $N_{h,P}(\tilde{\phi}_h) = 0$ instead, assuming that $\zeta_P$ is small enough to be neglected. 
The set of all such equations for all finite volumes forms an algebraic system which is solved using SIMPLE/multigrid in 
the present work. The solution $\phi^*_h$ thus obtained differs from the exact solution by the discretisation error 
$\epsilon_h = \phi_h - \phi^*_h$. 

If the exact solution $\phi$ were known, then the truncation error could be calculated from (\ref{eq: truncation 
error}). The exact solution is not known, but by using a solution on a finer grid, which is more accurate than 
$\tilde{\phi}_h$, an estimate of the truncation error can be obtained. This reasoning results in the following 
formula for estimating the truncation error, as is shown, for example, in \cite{Syrakos_12} or \cite{Fraysse_2012}:

\begin{equation} \label{eq: truncation error estimate}
 \zeta_h \approx -\frac{1}{2^p-1} \, I_{2h}^h \, N_{2h}(I_h^{2h}\tilde{\phi}_h)
\end{equation}
where now $\zeta_h$ is the vector of truncation errors in all finite volumes of grid $h$, $N_{2h}$ is the algebraic 
operator obtained with the same finite volume discretisation on a grid $2h$ which is twice as coarse as grid $h$, and 
$I_a^b$ are interpolation operators which transfer a grid function from grid $a$ to grid $b$. Also, $p$ is the order of 
the finite volume approximation, that is, it is the smallest power of the grid spacing  that appears among the terms 
that comprise the truncation error. This term reduces more slowly than the rest with grid refinement, and so at some 
point it becomes the dominant term of the truncation error; therefore $\zeta_h = O(h^p)$. The present finite volume 
method is second-order accurate, so $p=2$. In \cite{Syrakos_06a} it is demonstrated that, in the case of Newtonian 
flows, the present finite volume method works well with the estimate (\ref{eq: truncation error estimate}), which 
converges to the exact truncation error with grid refinement, provided that the restriction operator $I_h^{2h}$ is at 
least third-order accurate if $p=2$. If the truncation error estimate is used only as a local grid refinement criterion 
though, then its estimate need not be very accurate.

As the grid is refined, the truncation error converges to zero at a rate which is proportional to $h^p$, once the 
leading term has become much larger than the rest. However, the magnitude of the terms of the truncation error depends 
not only on the powers of $h$, but also on the higher-order derivatives of $\phi$. If the high-order derivatives are 
large, a very fine grid may be required for the leading term to become dominant and the truncation error to exhibit its 
asymptotic rate of convergence. In Bingham flows, these derivatives are discontinuous across the yield surfaces, since 
they are zero inside the unyielded zones and non-zero outside. If a regularised constitutive equation is used, then the 
derivatives are continuous, but they attain huge values near the yield surfaces, especially as the Bingham number and 
the parameter $M$ are increased. Therefore it is expected that the truncation error will be large there. In such cases, 
the truncation error can be reduced more efficiently by refining the grid locally at the high truncation error regions 
to counterbalance the high values of the derivatives, rather than using uniform grids. In \cite{Syrakos_12} it was 
shown that local grid refinement is very efficient in high-Reynolds number flows, which exhibit shear layers with large 
flow derivatives of high order. In the present Bingham flow case, the flow discontinuities are expected to make the 
gains from local refinement even more significant.

\subsection{Solution with local grid refinement}
\label{ssec: method with refinement}

\begin{figure}[b]
\centering
\includegraphics[scale=0.88]{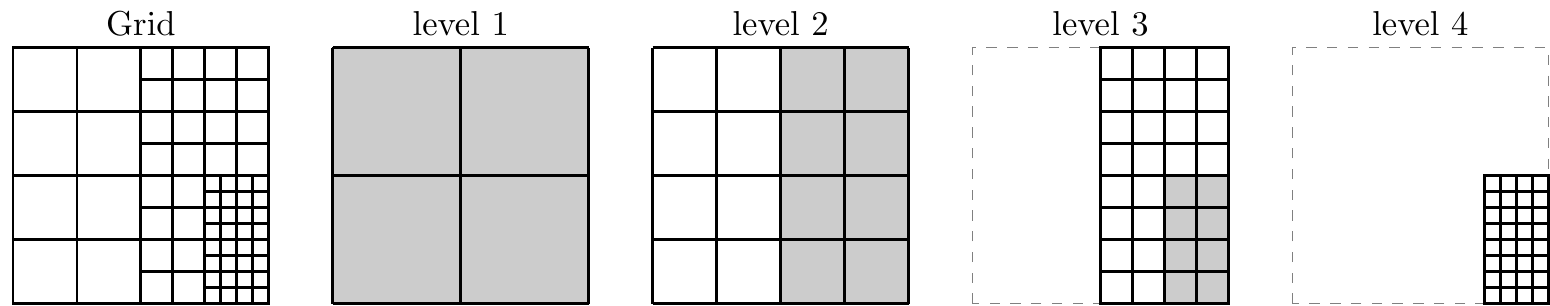}
\caption{An example of an organisation of a grid into levels. The local part of each level is shown in grey.}
\label{fig: grid levels}
\end{figure}

The local grid refinement scheme adopted here is that described in \cite{Syrakos_06b} and \cite{Syrakos_12}. After 
solving the problem on a given grid, those volumes which fulfil some criterion are marked for refinement. Refinement is 
performed by subdividing a volume (the \textit{parent}) into four smaller volumes (the \textit{children}) by joining the 
centre of the parent with the midpoints of its four faces. The volumes are organised into \textit{levels}, corresponding 
to the number of refinements performed to produce that particular volume. Therefore, if a parent volume is of level $k$ 
then its four children are of level $k+1$. Figure \ref{fig: grid levels} shows an example of the organisation of a 
locally refined grid into levels. When a volume is subdivided, its children are created and added to the data structure, 
but the parent is also retained in the data structure and not destroyed. Volumes that have children are characterised as 
\textit{local}; they have no effect on the final solution of the problem, but they are used by the multigrid, or more 
correctly \textit{multilevel}, procedure to accelerate algebraic convergence. Volumes that do not have children are 
characterised as \textit{global} and comprise the actual grid where the problem is solved.

\begin{figure}[t]
\centering
\includegraphics[scale=1.0]{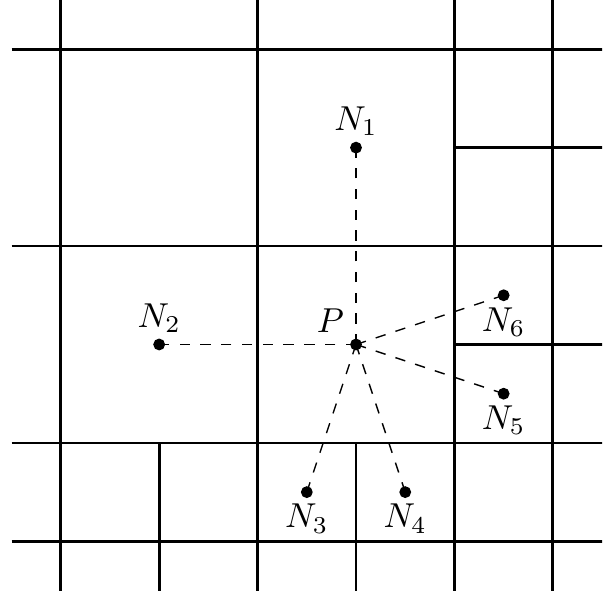}
\caption{On multilevel grids, volumes which lie at level interfaces may have more than four neighbours.}
\label{fig: level interface}
\end{figure}

The \textit{composite grid} consists of all global volumes of all levels, and it is the grid onto which the differential 
equations are actually discretised (shown on the left in Figure \ref{fig: grid levels}). Each volume is regarded as 
separated from its neighbouring volumes by its faces, and the momentum and mass fluxes through each face are discretised 
using central differences. Most volumes have four faces, but some volumes that are located at the interfaces between 
different grid levels may have more - for example volume $P$ of Figure \ref{fig: level interface} has six neighbours, 
and is separated from them by 6 corresponding faces. Despite the fact that, in the current study, all volumes have 
square shape, the central difference approximations of the fluxes through the faces which coincide with grid level 
interfaces would only be first-order accurate. This is because the line segment joining the centres of the volumes on 
either side of the face is not perpendicular to the face, is not bisected by the face, and does not pass through the 
face centre. To regain second-order accuracy, additional correction terms are incorporated into the central differencing 
scheme to account for these geometric irregularities. Full details of the discretisation scheme can be found in 
\cite{Syrakos_06b} or \cite{Syrakos_12}.

This discretisation procedure results in a non-linear algebraic system, which is formed using only the global volumes 
of each level. To solve this system, the local volumes of each level are also used, in order to accelerate algebraic 
convergence. The equations solved for these volumes are auxiliary equations which approximate the equations of the 
immediately finer grid, according to the multigrid philosophy. On the contrary, the equations solved for the global 
volumes of each level are the actual equations of the finite volume discretisation on the composite grid. The algorithm 
proceeds level-by-level; for example, if V-cycles were used, then the algorithm would proceed from the finest level down 
to the coarsest one, and then it would move up until the finest level. The fact that some levels do not extend 
throughout the domain is not a problem, as long as the mass and momentum fluxes through faces that separate global from 
local volumes are defined appropriately so that when the solution has been attained on the composite grid, and the 
residuals are zero, the multilevel algorithm does not produce any corrections. The full details of the algorithm can be 
found in \cite{Syrakos_06b}.

Using the experience gained in \cite{Syrakos_12}, we use the volume integral of the truncation error over each finite 
volume as the refinement criterion (it is calculated by multiplying the local truncation error estimate by the volume). 
In particular, after solving the equations on a given grid, this quantity is calculated at each volume of that grid. 
Then, the volumes are ordered according to the magnitude of this quantity, from highest to lowest. The 20\% of the 
volumes at the top of this list are selected for refinement. This selection procedure is performed for the $x-$ and 
$y-$momentum equations, but not for the continuity equation. The union of the two sets of volumes selected through the 
two momentum equations is the set of volumes which are refined. This results in a new composite grid, where the 
equations are again solved to obtain a more accurate solution than on the previous grid. The procedure can be repeated 
to obtain even more refined grids as many times as one wishes.

\clearpage
\begin{figure}[p]
\centering
\noindent\makebox[\textwidth]{
 \subfigure[{$Re = 0$}] {\label{sfig: streamlines Bn=0 Re=0}
  \includegraphics[scale=1.00]{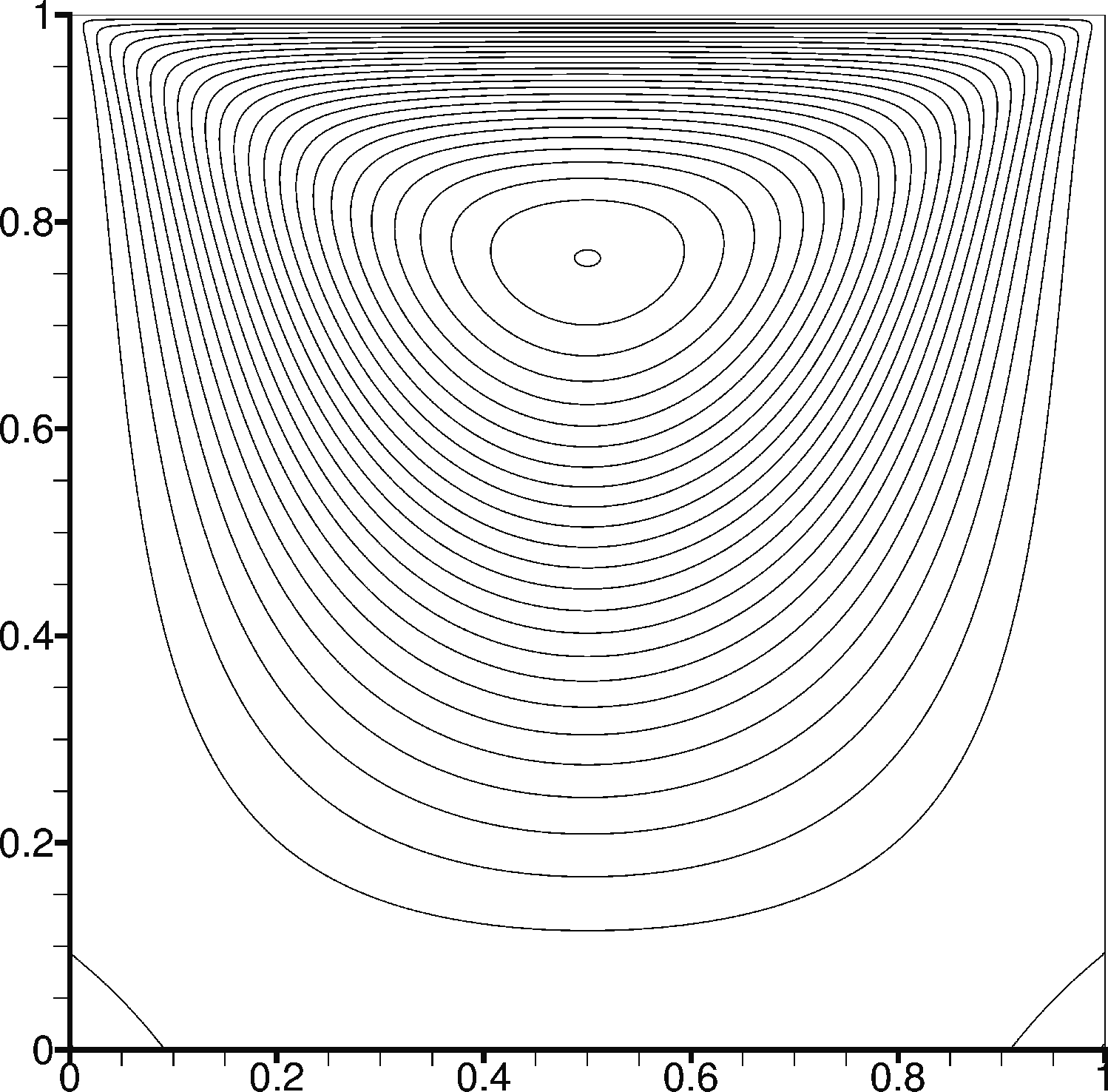}}
 \subfigure[{$Re = 1$}] {\label{sfig: streamlines Bn=0 Re=1}
  \includegraphics[scale=1.00]{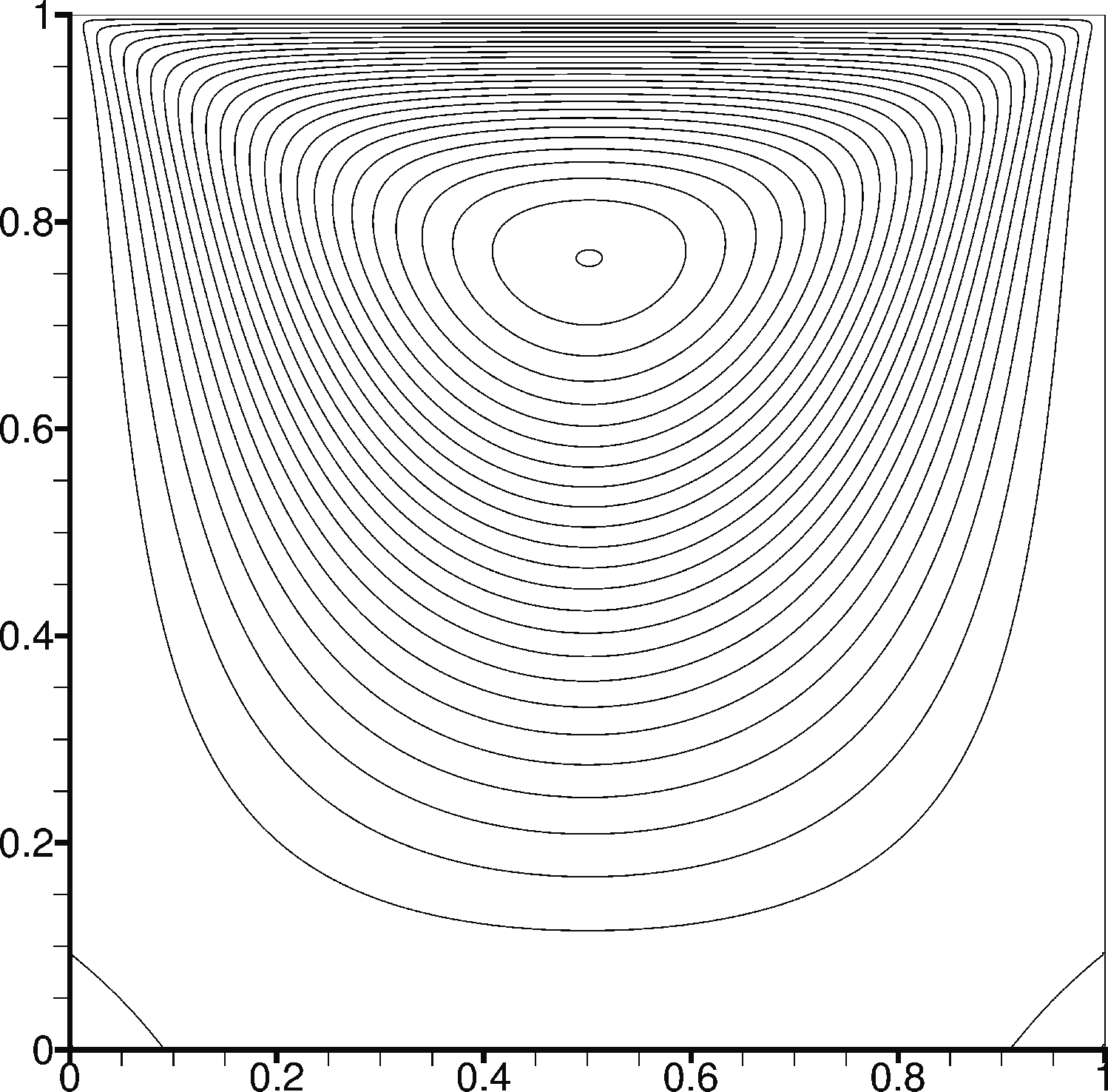}}
}
\noindent\makebox[\textwidth]{
 \subfigure[{$Re = 10$}] {\label{sfig: streamlines Bn=0 Re=10}
  \includegraphics[scale=1.00]{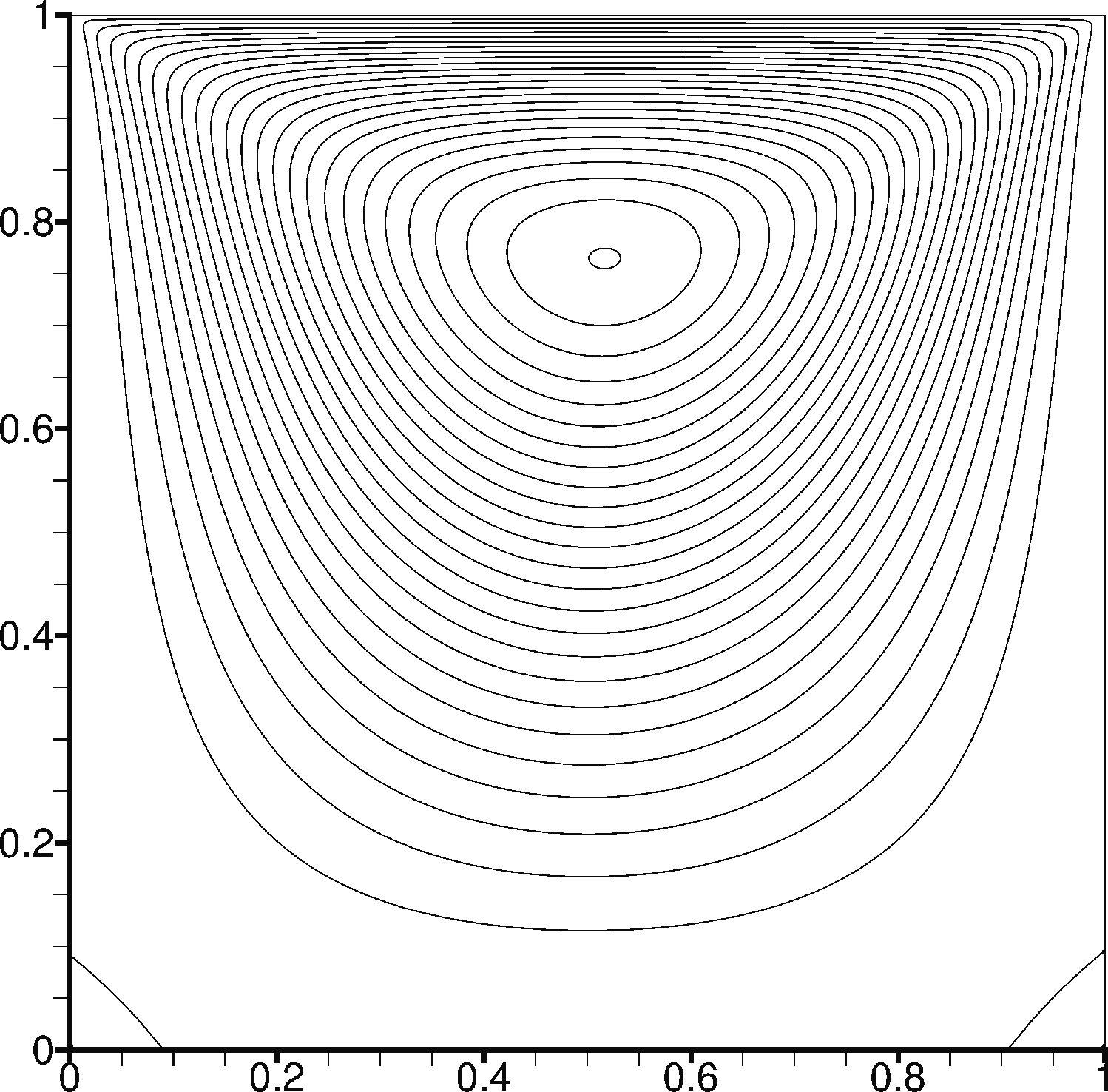}}
 \subfigure[{$Re = 100$}] {\label{sfig: streamlines Bn=0 Re=100}
  \includegraphics[scale=1.00]{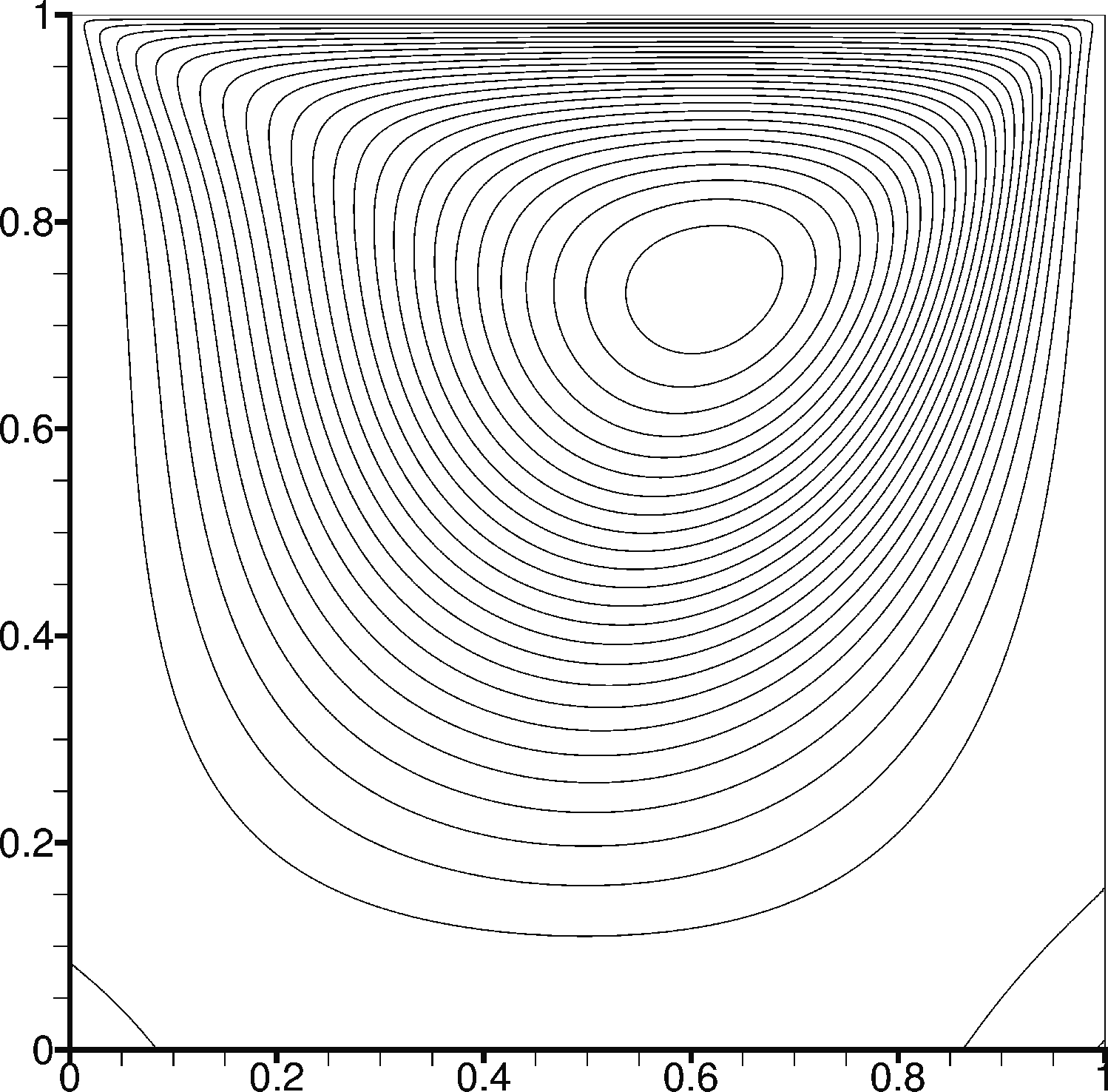}}
}
\noindent\makebox[\textwidth]{
 \subfigure[{$Re = 500$}] {\label{sfig: streamlines Bn=0 Re=500}
  \includegraphics[scale=1.00]{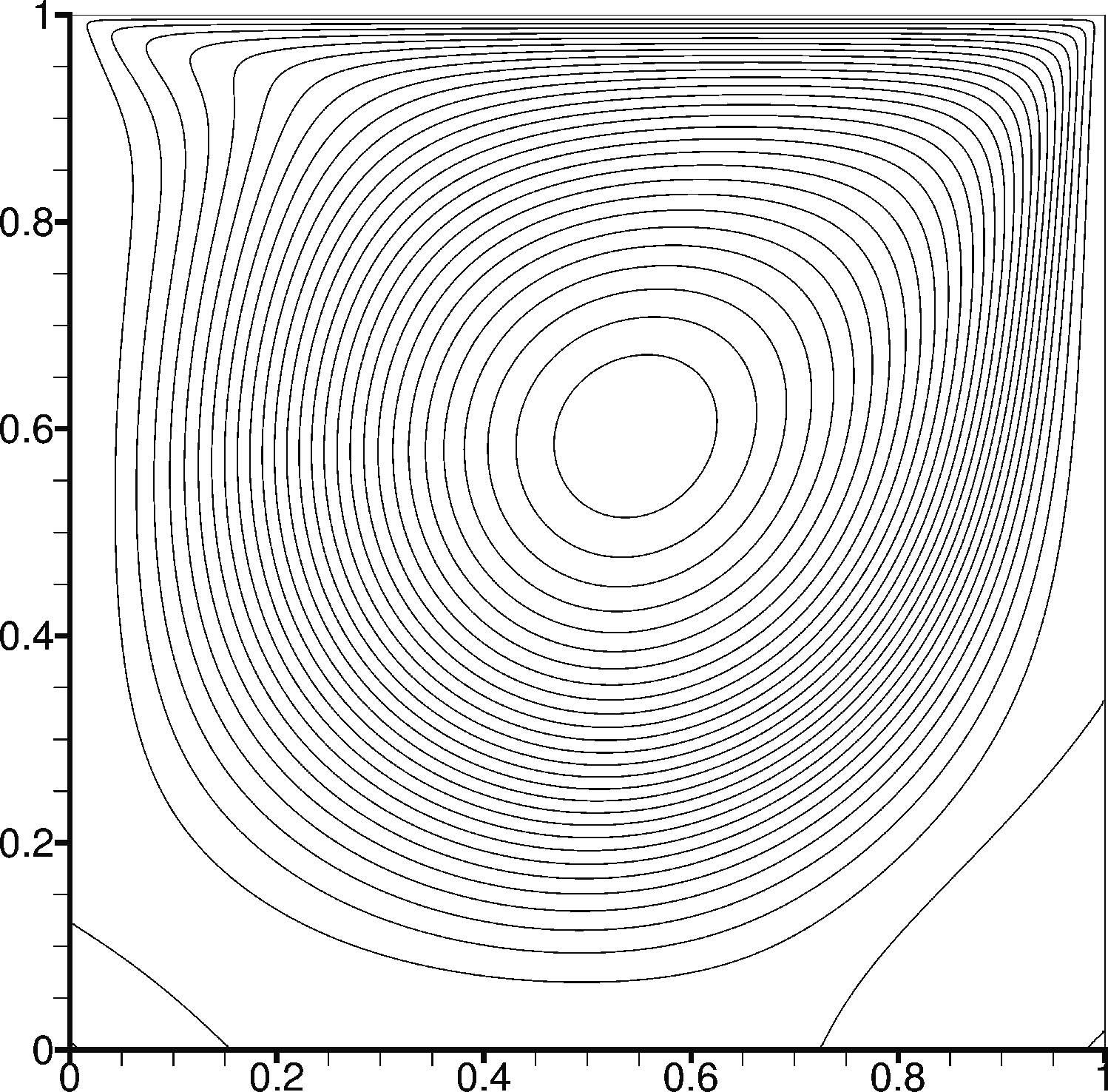}}
 \subfigure[{$Re = 1000$}] {\label{sfig: streamlines Bn=0 Re=1000}
  \includegraphics[scale=1.00]{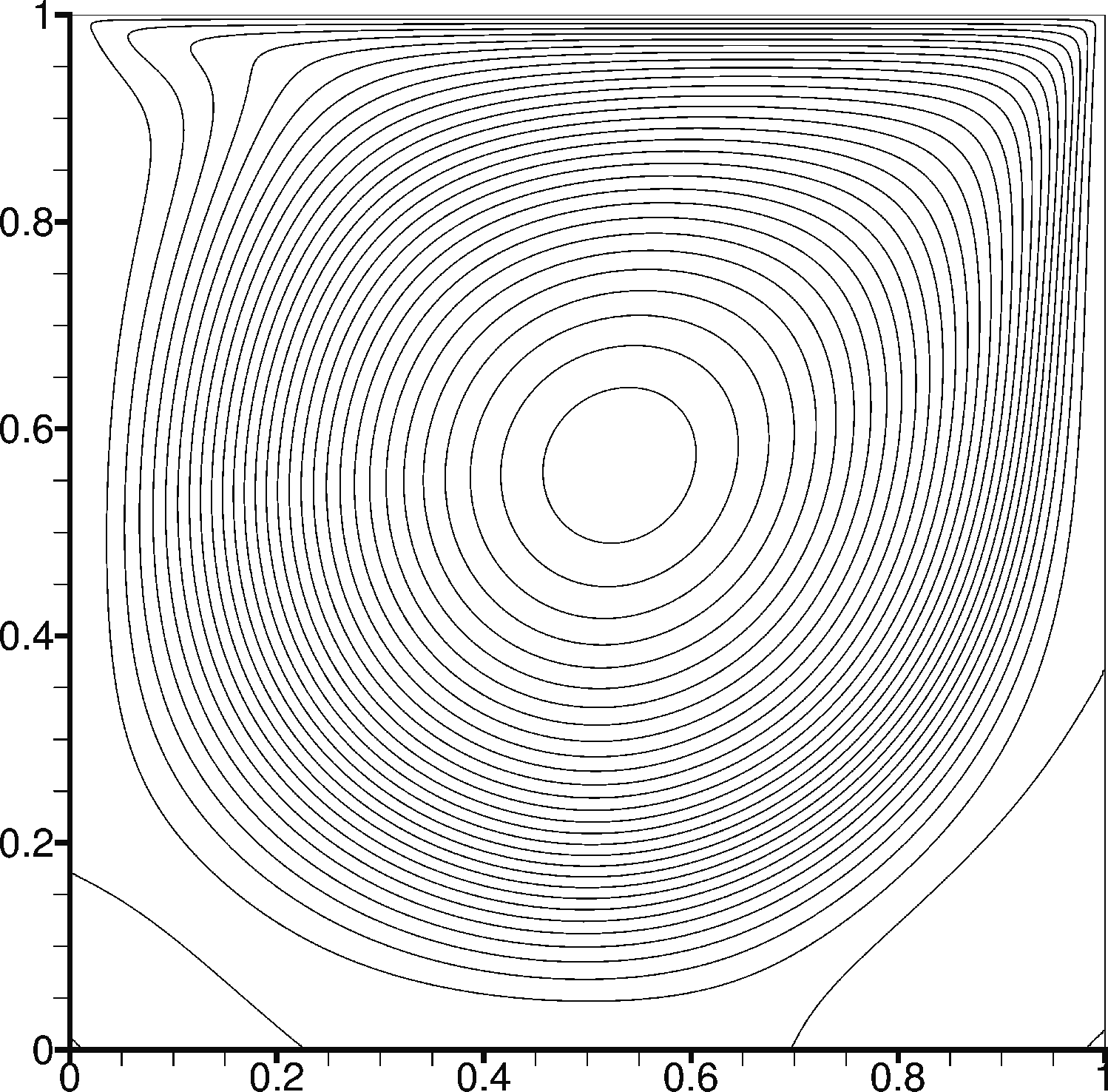}}
}
\caption{Streamlines in Newtonian flow ($Bn$ = 0), plotted at intervals of 0.004 starting from zero.}
\label{fig: streamlines Bn=0}
\end{figure}

\clearpage
\begin{figure}[p]
\centering
\noindent\makebox[\textwidth]{
 \subfigure[{$Re = 0$}] {\label{sfig: streamlines Bn=1 Re=0}
  \includegraphics[scale=1.00]{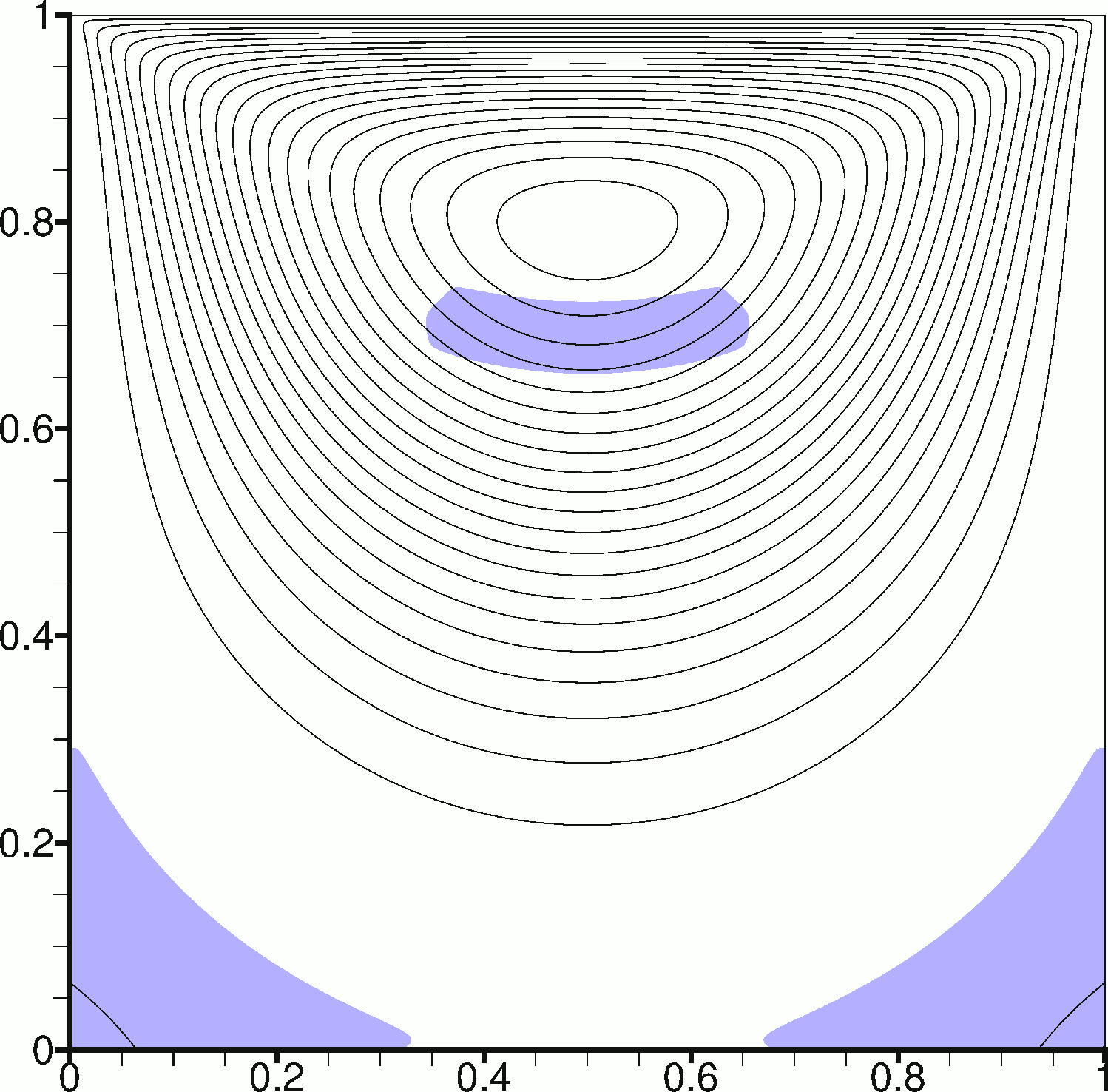}}
 \subfigure[{$Re = 1$}] {\label{sfig: streamlines Bn=1 Re=1}
  \includegraphics[scale=1.00]{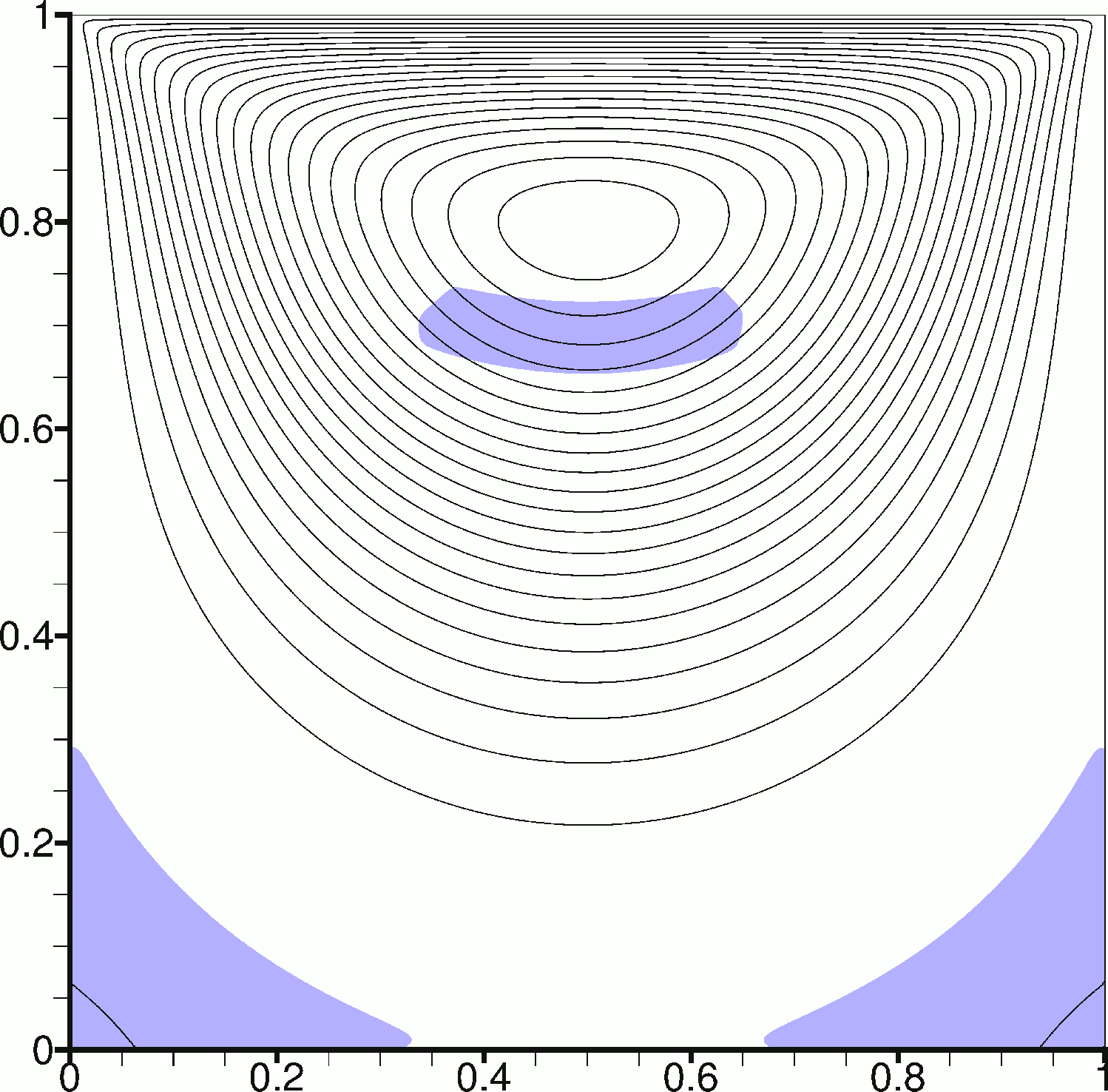}}
}
\noindent\makebox[\textwidth]{
 \subfigure[{$Re = 10$}] {\label{sfig: streamlines Bn=1 Re=10}
  \includegraphics[scale=1.00]{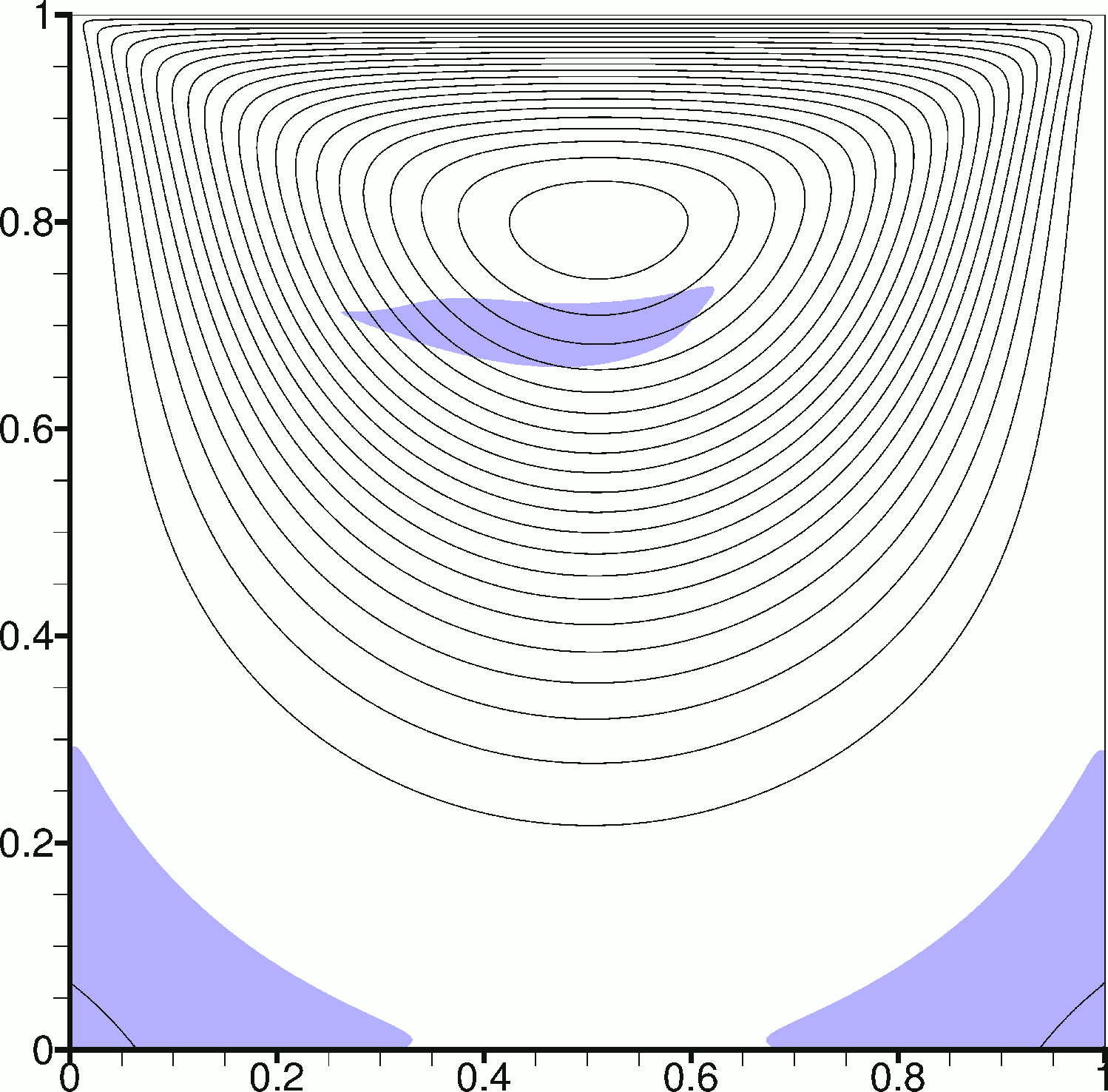}}
 \subfigure[{$Re = 100$}] {\label{sfig: streamlines Bn=1 Re=100}
  \includegraphics[scale=1.00]{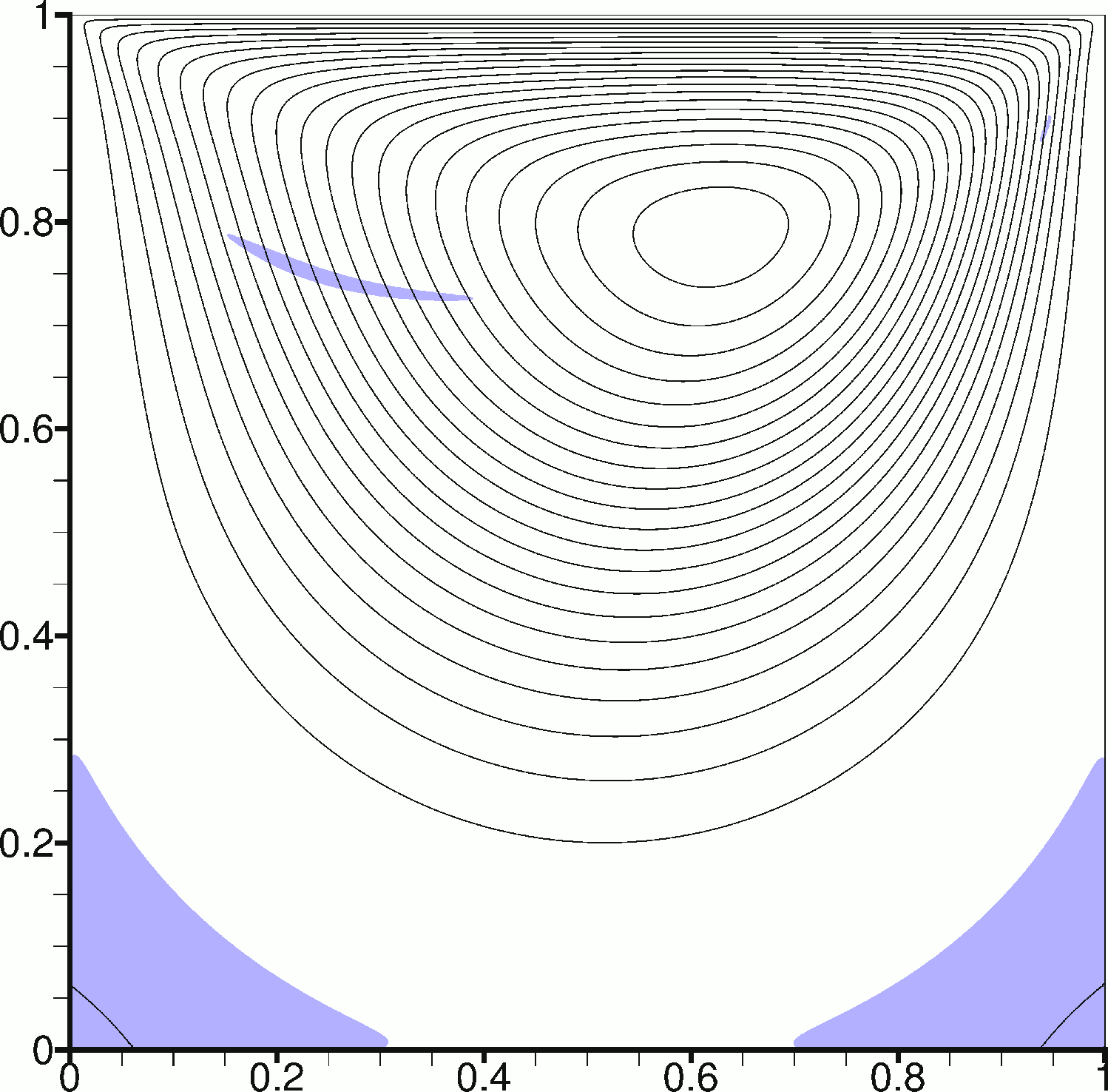}}
}
\noindent\makebox[\textwidth]{
 \subfigure[{$Re = 500$}] {\label{sfig: streamlines Bn=1 Re=500}
  \includegraphics[scale=1.00]{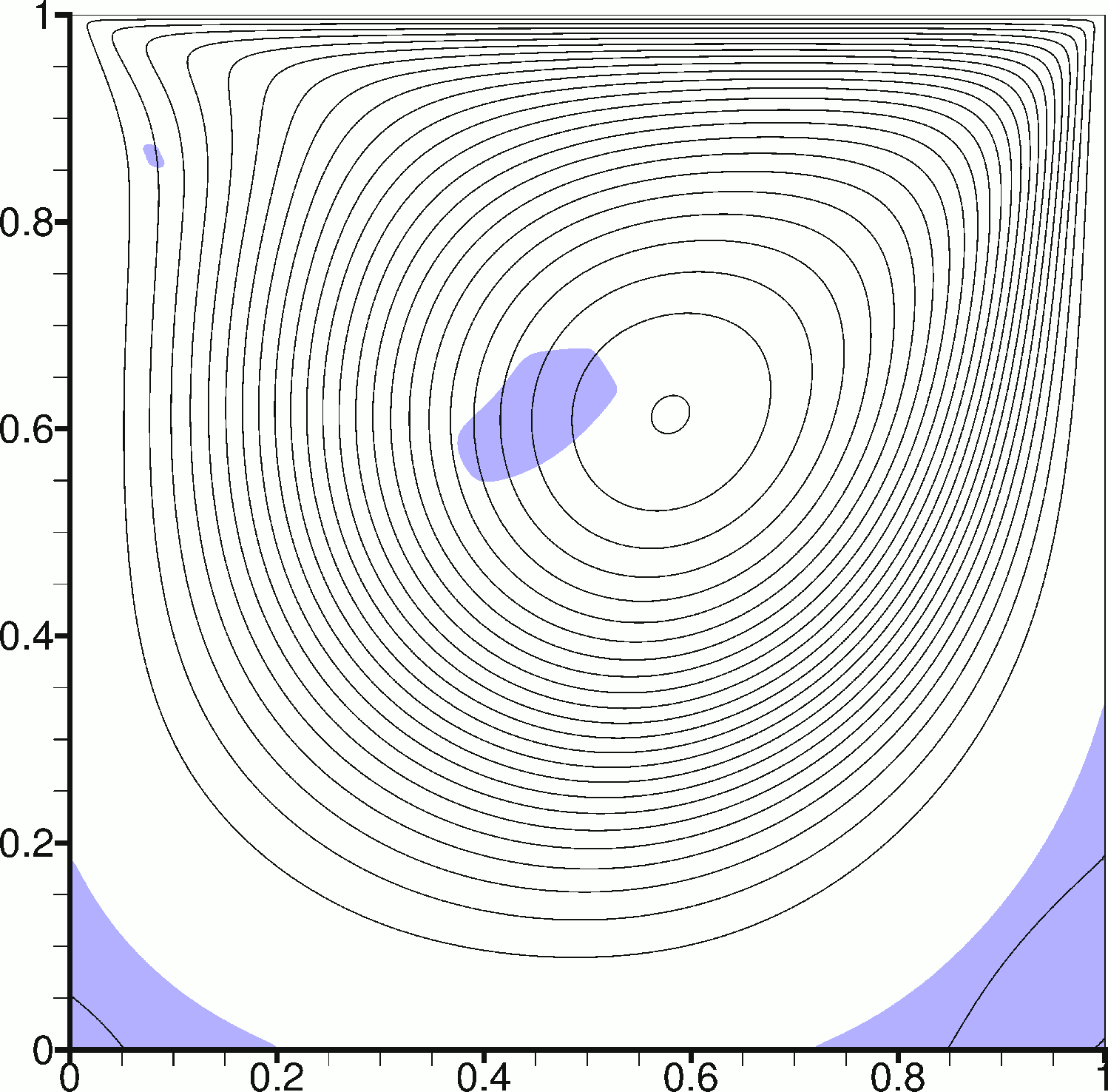}}
 \subfigure[{$Re = 1000$}] {\label{sfig: streamlines Bn=1 Re=1000}
  \includegraphics[scale=1.00]{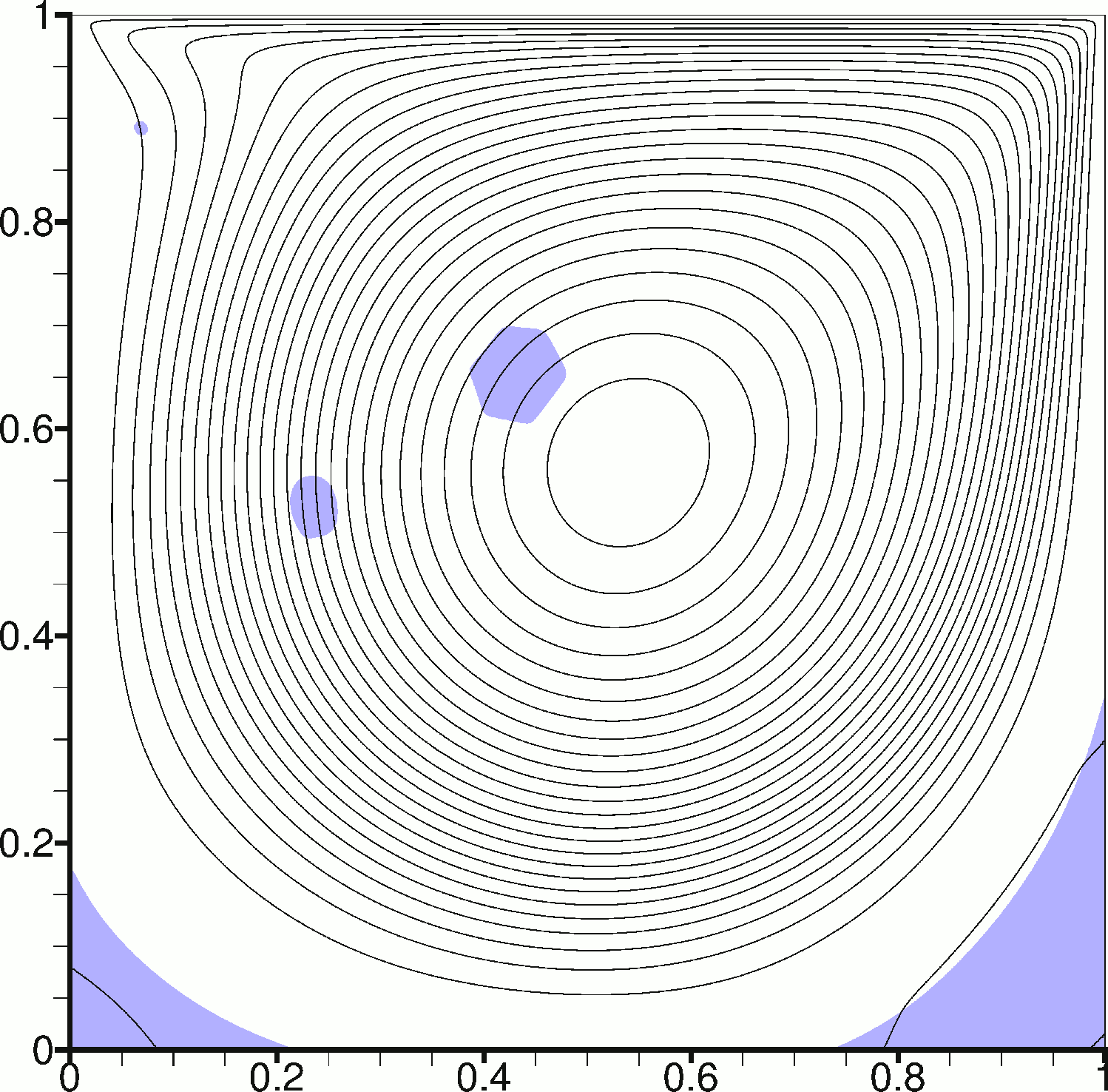}}
}
\caption{Streamlines in Bingham flow for $Bn$ = 1, plotted at intervals of 0.004 starting from zero. Unyielded areas 
($\tau < Bn$) are shown shaded.}
\label{fig: streamlines Bn=1}
\end{figure}

\clearpage
\begin{figure}[p]
\centering
\noindent\makebox[\textwidth]{
 \subfigure[{$Re = 0$}] {\label{sfig: streamlines Bn=10 Re=0}
  \includegraphics[scale=1.00]{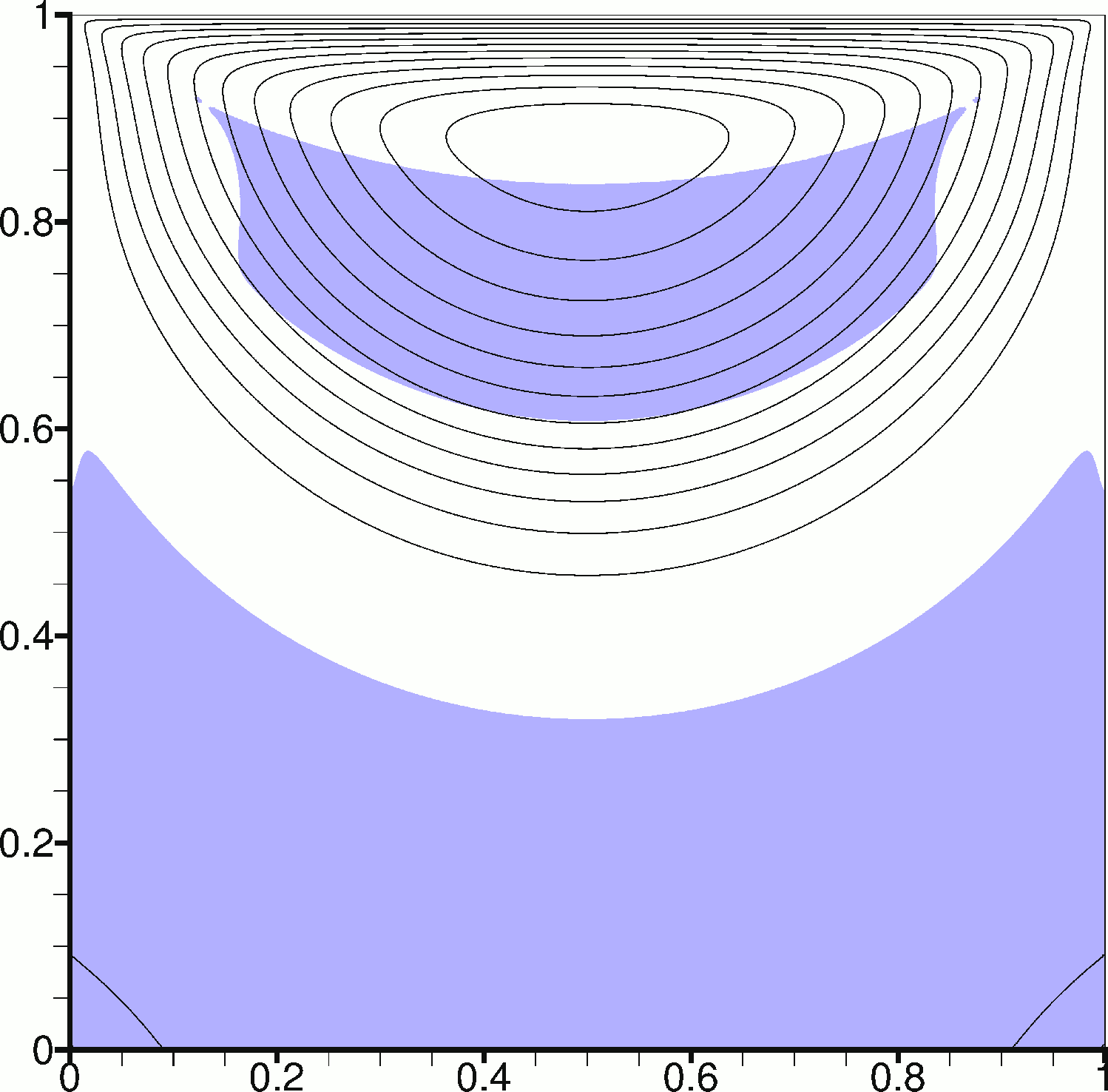}}
 \subfigure[{$Re = 1$}] {\label{sfig: streamlines Bn=10 Re=1}
  \includegraphics[scale=1.00]{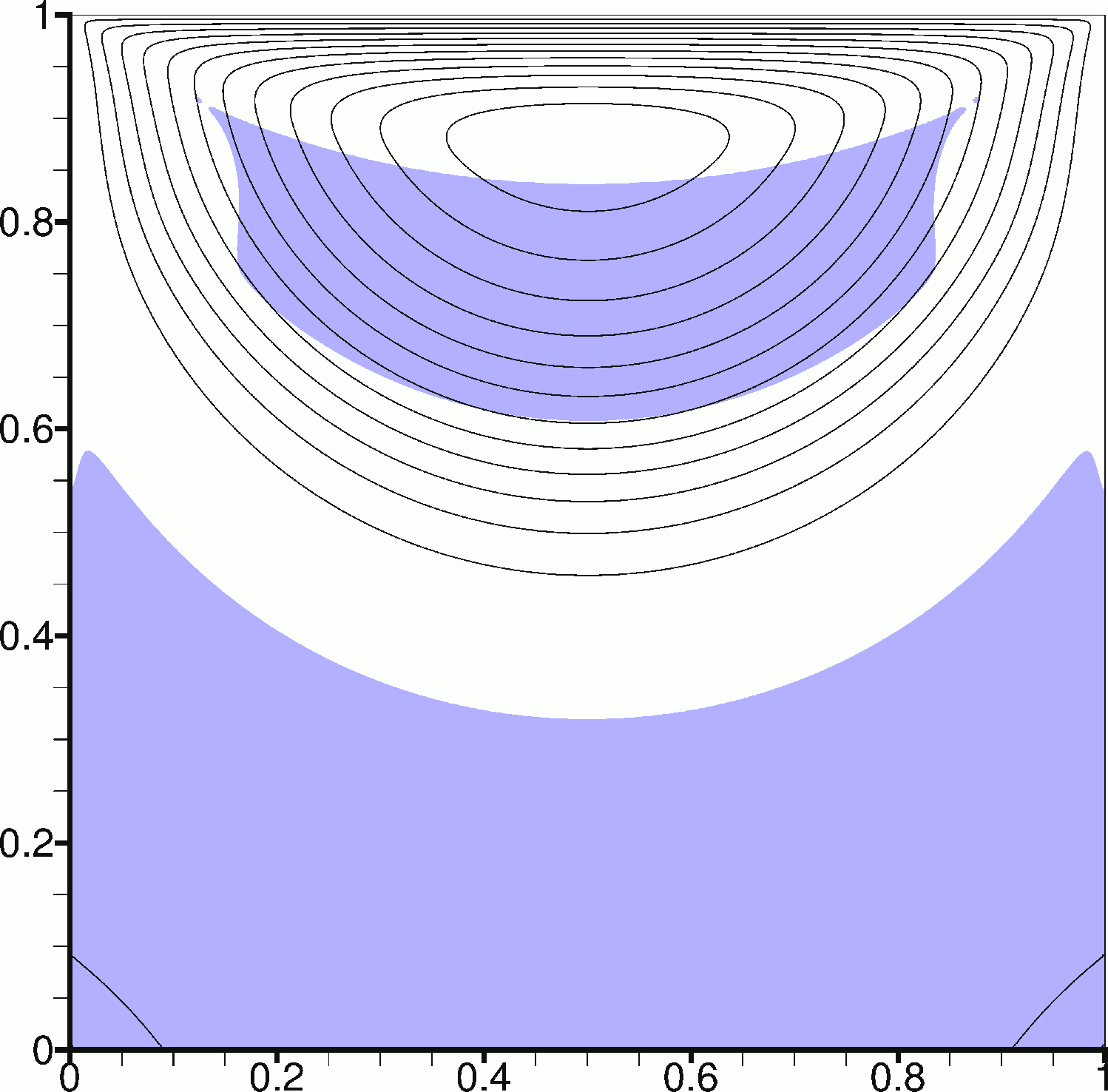}}
}
\noindent\makebox[\textwidth]{
 \subfigure[{$Re = 10$}] {\label{sfig: streamlines Bn=10 Re=10}
  \includegraphics[scale=1.00]{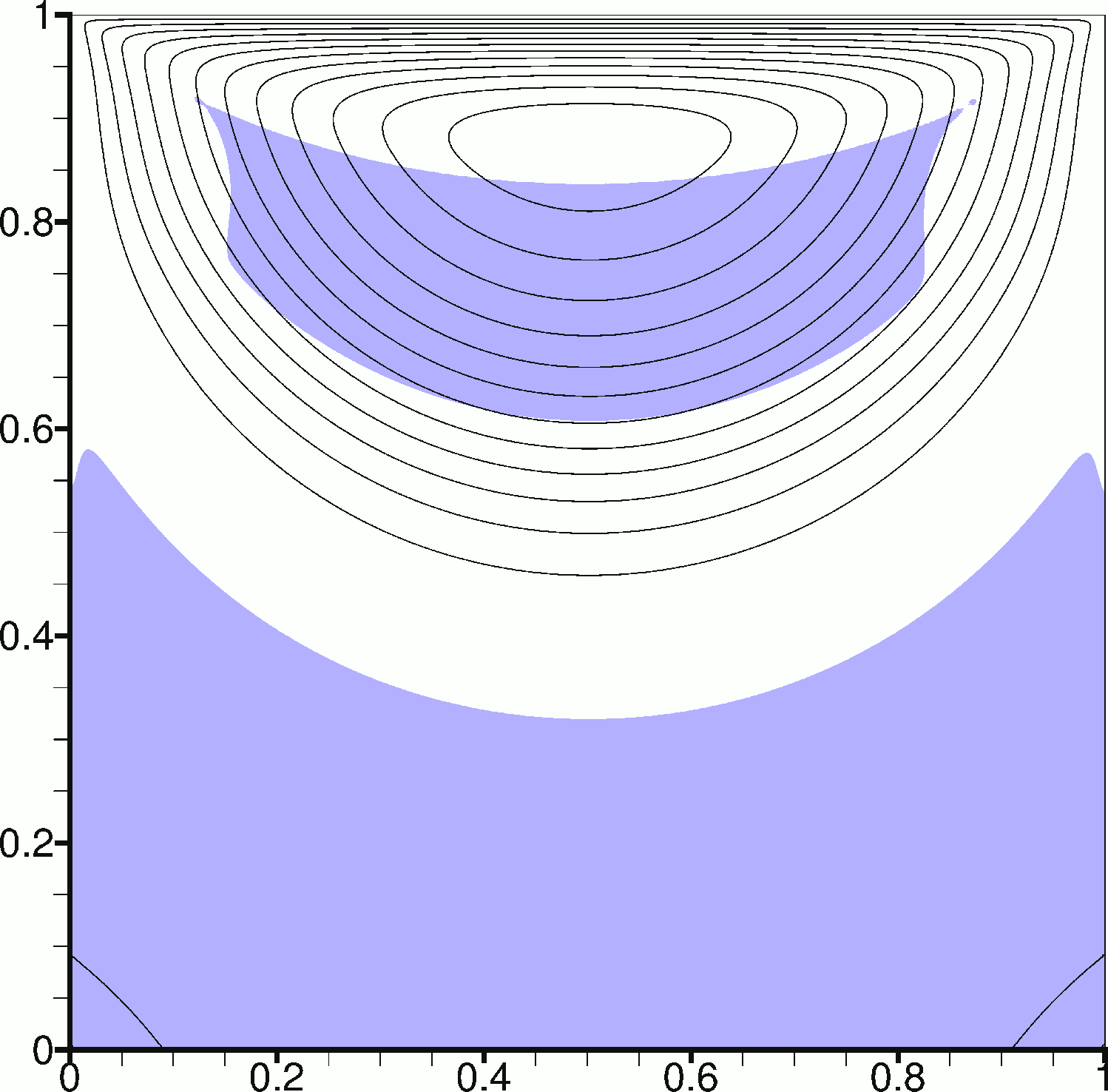}}
 \subfigure[{$Re = 100$}] {\label{sfig: streamlines Bn=10 Re=100}
  \includegraphics[scale=1.00]{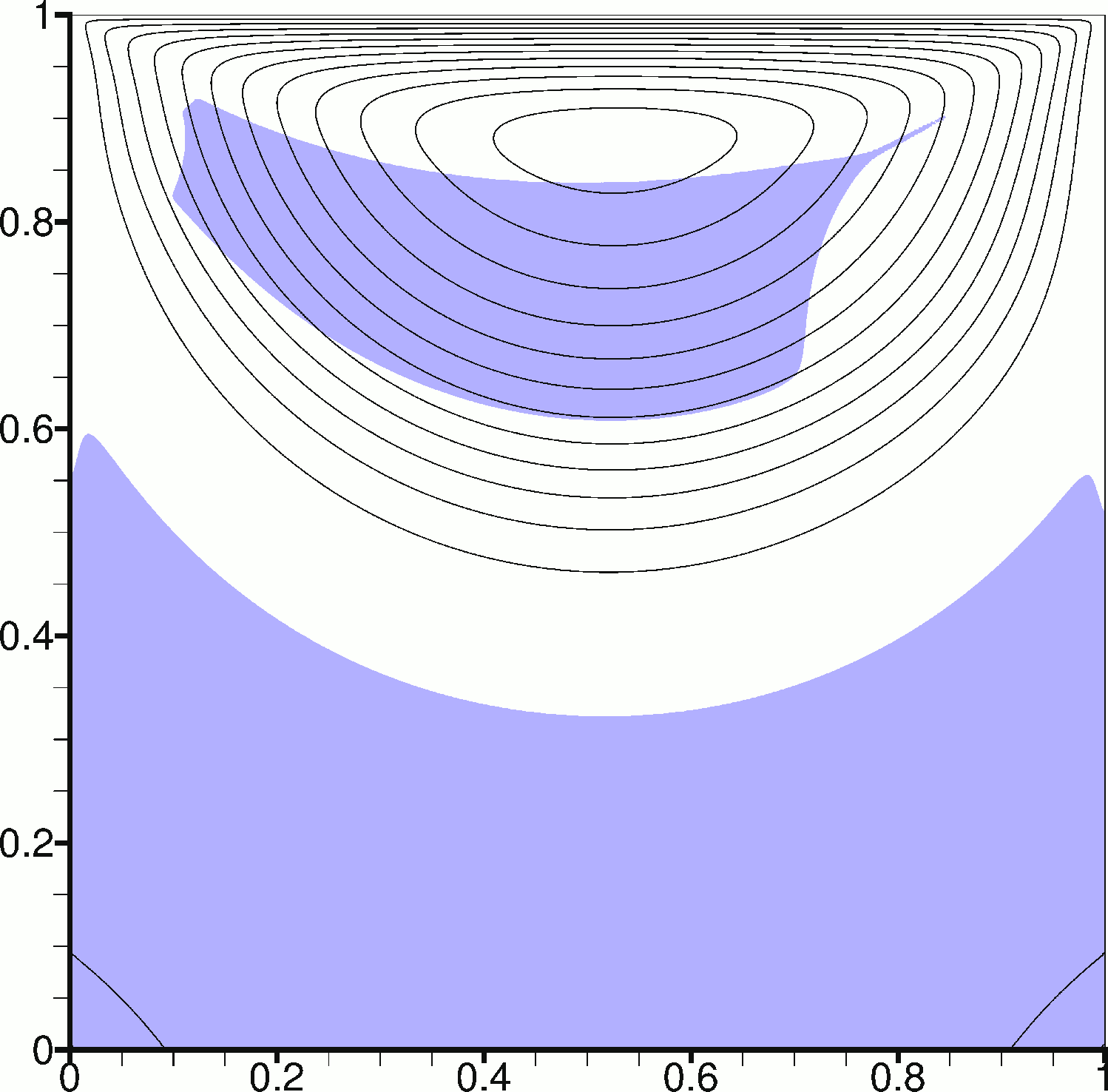}}
}
\noindent\makebox[\textwidth]{
 \subfigure[{$Re = 500$}] {\label{sfig: streamlines Bn=10 Re=500}
  \includegraphics[scale=1.00]{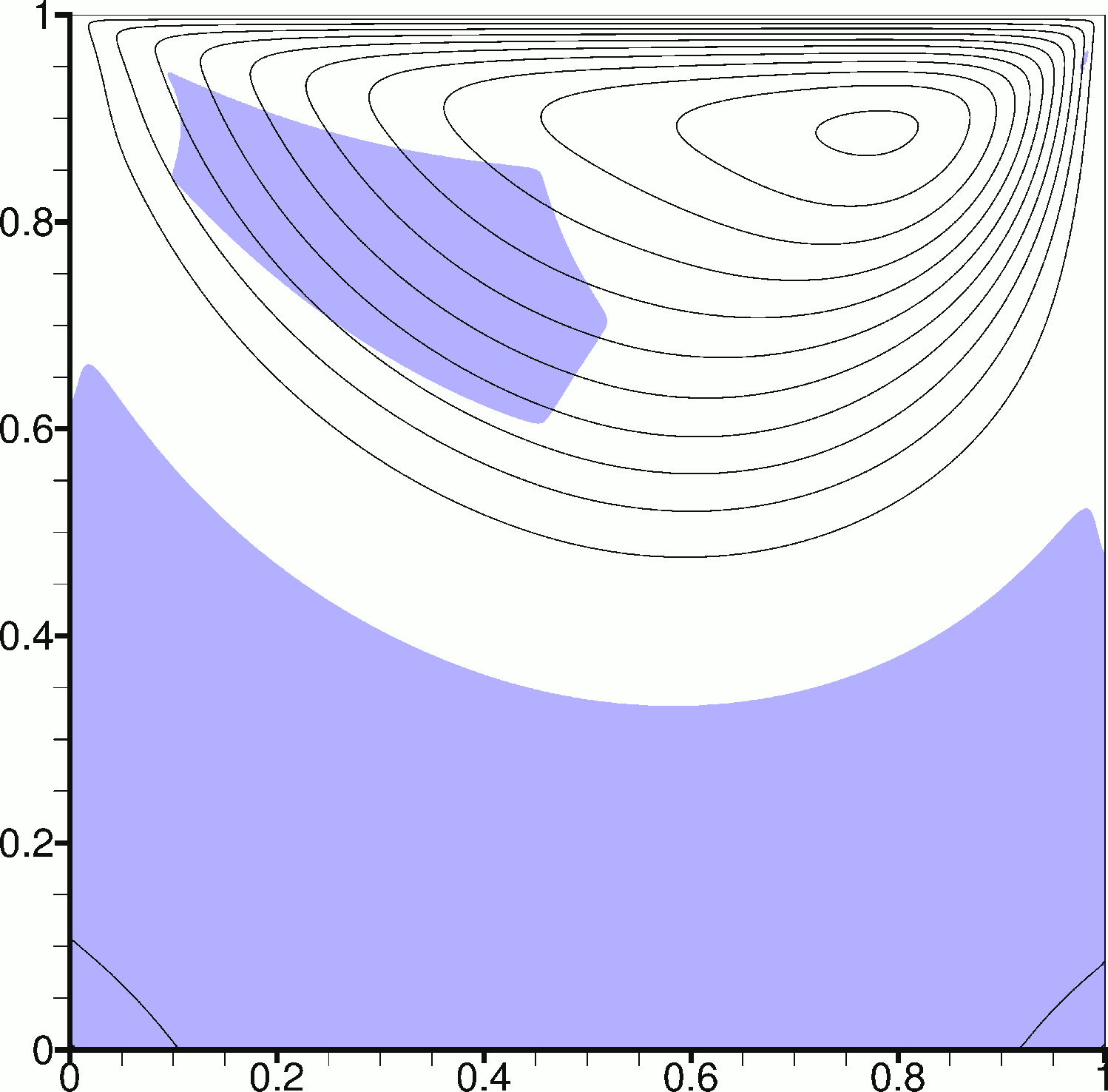}}
 \subfigure[{$Re = 1000$}] {\label{sfig: streamlines Bn=10 Re=1000}
  \includegraphics[scale=1.00]{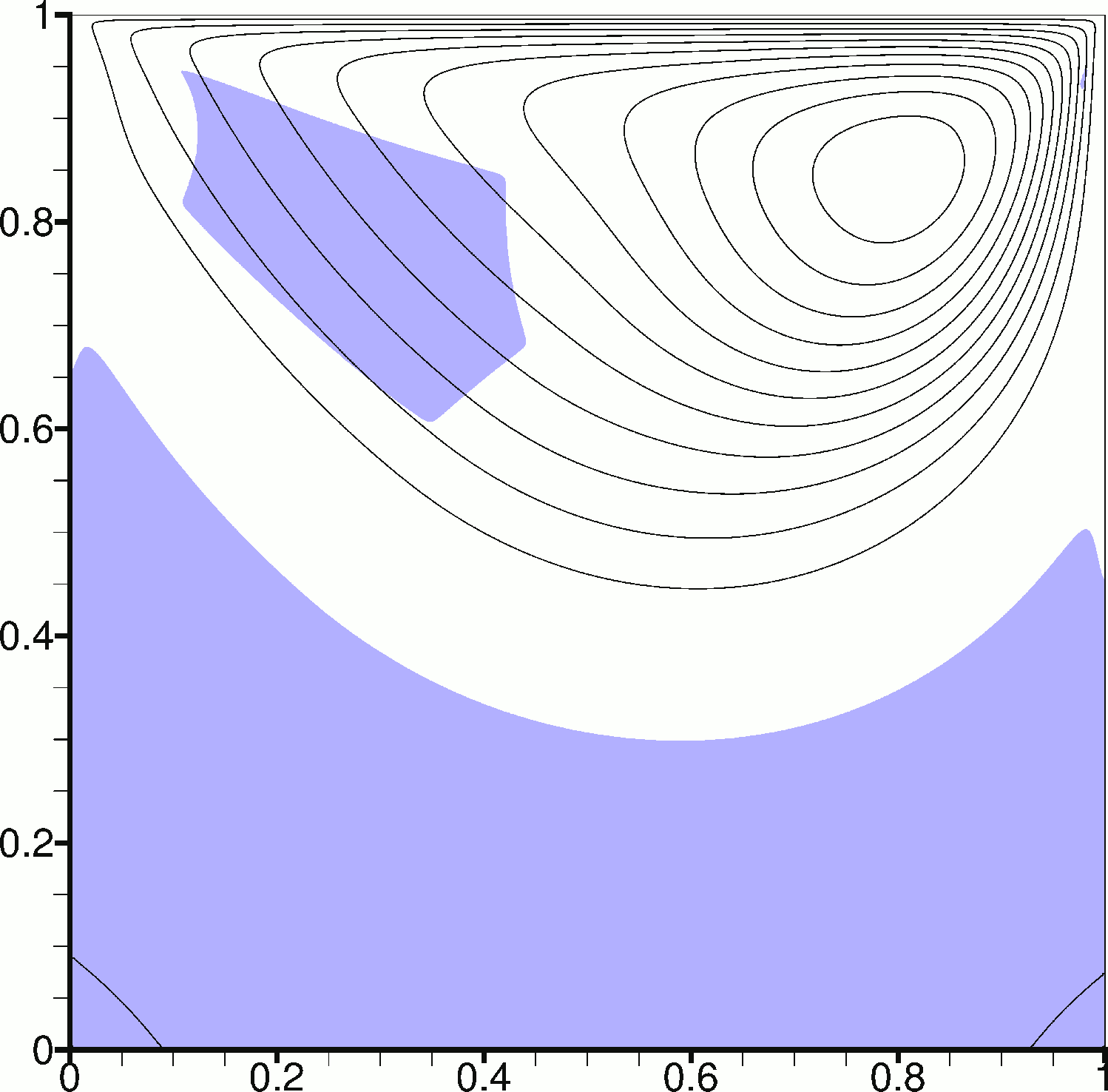}}
}
\caption{Streamlines in Bingham flow for $Bn$ = 10, plotted at intervals of 0.004 starting from zero. Unyielded areas 
($\tau < Bn$) are shown shaded.}
\label{fig: streamlines Bn=10}
\end{figure}

\clearpage
\begin{figure}[p]
\centering
\noindent\makebox[\textwidth]{
 \subfigure[{$Re = 0$}] {\label{sfig: streamlines Bn=100 Re=0}
  \includegraphics[scale=1.00]{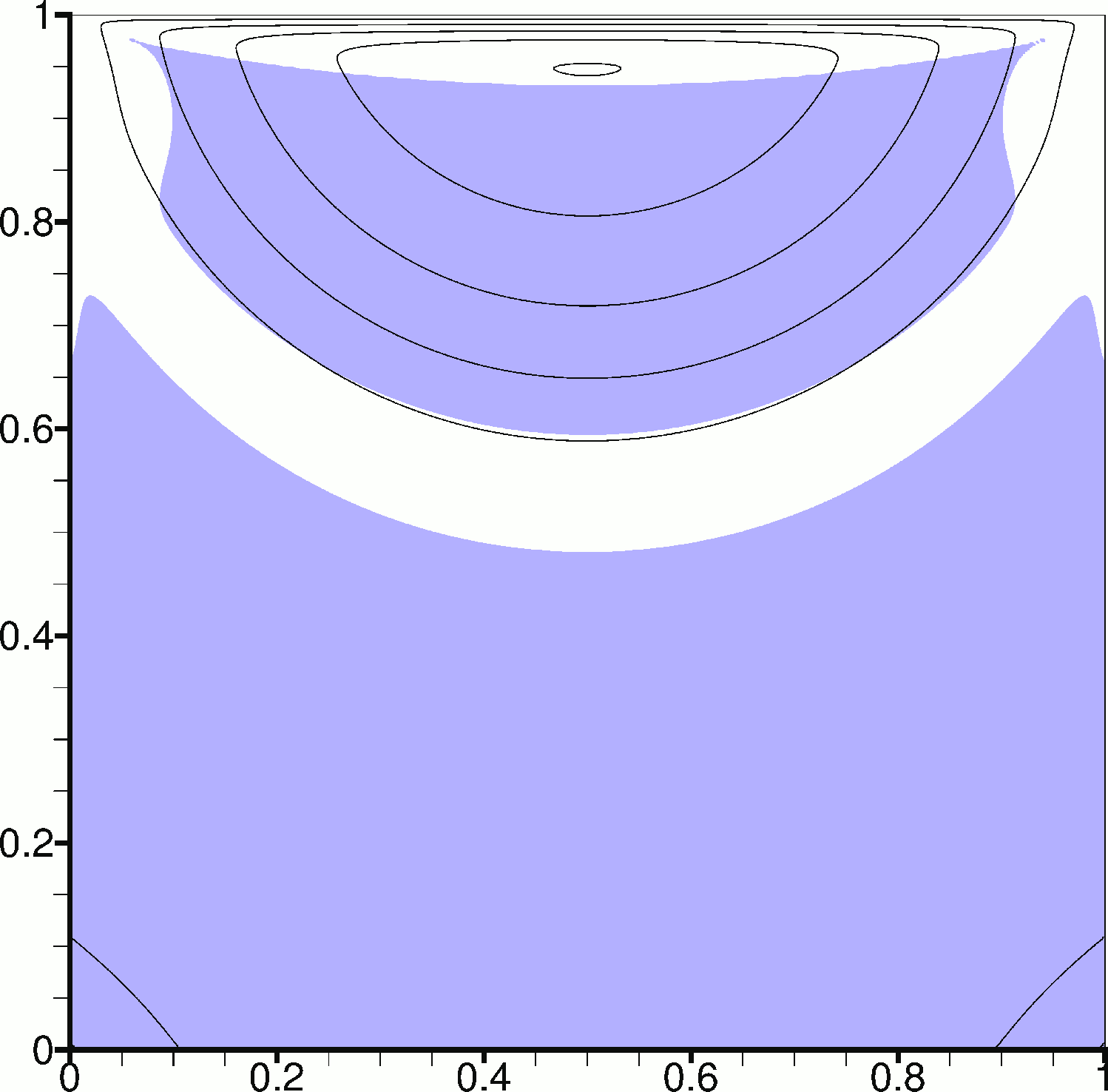}}
 \subfigure[{$Re = 1$}] {\label{sfig: streamlines Bn=100 Re=1}
  \includegraphics[scale=1.00]{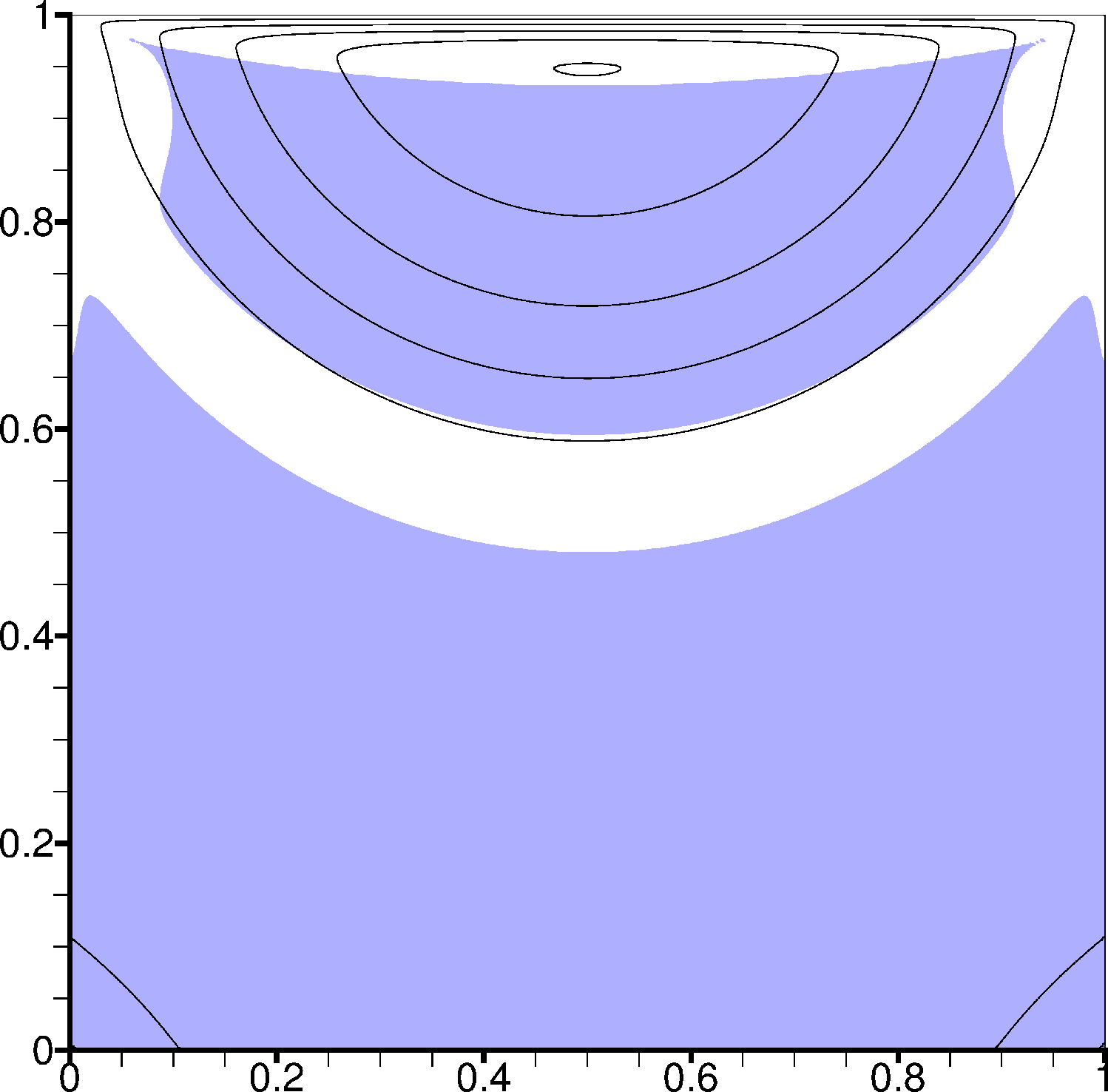}}
}
\noindent\makebox[\textwidth]{
 \subfigure[{$Re = 10$}] {\label{sfig: streamlines Bn=100 Re=10}
  \includegraphics[scale=1.00]{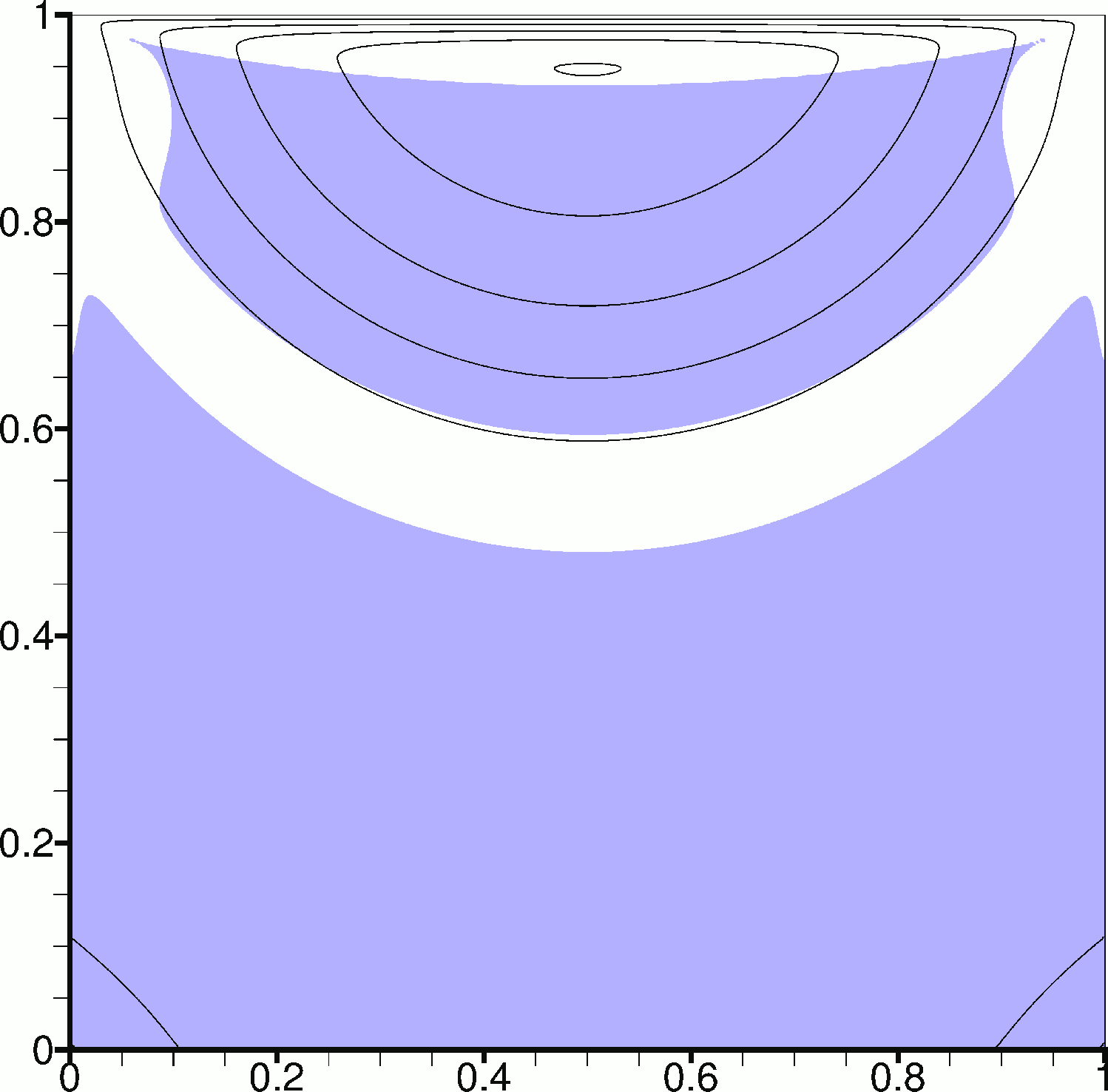}}
 \subfigure[{$Re = 100$}] {\label{sfig: streamlines Bn=100 Re=100}
  \includegraphics[scale=1.00]{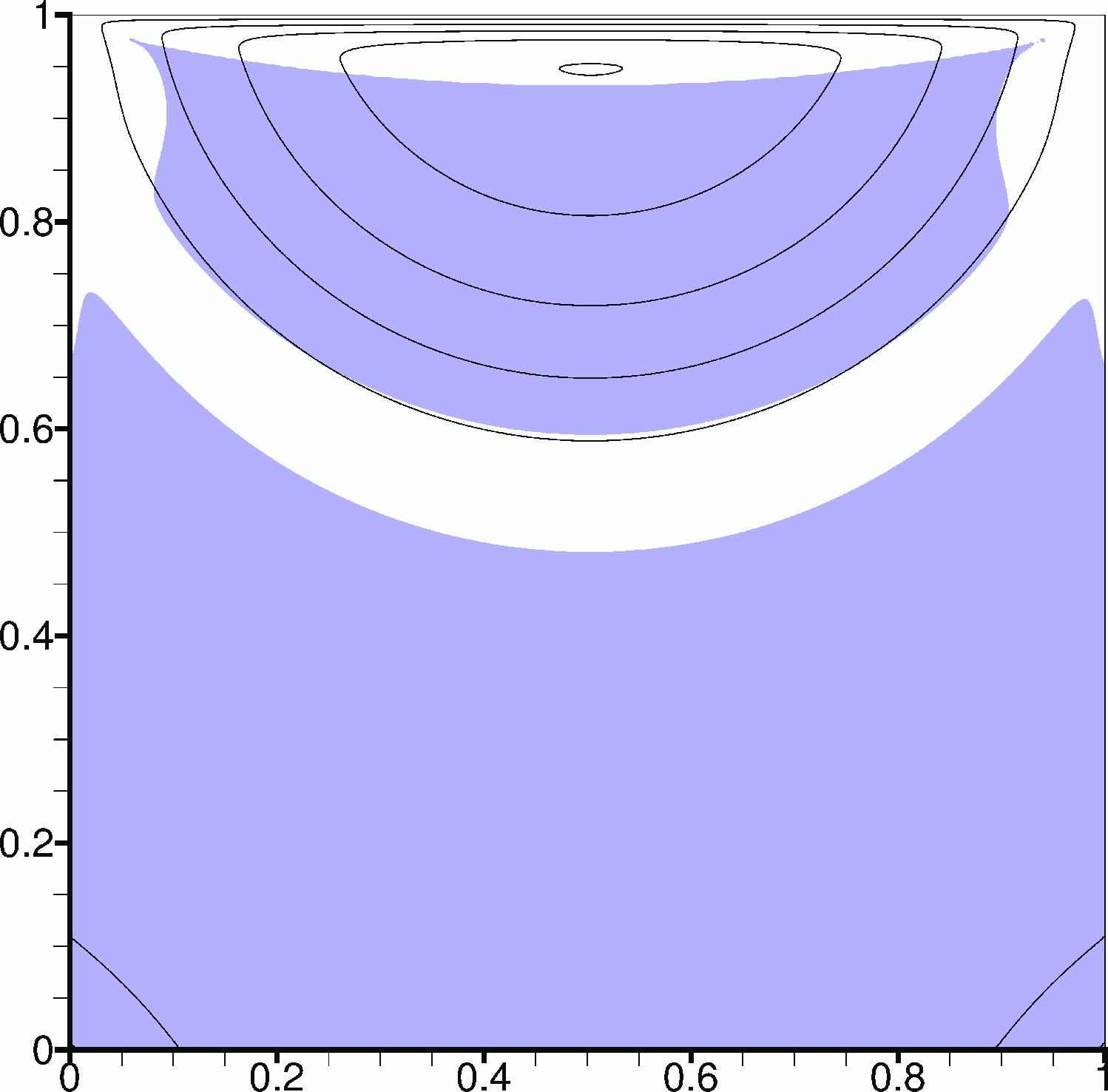}}
}
\noindent\makebox[\textwidth]{
 \subfigure[{$Re = 500$}] {\label{sfig: streamlines Bn=100 Re=500}
  \includegraphics[scale=1.00]{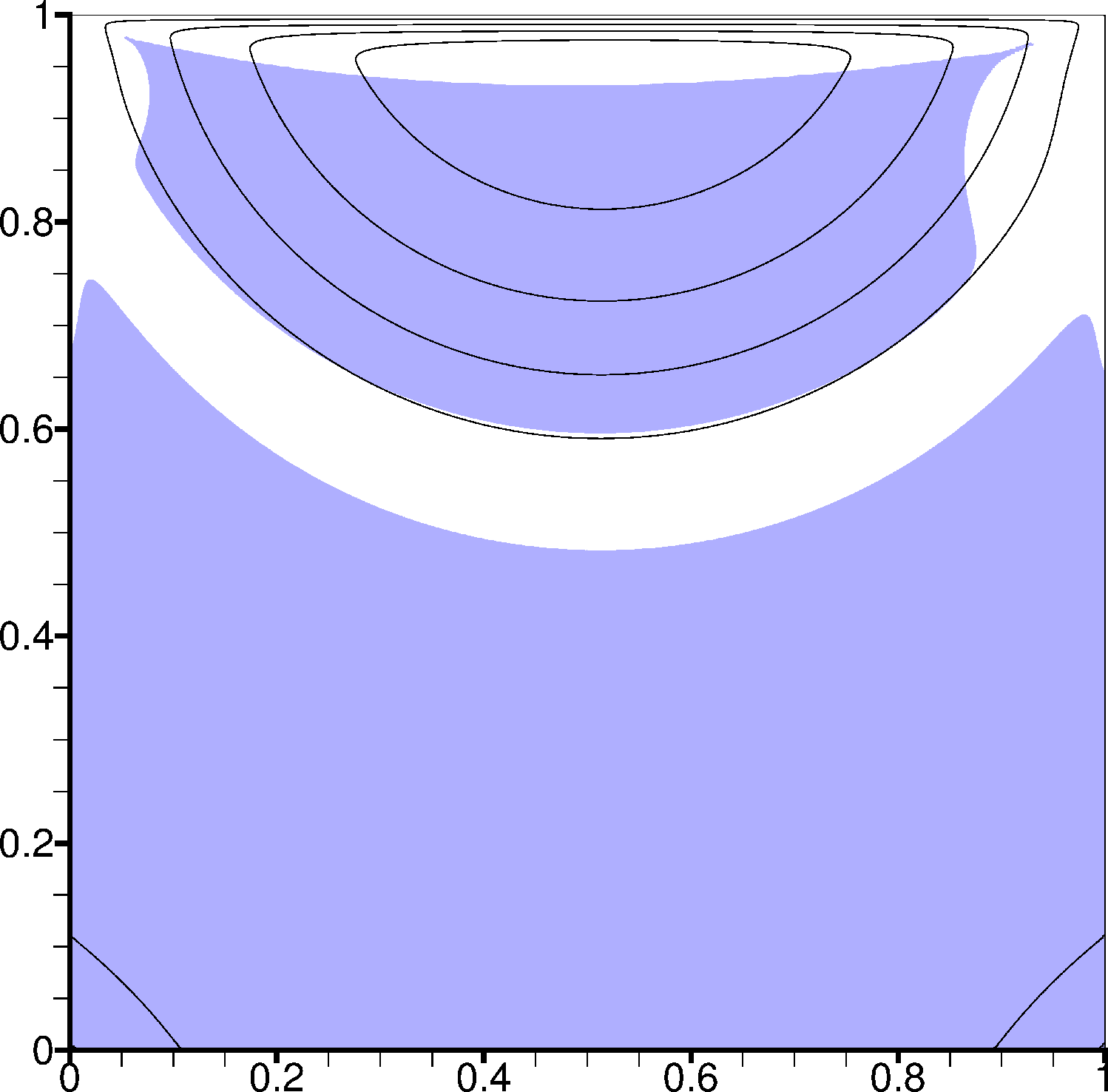}}
 \subfigure[{$Re = 1000$}] {\label{sfig: streamlines Bn=100 Re=1000}
  \includegraphics[scale=1.00]{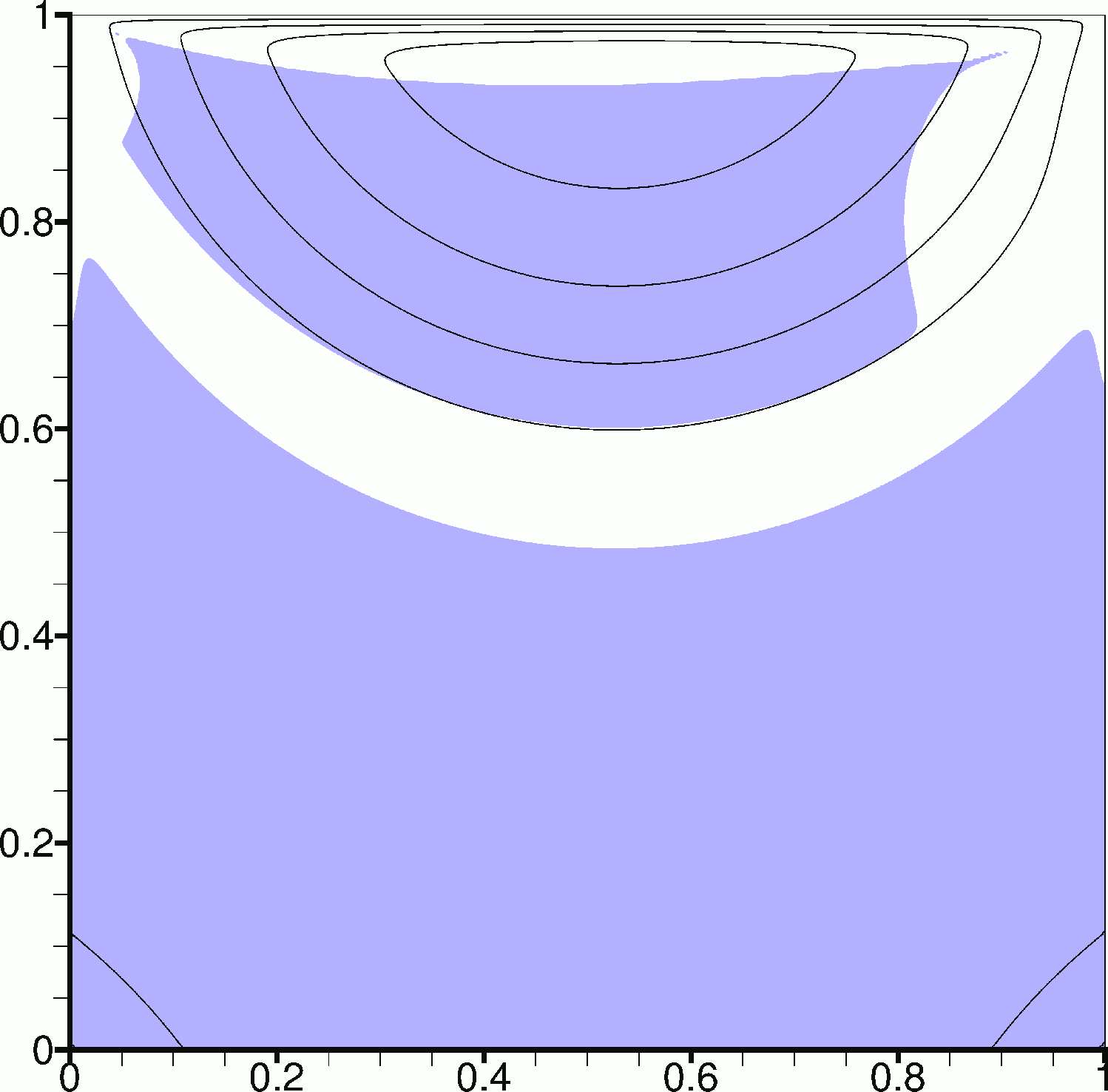}}
}
\caption{Streamlines in Bingham flow for $Bn$ = 100, plotted at intervals of 0.004 starting from zero. Unyielded areas 
($\tau < Bn$) are shown shaded.}
\label{fig: streamlines Bn=100}
\end{figure}

\clearpage
\section{Numerical results}
\label{sec: results}

Using the method described in the previous section, the lid-driven cavity problem has been solved for Reynolds numbers 
up to 5000, and for Bingham numbers up to 100. Unless otherwise stated, the results presented were obtained on the $512 
\times 512$ uniform grid, using $M=400$. For $Bn$ = 100 and Reynolds numbers other than 1000, a lower value of $M$ = 200 
was used to shorten the computational time, as the SIMPLE/multigrid method converges very slowly at such a high Bingham 
number when the value of $M$ is also high. The results of the simulations are presented in subsection \ref{ssec: 
description of flow field}. Then, in subsections \ref{ssec: accuracy of yield surfaces} and \ref{ssec: accuracy of flow 
field}, the accuracy of the results is examined concerning the calculation of the yield surfaces and the velocity field 
respectively. Finally, in section \ref{ssec: algebraic convergence} the performance of the SIMPLE/multigrid algebraic 
solver is discussed.

\subsection{Description of the flow field}
\label{ssec: description of flow field}

Figures \ref{fig: streamlines Bn=0} -- \ref{fig: streamlines Bn=100} describe the flow field as the Reynolds number 
increases, for $Bn$ = 0 (Newtonian flow), 1, 10, and 100, respectively. In the Newtonian case (Fig.\ \ref{fig: 
streamlines Bn=0}), the flow field is initially symmetric (Figs. \ref{sfig: streamlines Bn=0 Re=0} - \ref{sfig: 
streamlines Bn=0 Re=10}) but as the Reynolds number increases the main vortex shifts to the right (Fig.\ \ref{sfig: 
streamlines Bn=0 Re=100}), and then towards the centre of the cavity (Figs. \ref{sfig: streamlines Bn=0 Re=500}, 
\ref{sfig: streamlines Bn=0 Re=1000}). The same phenomena are observed also in Bingham flow, but they are postponed to 
larger Reynolds numbers as the Bingham number is increased. This will be discussed in more detail later on.

In the Bingham flow cases, unyielded zones form at the bottom of the cavity because the stresses are low there, due to 
the distance from the source of motion (the lid). These zones expand as the Bingham number is increased, and leave less 
space for flow to take place, thus pushing the vortex upwards towards the lid. They are in contact with the side and 
bottom walls which are motionless, and thus the material in contact is also motionless due to the no-slip boundary 
condition. This implies, due to the zero rate-of-strain condition within an uyielded zone, that the material is 
motionless throughout these unyielded zones. Actually, the regularisation method employed allows for weak non-zero 
deformation rates within unyielded zones, and thus for example one can observe extremely weak vortices at the lower 
corners, within the unyielded zones. Such features must be regarded as artefacts of regularisation, and discarded in 
order to get a more accurate picture of the actual Bingham flow.

Figures \ref{fig: streamlines Bn=1} -- \ref{fig: streamlines Bn=100} show also the existance of one more unyielded zone 
(in a few cases more than one) which is usually located just below the vortex, or to the left of the vortex when the 
latter is shifted towards the right. These zones do not touch the cavity walls and are not motionless, as implied from 
the spacing of the streamlines inside these regions, but move as solid bodies. Since the flow is steady-state, their 
locations and shapes are fixed, which means that they lose mass on their downstream boundary at a rate equal to that at 
which they gain mass on their upstream boundary.

\begin{figure}[!b]
\centering
\includegraphics[scale=1.00]{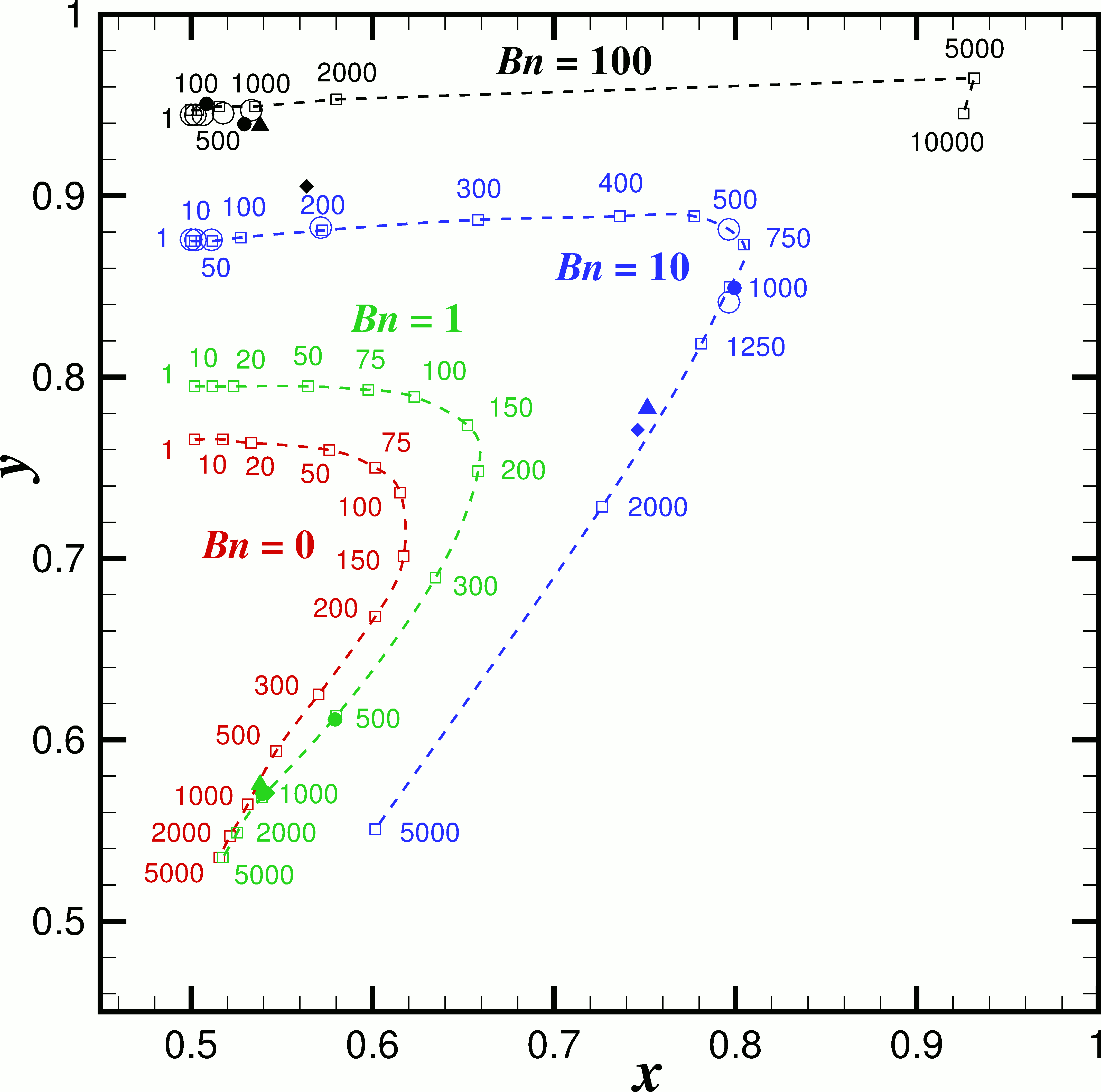}
\caption{The position of the vortex centre, for various $Re$ and $Bn$ numbers. The results of the present study are
shown as empty squares ($\square$), with the Reynolds number written next to each square. Results of other researchers 
are also included for comparison: Results of Vola et al. \cite{Vola_03} ($Re$~= 1000: $Bn$~= 1, 10, 100) are indicated 
by filled triangles ($\blacktriangle$); results of Elias et al. \cite{Elias_06} ($Re$~= 1000: $Bn$~= 1, 10, 100), are 
indicated by filled diamonds ($\blacklozenge$); results of Frey et al. \cite{Frey_10} ($Re$~= 500: $Bn$~= 1, 100; $Re$~= 
1000: $Bn$~= 1, 10, 100) are indicated by filled circles ($\bullet$); and results of Prashant \& Derksen 
\cite{Prashant_11} ($Re$~= 0.5, 10, 50, 200, 600, 1000: $Bn$~= 10, 100), are indicated by empty circles ($\bigcirc$).}
\label{fig: vortex centre}
\end{figure}

\begin{figure}[t]
\centering
\includegraphics[scale=1.00]{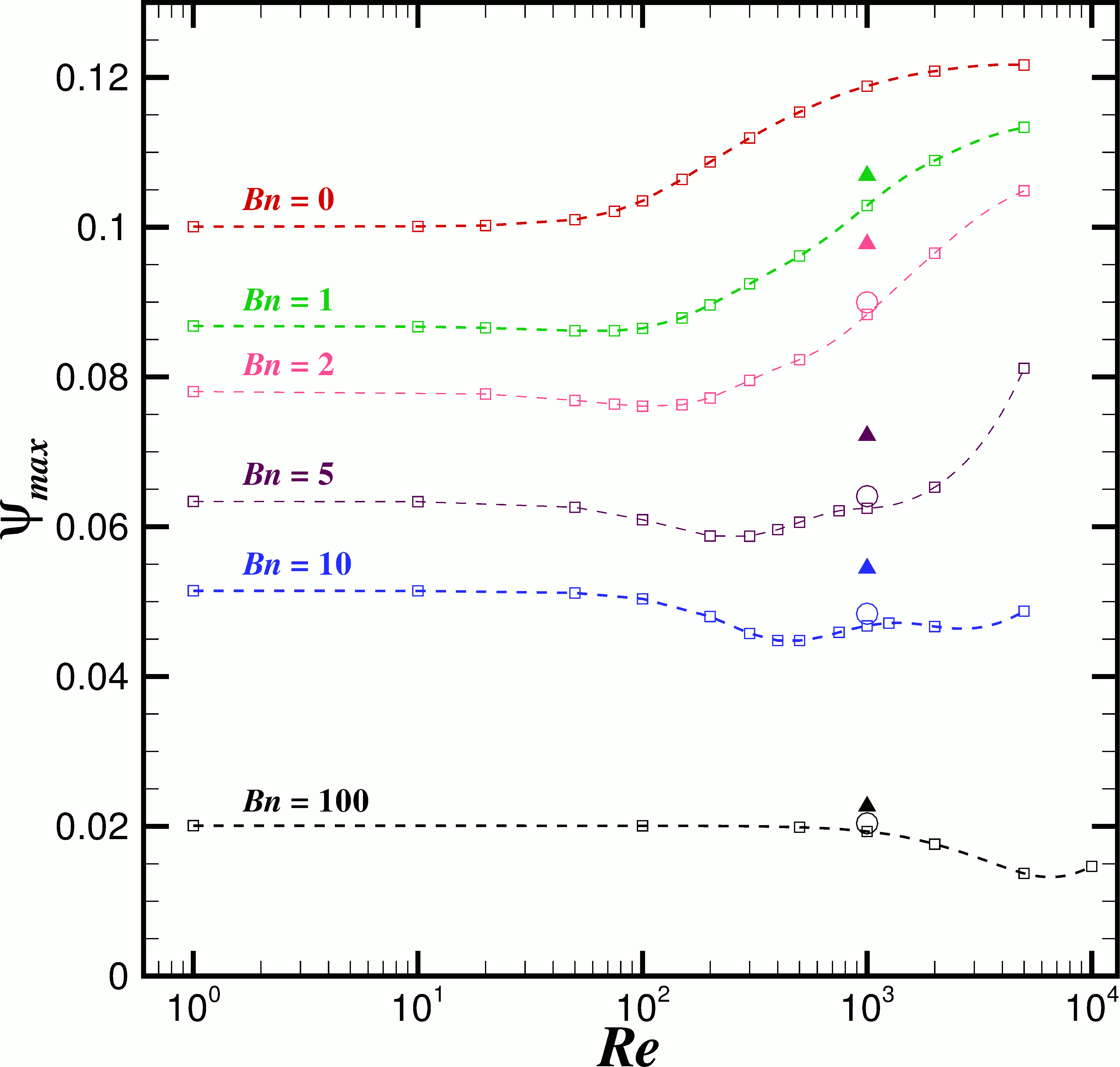}
\caption{The strength of the vortex, dedimensionalised by $U\!\cdot\!L$, as a function of $Re$ for various Bingham 
numbers. The results of the present study are shown as empty squares ($\square$). The results of Vola et al. 
\cite{Vola_03} ($\blacktriangle$) and Prashant \& Derksen \cite{Prashant_11} ($\bigcirc$) for $Re$~= 1000 are also 
shown.}
\label{fig: vortex strength}
\end{figure}

As noted, Figs. \ref{fig: streamlines Bn=0} -- \ref{fig: streamlines Bn=100} reveal that the effect on the flow of 
increasing the Reynolds number is similar for all Bingham numbers. This can be explored in greater detail with the aid 
of the plots of the vortex position and strength, Figs. \ref{fig: vortex centre} and \ref{fig: vortex strength}, 
respectively. Three flow regimes are discernible:

\begin{enumerate}

\item Up to a certain Reynolds number the vortex is approximately fixed in space, and its strength is constant ($Re 
\approx 20$ for $Bn = 0$; $Re \approx 50$ for $Bn = 10$; $Re \approx 500$ for $Bn = 100$).

\item Beyond that Reynolds number, the vortex moves towards the right (in the same direction as the lid), until a 
second critical Reynolds number is reached ($Re \approx 75$ for $Bn = 0$; $Re \approx 500$ for $Bn = 10$; $Re \approx 
5000$ for $Bn = 100$)\footnote{For $Bn$ = 100, to find this second critical Reynolds number it was necessary to perform 
a simulation also for $Re = 10000$. We note that these $Re$ = 10000 results were obtained on the 256 $\times$ 256 grid 
due to convergence problems on the 512 $\times$ 512 grid.}. If the Bingham number is large enough ($Bn \geq 2$) so that 
the the lower unyielded region has pushed the vortex close to the lid, then this motion of the vortex towards the right 
is accompanied by a weakening of the vortex, due to geometric restrictions.

\item Beyond the second critical Reynolds number, the vortex moves towards the centre of the cavity. This is 
accompanied by a strengthening of the vortex.

\end{enumerate}

\begin{figure}[t]
\centering
\noindent\makebox[\textwidth]{
 \subfigure[{$Bn = 1$}] {\label{sfig: Reynolds Bn=1 Re=1000}
  \includegraphics[scale=0.80]{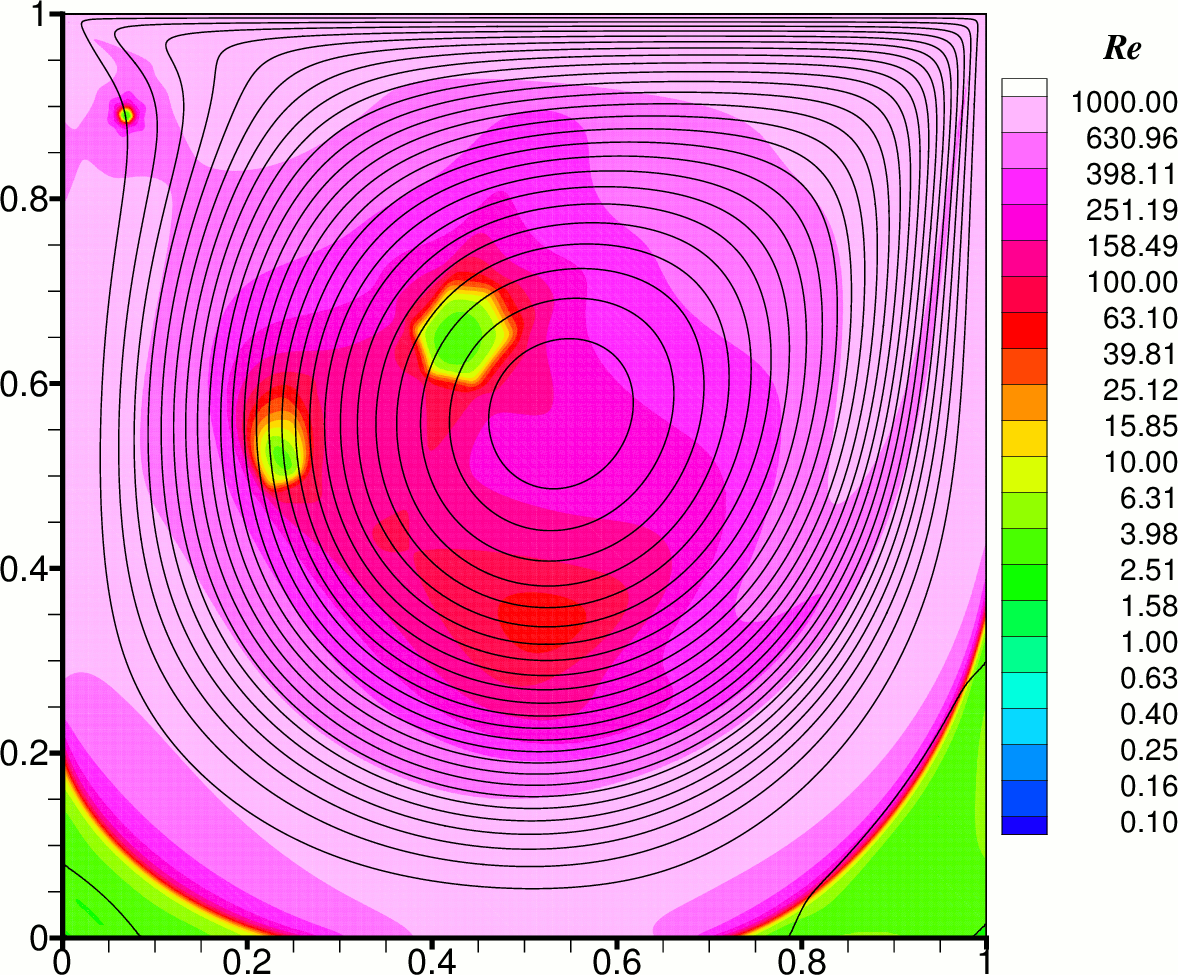}}
 \subfigure[{$Bn = 10$}] {\label{sfig: Reynolds Bn=10 Re=1000}
  \includegraphics[scale=0.80]{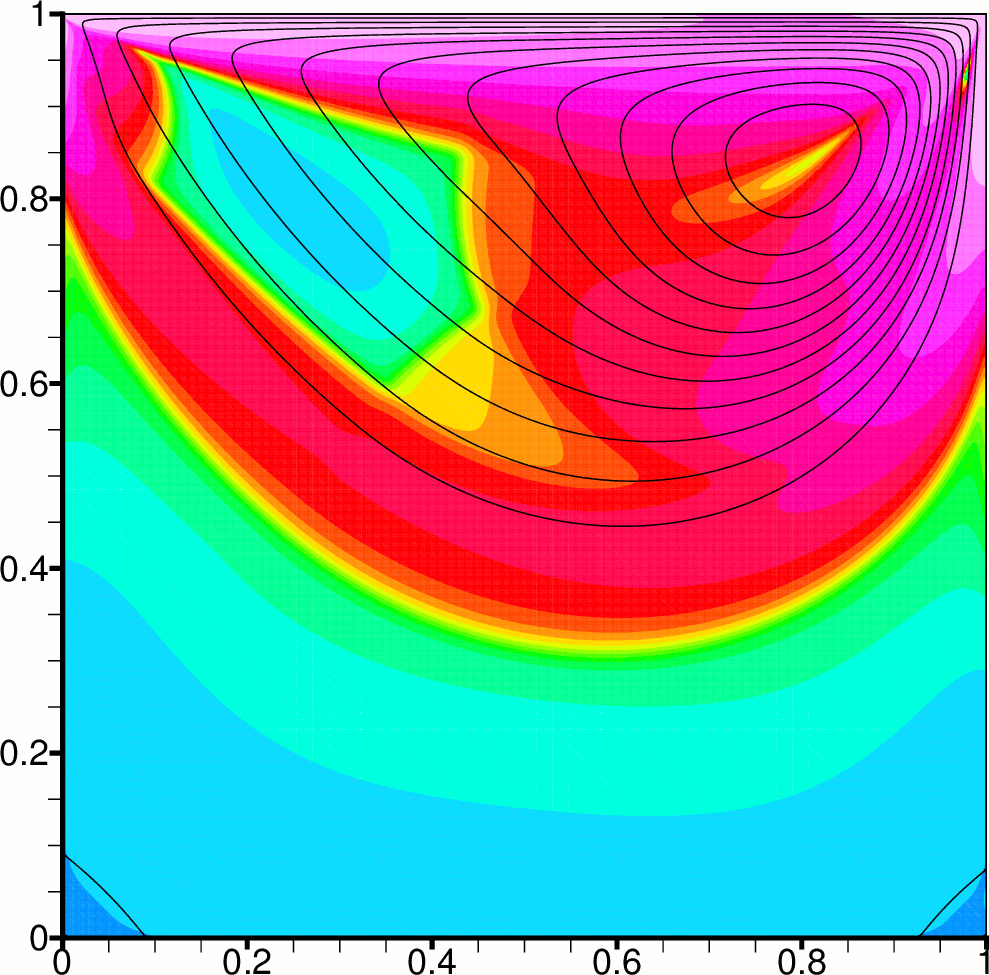}}
}
\noindent\makebox[\textwidth]{
 \subfigure[{$Bn = 100$}] {\label{sfig: Reynolds Bn=100 Re=1000}
  \includegraphics[scale=0.80]{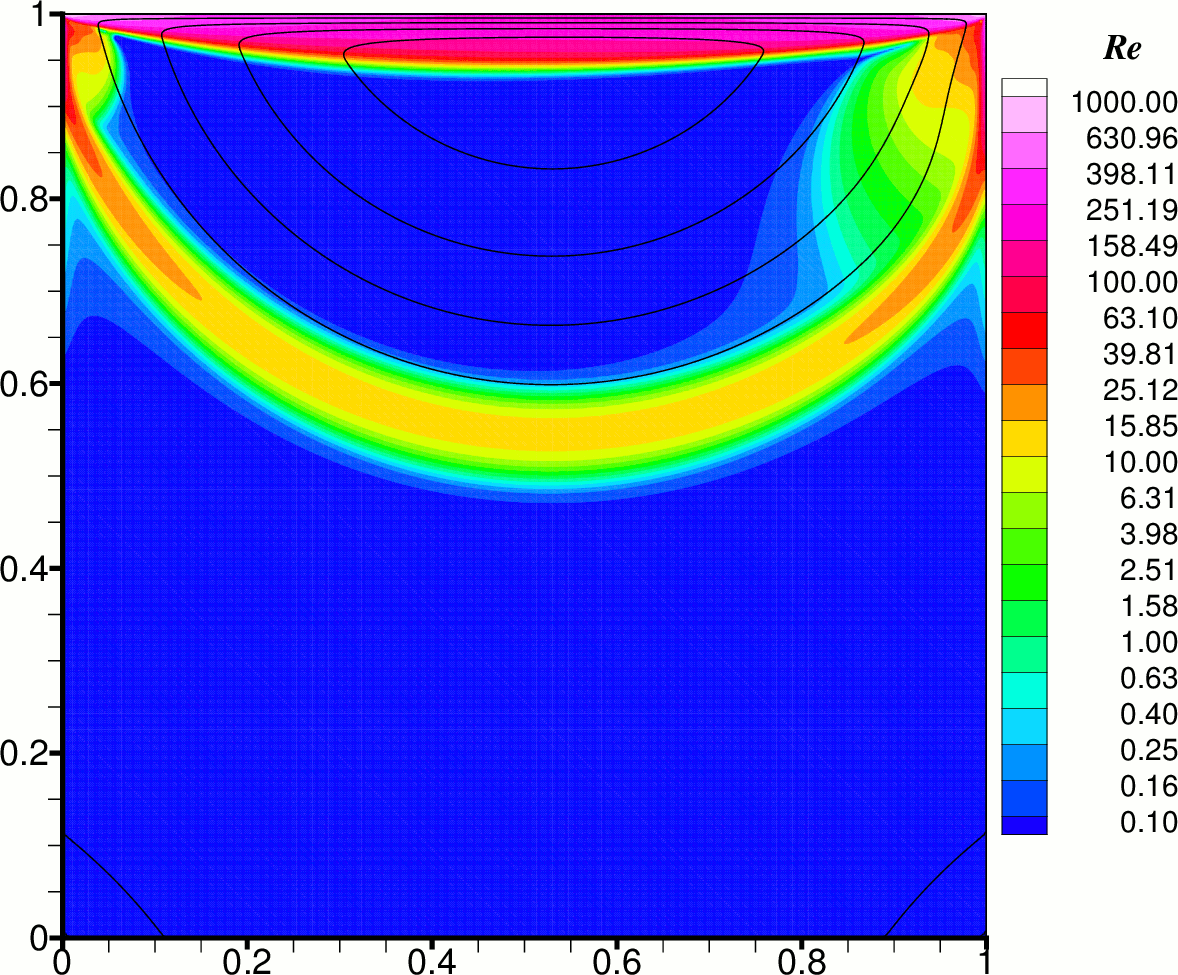}}
}
\caption{Colour contours of the local Reynolds number $Re_l$ (equation (\ref{eq: Re_l})), when $Re$ = 1000, for various
Bingham numbers. Streamlines are also shown (black lines).}
\label{fig: Reynolds Re=1000}
\end{figure}

The main conclusion from these results is that the flow field at a certain combination of $Bn$ and $Re$ numbers 
resembles to some extent that of any lower $Bn$ number if the $Re$ number is also sufficiently lowered. To investigate 
this, we calculated the ``local Reynolds number'', $Re_l$, which is defined based on the apparent viscosity $\eta(x,y)$ 
(\ref{eq: viscosity}) at each point, instead of the plastic viscosity $\mu$:

\begin{equation} \label{eq: Re_l}
 Re_l(x,y) \equiv \frac{\rho U L}{\eta(x,y)}
\end{equation}
Figure \ref{fig: Reynolds Re=1000} shows the contours of $Re_l$ for $Re=1000$ and $Bn$ = 1, 10, and 100 (for $Bn=0$ it 
is clear that $Re_l=Re=1000$). Since $\rho$, $U$ and $L$ are fixed, $Re_l$ is simply proportional to the reciprocal of 
the effective viscosity. However, the plots allow a comparison between these flows and Newtonian flows of a similar 
Reynolds number. Indeed, one notices that for $Bn=1$ (Fig.\ \ref{sfig: Reynolds Bn=1 Re=1000}) the local Reynolds number 
is in the range 200 -- 1000 in most of the cavity, and there is not much difference between this flow field and 
Newtonian flow at $Re=1000$ (Fig.\ \ref{sfig: streamlines Bn=0 Re=1000}). For $Bn=10$ (Fig.\ \ref{sfig: Reynolds Bn=10 
Re=1000}), $Re_l$ is less than 100 almost everywhere, except near the top and the upper right of the cavity, and the 
flow resembles a Newtonian flow at $Re=100$ (Fig.\ \ref{sfig: streamlines Bn=0 Re=100}), where the vortex has moved 
towards the right. Finally, at $Bn=100$ (Fig.\ \ref{sfig: Reynolds Bn=100 Re=1000}) it can be seen that $Re_l$ is well 
below 1 in most of the cavity, and below 10 in most of the yielded area - the flow resembles a Newtonian flow at $Re$ = 
1 or 10 (Figs. \ref{sfig: streamlines Bn=0 Re=1}, \ref{sfig: streamlines Bn=0 Re=10}), being nearly symmetric with 
respect to the vertical centreline.

Finally, we note that Figure \ref{fig: vortex strength} shows that increasing the Bingham number causes significant 
weakening of the flow.

\subsection{Accuracy of the yield surfaces}
\label{ssec: accuracy of yield surfaces}

Regularisation methods produce results which do not contain truly unyielded regions. The question is then, how to 
deduce the unyielded regions from these results. The most common approach is to identify the yield surfaces with the 
contours of $\tau$ = $\tau_y$ ($\tau = Bn$ in the non-dimensional case). This is the method used in Figures \ref{fig: 
streamlines Bn=1} - \ref{fig: streamlines Bn=100}. It is reasonable to assume that the yield surfaces calculated in this 
manner will converge to the true yield surfaces as $M \rightarrow \infty$, but a theoretical proof of this does not 
exist, unlike for the velocity field which is known to converge to the exact solution \cite{Frigaard_05}. There do exist 
a few studies where comparisons against analytical solutions or results obtained with augmented Lagrangian methods have 
shown that, for the particular test cases studied, yield surfaces calculated with regularisation methods do converge to 
the true surfaces as the regularisation parameter is increased (e.g.\ Burgos et al. \cite{Burgos_99}, Dimakopoulos et al. 
\cite{Dimakopoulos_2013}). Unfortunately, the minimum value of the regularisation parameter $M$ which is necessary to 
accurately predict the yield surfaces is problem-dependent. In cases where the stress is close to the yield stress over 
a relatively large region, very large values of the regularisation parameter may be needed \cite{Frigaard_05}. In our 
case, using the SIMPLE/multigrid algebraic solver we were unable to obtain solutions with $M$ larger than about 400, 
but because the flow domain is confined, the stress variation is rather rapid and therefore good approximations of the 
yield surfaces can be obtained with low values of $M$.

\begin{figure}[!t]
\centering
\noindent\makebox[\textwidth]{
 \subfigure[{$Bn$ = 1; $Re$ = 0}] {\label{sfig: M-convergence Bn=1 Re=0}
  \includegraphics[scale=1.00]{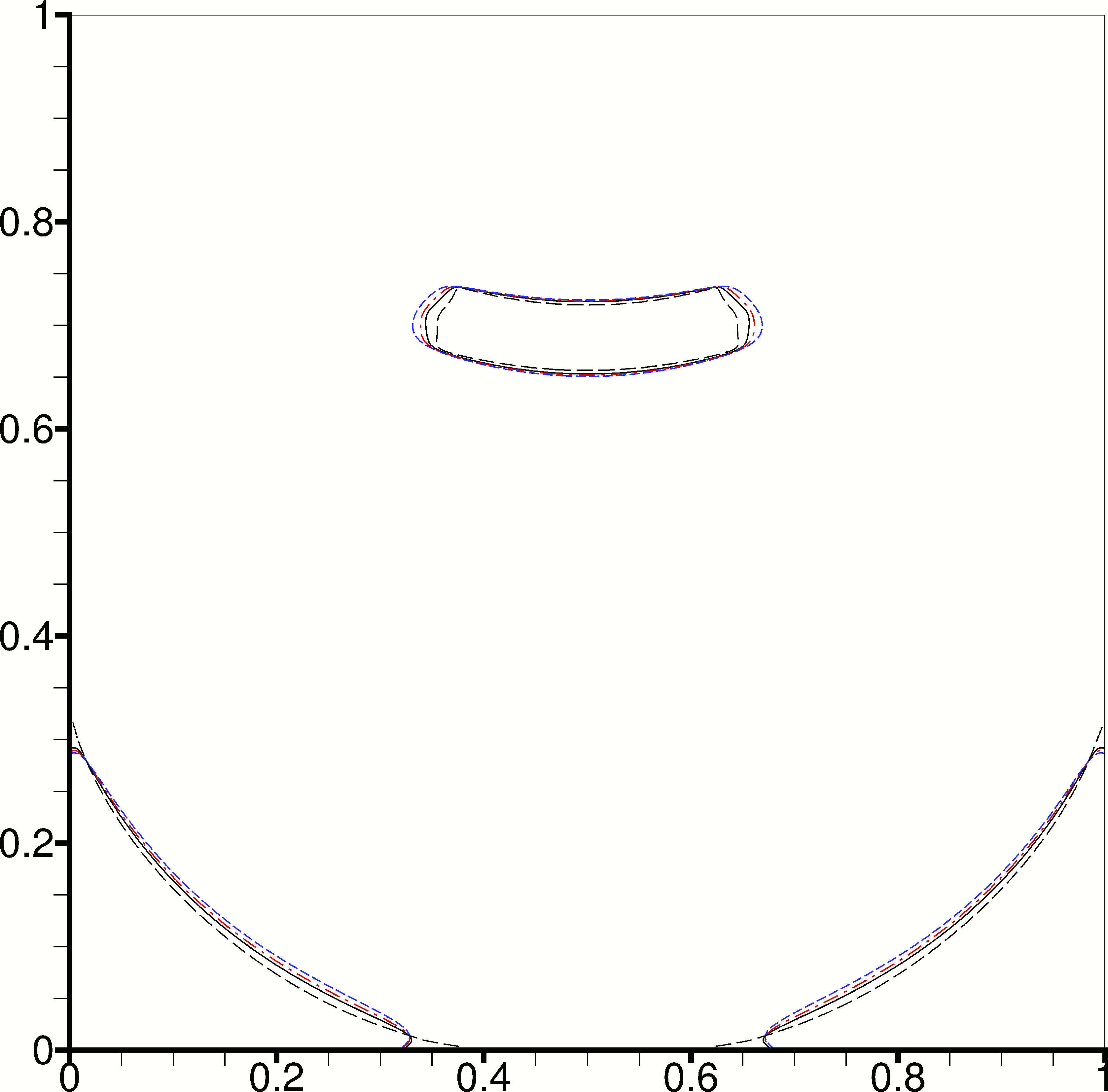}}
 \subfigure[{$Bn$ = 10; $Re$ = 0}] {\label{sfig: M-convergence Bn=10 Re=0}
  \includegraphics[scale=1.00]{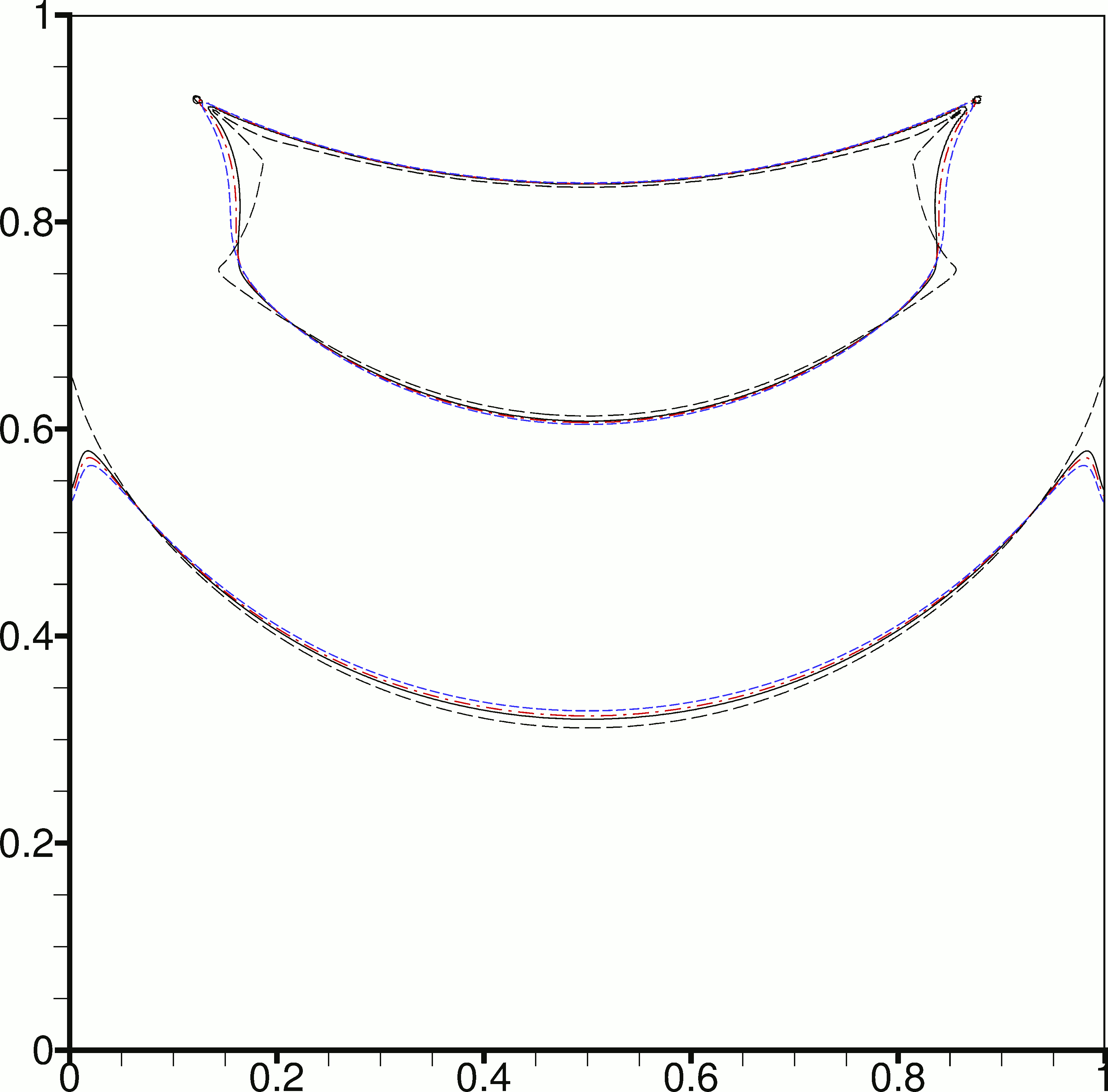}}
}
\noindent\makebox[\textwidth]{
 \subfigure[{$Bn$ = 1; $Re$ = 1000}] {\label{sfig: M-convergence Bn=1 Re=1000}
  \includegraphics[scale=1.00]{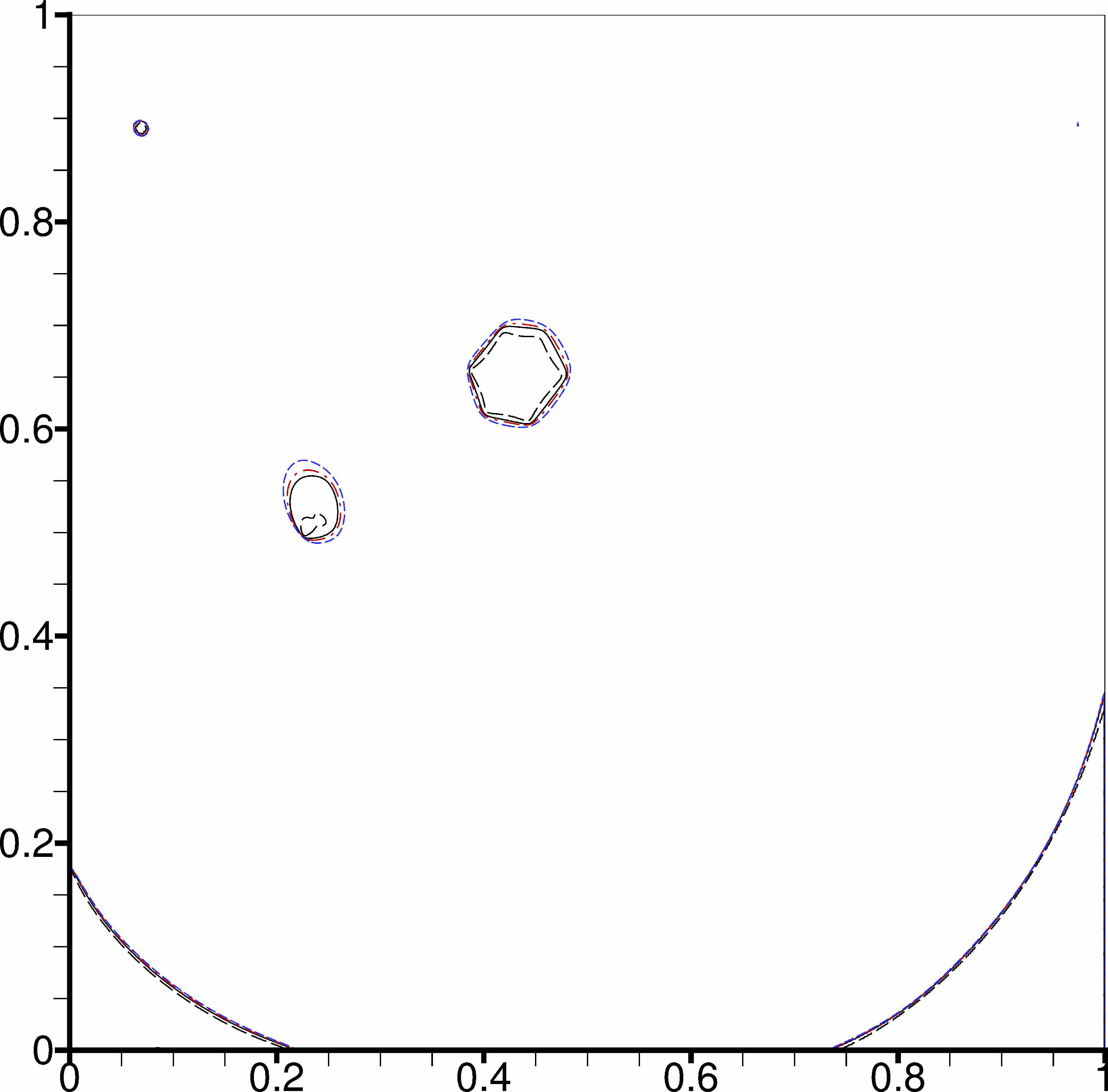}}
 \subfigure[{$Bn$ = 10; $Re$ = 1000}] {\label{sfig: M-convergence Bn=10 Re=1000}
  \includegraphics[scale=1.00]{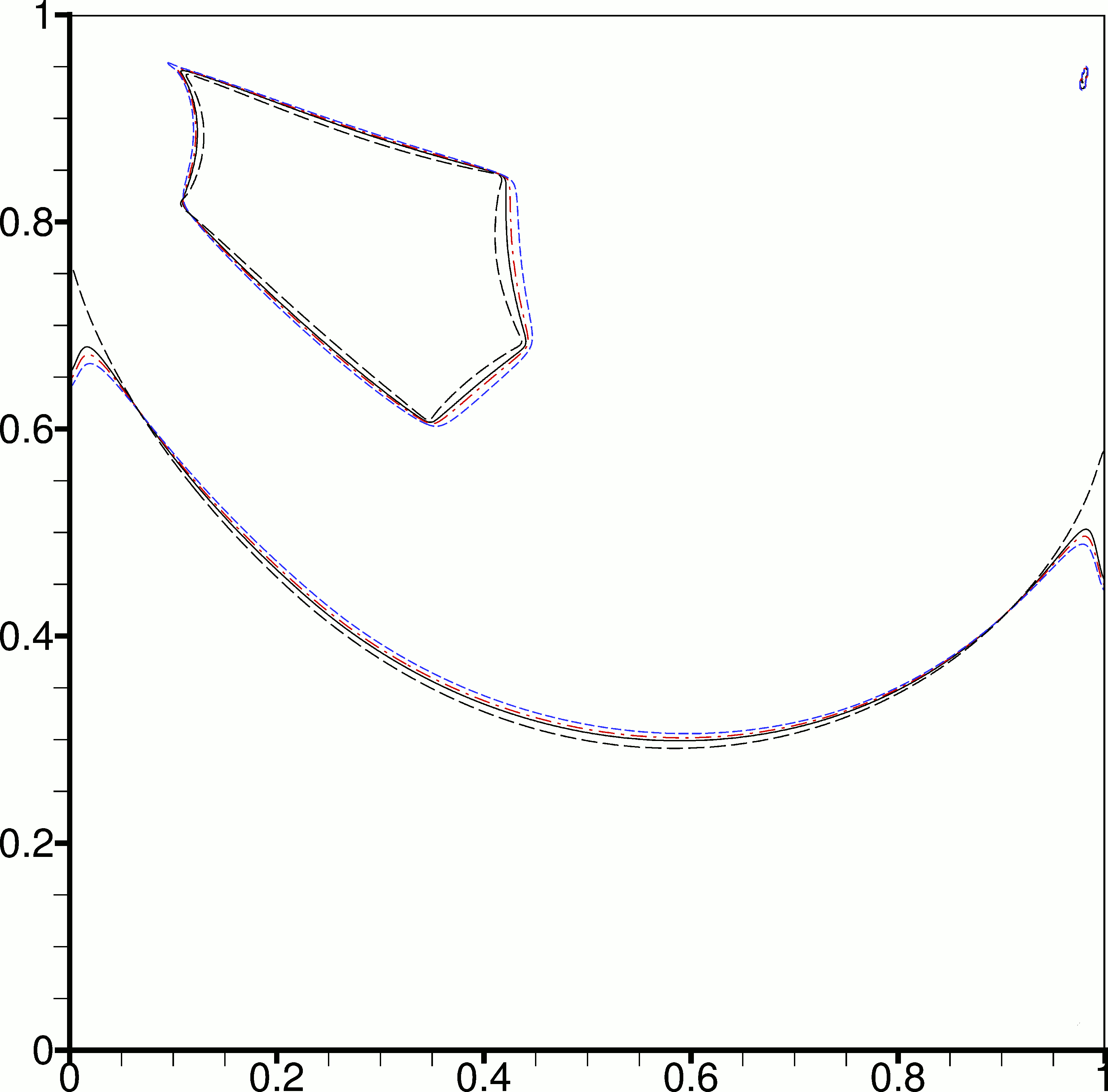}}
}
\caption{Contours of $\tau = Bn$ calculated with $M$ = 100 (blue dashed lines), $M$ = 200 (red chained lines), and $M$ 
= 400 (black solid lines). Also shown, with black dashed lines (long dashes) are estimates of the true yield surfaces 
according to the extrapolation technique of Liu et al. \cite{Liu_2002}.} 
\label{fig: M-convergence of yield surfaces}
\end{figure}

Figure \ref{fig: M-convergence of yield surfaces} shows how the contours $\tau = Bn$ vary as $M$ is increased, for some 
sample cases. In general, the variation is small, so that one can be confident that the general shape has been captured 
well. There are some inaccuracies in the fine details though, such as the fact that there is a concavity inversion of 
the $\tau = Bn$ contours where they meet the cavity walls, expecially for high $Bn$, giving the impression that the 
unyielded zones exhibit ``tips'' near the walls. Creeping flow results obtained with augmented Lagrangian methods 
\cite{Yu_07, Zhang_10, Muravleva_08, Glowinski_2011}, and the limited results for $Re$ = 1000 of Vola et al. 
\cite{Vola_03}, suggest that this is not a physically correct result. Increasing $M$ improves the results, but the 
problem has not completely disapeared at the maximum value of $M$ = 400 used. For engineering applications such small 
inaccuracies would probably be unimportant - the errors introduced by the deviation of the chosen mathematical model 
(e.g.\ Bingham model) from the behaviour of a real material would be much greater. However, we discuss below a couple of 
techniques which can be applied at the postprocessing stage to improve the results.

\begin{figure}[t]
\centering
\noindent\makebox[\textwidth]{
 \subfigure[{$Bn=10$; $Re = 0$}] {\label{sfig: ty+-e Bn=10 Re=0}
  \includegraphics[scale=1.0]{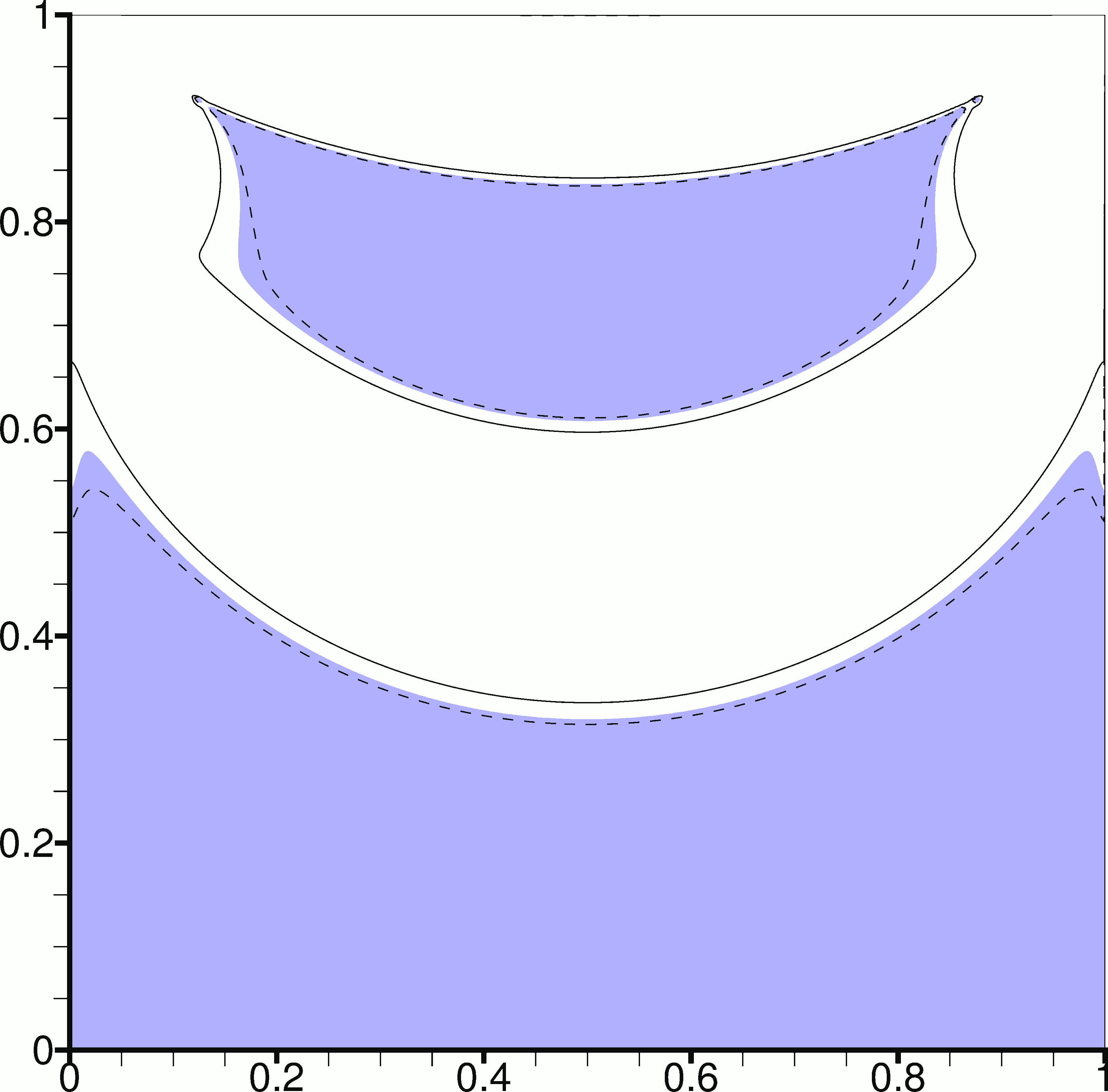}}
 \subfigure[{$Bn=10$; $Re = 1000$}] {\label{sfig: ty+-e Bn=10 Re=1000}
  \includegraphics[scale=1.0]{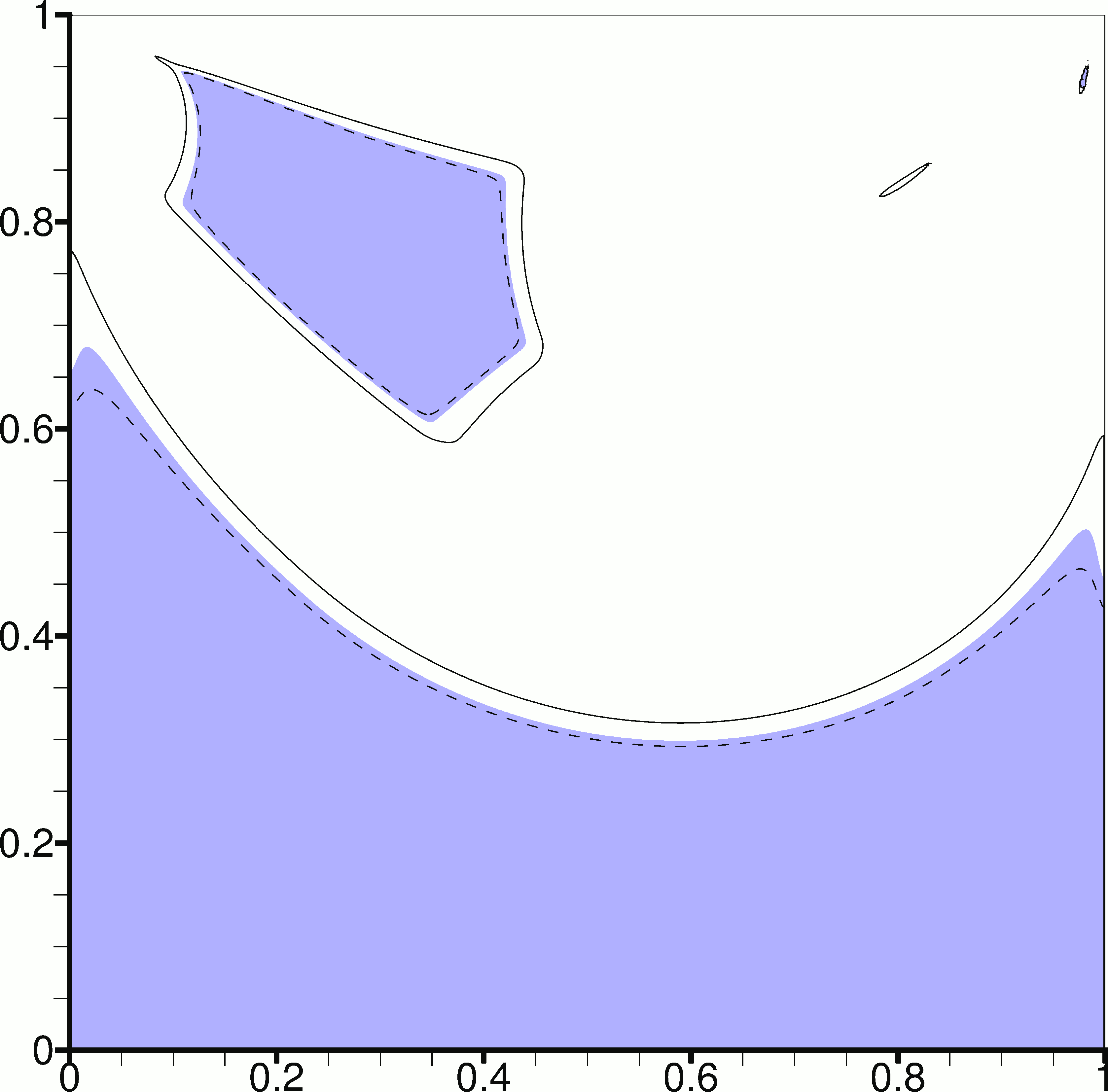}}
}
\caption{Shown shaded are the unyielded regions ($\tau \leq Bn$), for $Bn$ = 10, and $Re$ = 0 and 1000. The contours 
(black lines) correspond to $\tau = 0.99 Bn$ (dashed lines) and $\tau = 1.01 Bn$ (solid lines).}
\label{fig: ty+-e}
\end{figure}

It was suggested by Burgos et al. \cite{Burgos_99} that the shape of the yield surfaces is described better by contours 
of $\tau = (1 + \epsilon) Bn$ where $\epsilon \ll 1$ is a small positive number, than by the contours $\tau = Bn$. This 
is because regularised constitutive equations converge very rapidly to the Bingham equation as $\tau$ increases beyond 
$Bn$, but they diverge from the Bingham constitutive equation when $\tau$ drops below $Bn$ because the regularised 
$\tau$ -- $\dot{\gamma}$ graph must pass through the origin. Therefore, the contour $\tau = (1 + \epsilon) Bn$ of the 
Bingham flow field is better approximated by the corresponding regularised contour, than is the contour $\tau = Bn$. And 
if $\epsilon$ is small enough, then the $\tau = (1 + \epsilon) Bn$ contour will not be much different than the $\tau = 
Bn$ contour. Figure \ref{fig: ty+-e} shows that for $Bn$=10 the $\tau = 1.01 Bn$ contour in the proximity of the lower 
unyielded region continues straight up to meet the side walls, in contrast to the $\tau = Bn$ contour which inverts its 
concavity near the walls, exhibiting a pair of tips. In fact, near the walls it can be seen that the stress is nearly 
equal to the yield stress ($0.99 Bn \leq \tau \leq 1.01 Bn$) over relatively large regions, which is precisely the 
problematic condition described by Frigaard and Nouar \cite{Frigaard_05} under which the yield surface is difficult to 
compute, and very large $M$ parameters are required. A similar situation appears also at the lower corners of the upper 
unyielded region of the case \{$Bn$ = 10, $Re$ = 0\}.

Liu et al. \cite{Liu_2002} have proposed an interesting extrapolation procedure to approximate the yield surfaces using 
a number of solutions corresponding to different values of the regularisation parameter. Their method uses not the 
stress, but the strain rate. In a real Bingham fluid, when the magnitude of the stress tensor becomes equal to the yield 
stress then the material is at the onset of yielding, and the strain rate is zero. However, regularisation results in a 
non-zero critical strain rate $\dot{\gamma}_y$ at the $\tau = Bn$ surface. By writing equation (\ref{eq: Papanastasiou, 
dimensionless}) in terms of the tensor magnitudes, and setting $\tau = Bn$, one obtains:

\begin{equation} \label{eq: gamma_y}
 \dot{\gamma}_y \;-\; Bn\cdot\exp(-M \dot{\gamma}_y) \;=\; 0
\end{equation}
This equation can be solved numerically for any given $Bn$ to obtain the corresponding $\dot{\gamma}_y$. Identifying 
the yield surfaces with the contours of $\dot{\gamma} = \dot{\gamma}_y$ is exactly equivalent to the previous criterion,
$\tau = Bn$. Instead, Liu et al. \cite{Liu_2002} considered the ratio $\tilde{\gamma} \equiv \dot{\gamma} / 
\dot{\gamma}_y$. At a certain point in the flow field, as $M$ is increased and the exact Bingham flow field is 
approached, $\dot{\gamma}_y$ tends to zero, while $\dot{\gamma}$ tends either also to zero, if the point is unyielded, 
or to a specific non-zero value, if the point is yielded. So, the ratio $\tilde{\gamma} = \dot{\gamma} / 
\dot{\gamma}_y$ tends to infinity in yielded regions, whereas it does not tend to infinity in unyielded regions (where 
in fact one would expect that $\dot{\gamma} / \dot{\gamma}_y < 1$). Liu et al. \cite{Liu_2002} noticed in their 
investigation of the creeping flow of a Bingham material about a sphere that there existed surfaces within the domain 
where the value of the ratio $\tilde{\gamma}(M) = \dot{\gamma}(M) / \dot{\gamma}_y(M)$ is constant, independent of $M$. 
Across such a surface, on one side of the surface the function $\tilde{\gamma}(M)$ increases with increasing $M$, and 
thus the material there appears more  and more fluid-like, while on the other side of the surface the function 
$\tilde{\gamma}(M)$ decreases with increasing $M$ (although it may increase again further inside the solid region), and 
thus the material there appears more and more solid-like. So, Liu et al. \cite{Liu_2002} suggested that these surfaces 
coincide with the yield surfaces.

\begin{figure}[!b]
\centering
 \noindent\makebox[\textwidth]{
 \subfigure[{$Bn=10$; $Re = 0$}] {\label{sfig: Liu Bn=10 Re=0}
  \includegraphics[scale=1.0]{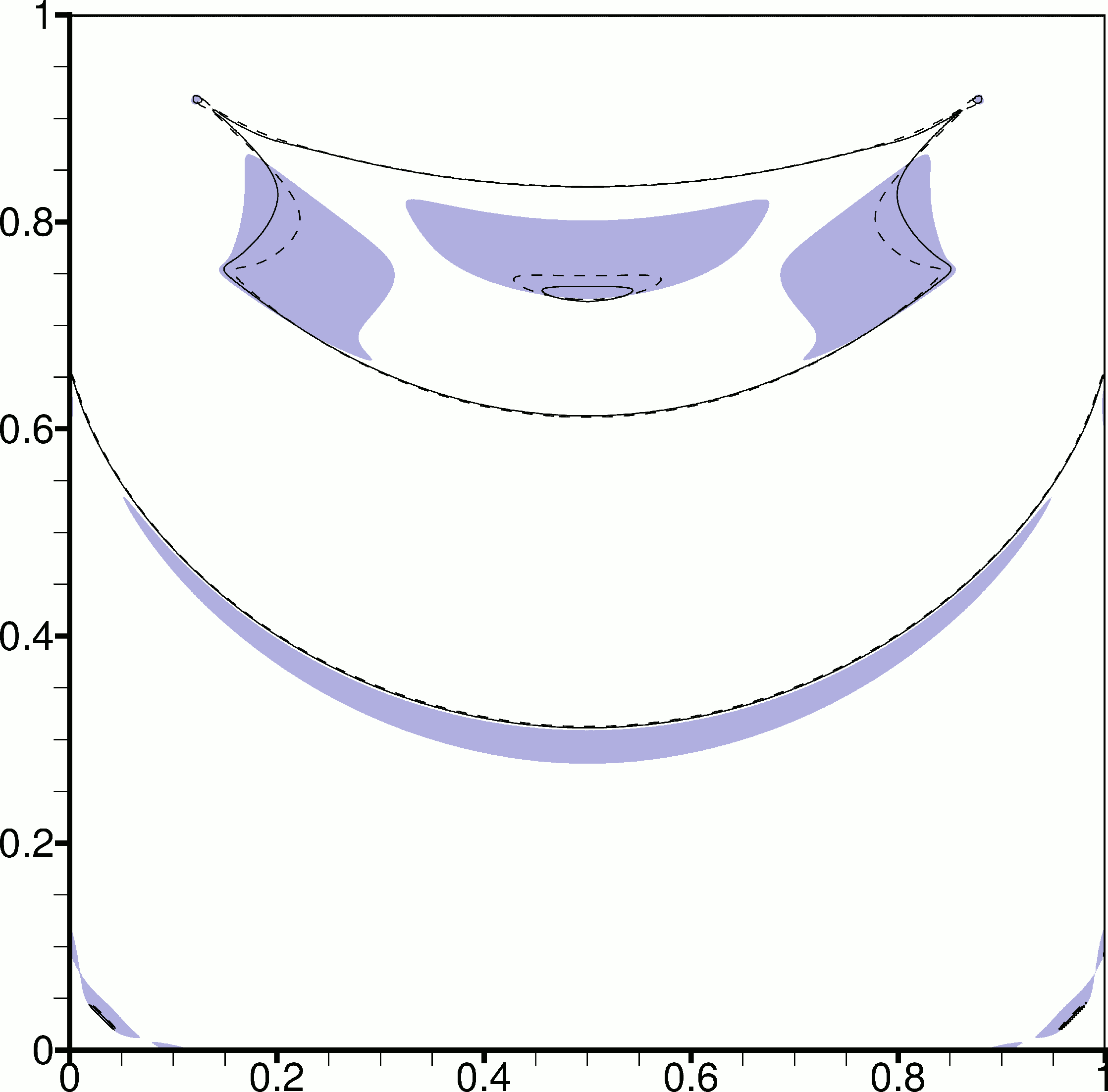}}
 \subfigure[{$Bn=10$; $Re = 1000$}] {\label{sfig: Liu Bn=10 Re=1000}
  \includegraphics[scale=1.0]{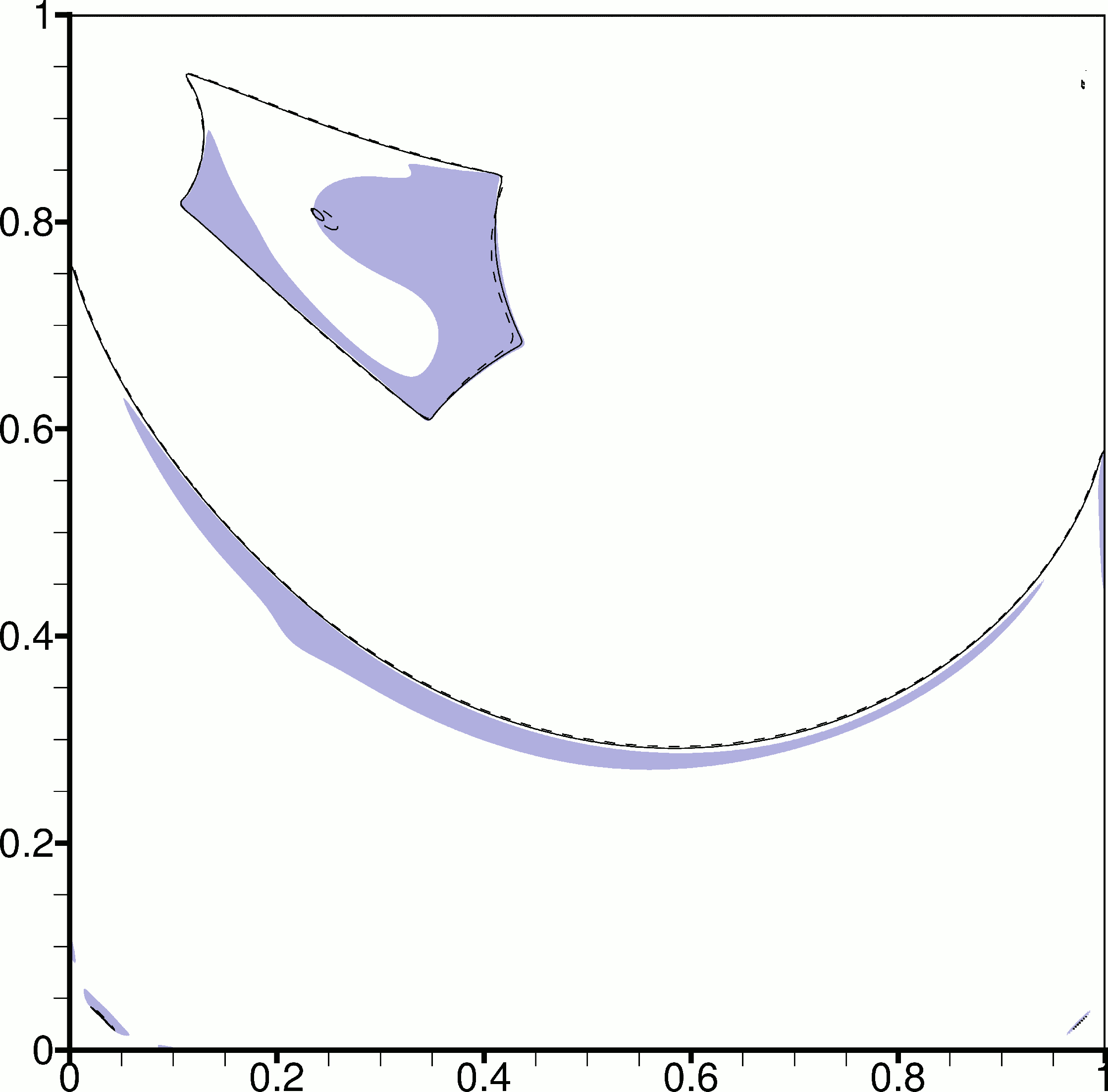}}
} 
\caption{Contours of $c(400,200)=0$ (solid lines) and $c(200,100)=0$ (dashed lines), which approximate the yield 
surfaces. See paragraph \ref{ssec: accuracy of yield surfaces} for definition of the function $c(M_1,M_2)$. Shown 
shaded are the regions where $c(400,200) - c(200,100) < 0$, which reveal some uncertainty where they cross the $c=0$ 
contours into the yielded region.}
\label{fig: Liu extrapolation}
\end{figure}

These surfaces can be sought by solving the problem for two values of $M$ and then plotting the difference 
$\tilde{\gamma}(M_1) - \tilde{\gamma}(M_2)$. The contour $c(M_1,M_2) = \tilde{\gamma}(M_1) - \tilde{\gamma}(M_2) = 0$ 
is potentially such a surface, because there $\tilde{\gamma}(M_1) = \tilde{\gamma}(M_2)$. To check that this is the 
sought surface, one can repeat the calculations with one or more different values of $M$. If for a different value $M_3$ 
the contour $c(M_2,M_3) = 0$ coincides with the contour $c(M_1,M_2) = 0$ then this increases the confidence that these 
contours are the sought surfaces, because there $\tilde{\gamma}(M_1) = \tilde{\gamma}(M_2) = \tilde{\gamma}(M_3)$. For 
the present work we used three values of $M$: 100, 200 and 400. Figure \ref{fig: Liu extrapolation} shows the computed 
surfaces for sample cases. It can be seen that despite the relatively low values of $M$, the surfaces $c(400,200)$ 
= 0 and $c(200,100)$ = 0 coincide for the most part, in agreement with the observations of Liu et al. The contours have 
not converged at the sides of the upper unyielded region of the \{$Re$ = 0, $Bn$ = 10\} case, but in the rest of the 
domain there is perfect coincidence of the contours, while the tips of the lower unyielded regions have disapeared. 
In Figure \ref{fig: Liu extrapolation}, the regions where $c(400,200) - c(200,100) < 0$ are shown shaded, as an 
additional means of investigation, because in yielded regions as $\tilde{\gamma} \rightarrow \infty$, $c(M_1,M_2)$ 
should be larger than $c(M_2,M_3)$ if $M_1 > M_2 > M_3$ and the parameters are increased in a consistent manner. So, 
where the shaded regions cross the $c$ = 0 contours into the yielded zone, there is uncertainty as to where the yield 
surface actually lies. This happens for example at the side walls of the upper unyielded region of the \{$Bn$ = 10, 
$Re$ = 0\} case, where the $c(400,200)=0$ and $c(200,100)=0$ contours have not yet converged. We note also that there 
appear $c=0$ contours inside the upper unyielded regions, near their centres, but they diminish as $M$ increases, and 
they clearly do not represent yield surfaces. The $c=0$ contours are plotted also in Figure \ref{fig: M-convergence of 
yield surfaces}, for direct comparison with the $\tau = Bn$ criterion. To give a better estimate of the yield surface, 
in the \{$Bn$ = 10, $Re$ = 0\} case the sides of the upper unyielded region have been corrected by joining the lower 
corners with the points where the contours $c(400,200)=0$ and $c(200,100)=0$ intersect.

Thus, the technique of Liu et al. \cite{Liu_2002} appears to yield more accurate yield surfaces. However, for 
convenience, in the rest of this paper the yield surfaces are taken to be the $\tau = Bn$ (or $\dot{\gamma} = 
\dot{\gamma}_y$) contours.

Finally, we investigate the effect of grid density on the yield surfaces, calculated as $\tau = Bn$. Figure \ref{fig: 
Grid convergence of yield surfaces} shows the contours $\tau = Bn$ for the \{$Bn$ = 10, $Re$ = 1000\} case, on various 
grids. Interestingly, there is observable improvement of the yield surface as the grid density is increased, even up to 
the 2048$\times$2048 grid: the tips of the boundary of the lower unyielded region move closer to the walls and upwards.
The benefit from local refinement is visible, since that particular locally refined grid has approximately the same 
number of volumes as the 512$\times$512 grid (see the next section for details). Overall though, for the yield surfaces, 
the effect of $M$ is more significant than the grid density.

\begin{figure}[t]
\centering
\includegraphics[scale=1.0]{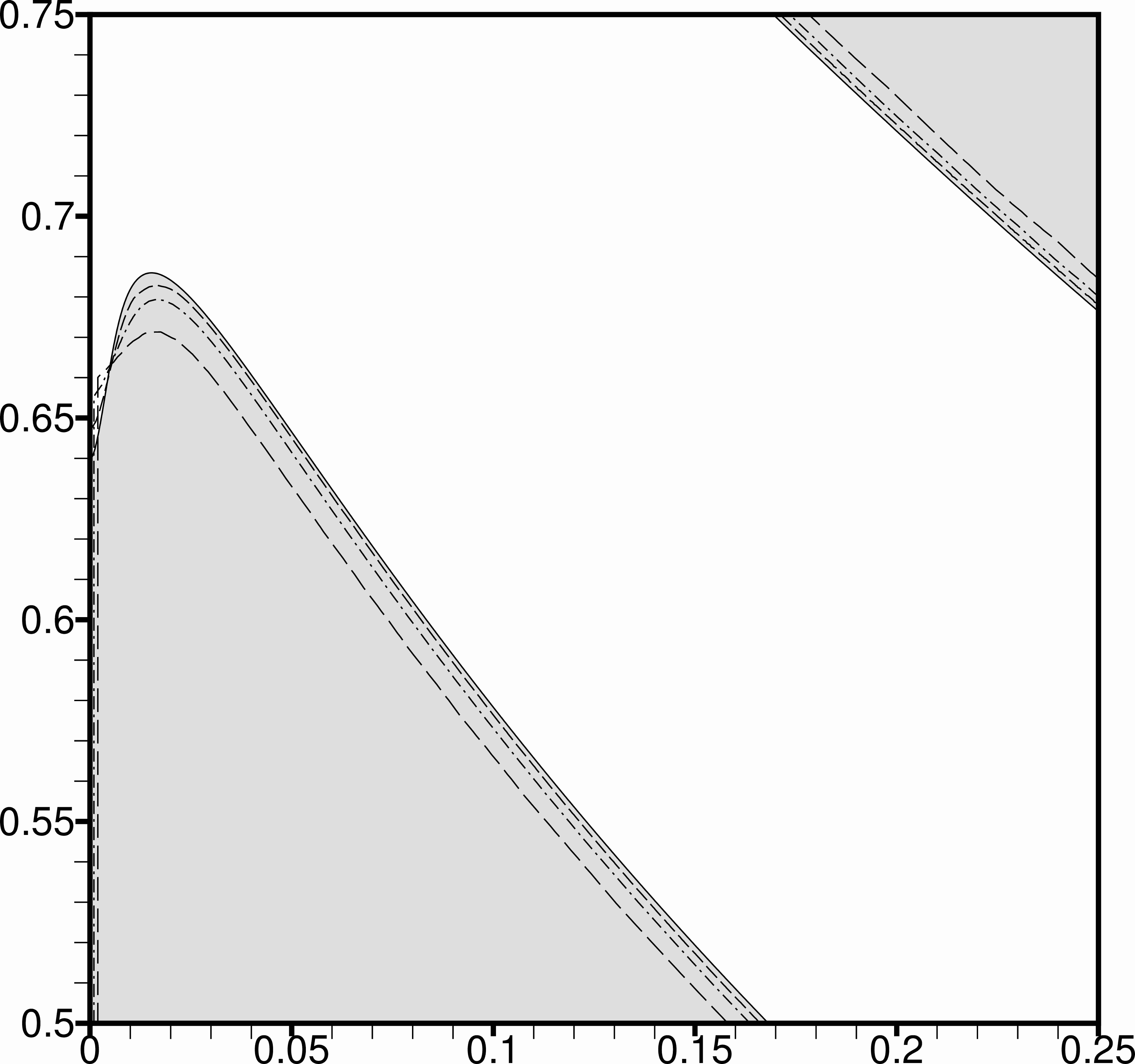}
\caption{Contours of $\tau = Bn$ on various grids, for the \{$Bn$ = 10, $Re$ = 1000\} case, at a subregion of the 
domain. The lines correspond to grids 256$\times$256 (long dashes), 512$\times$512 (chained), the locally refined grid 
of Figure \ref{sfig: LR grid Bn=10} (short dashes), and 2048$\times$2048 (solid).} 
\label{fig: Grid convergence of yield surfaces}
\end{figure}

\subsection{Accuracy of the flow field}
\label{ssec: accuracy of flow field}

Figures \ref{fig: vortex centre} and \ref{fig: vortex strength} include results by other researchers for validation 
purposes. In general, there is a good agreement with the present results. There is some discrepancy concerning the 
vortex position for \{$Re$ = 1000, $Bn$ = 10\} with some of the other publications, while we should also note that Vola 
et al. \cite{Vola_03} predict somewhat stronger vortices. In addition to these results, for creeping flow our previous 
study \cite{Syrakos_13} showed that the present method produces results that are in very good agreement with the 
literature.

The rest of this paragraph investigates the effect of $Bn$, $M$, and the grid spacing $h$ on the accuracy. To confine 
the investigation, three cases were selected: $Bn$ = 1, 10 and 100, with $Re$ fixed at $Re$ = 1000. To aid the 
investigation, grid-independent solutions were sought by solving the cases also on finer grids of 1024$\times$1024 and 
2048$\times$2048 volumes, and performing Richardson extrapolation (see, for example, \cite{Ferziger_02}), assuming 
second-order convergence. For the $Bn$ = 100 case, results were not obtained on the 2048$\times$2048 grid because this 
would require a computing time of a few months with our present serial code. Instead, Richardson extrapolation was 
performed using the 512$\times$512 and 1024$\times$1024 grids. The present finite volume method adopts a cell-centred 
strategy for storing the variables, which implies that cell centres of different grids do not coincide. Therefore, 
interpolation is needed in order to compare the solutions of two different grids and perform Richardson extrapolation. 
We use a third-order accurate interpolation scheme which is described in \cite{Syrakos_12}, so that the errors 
introduced by this interpolation are smaller than those of the finite volume discretisation.

\begin{table}[t]
\caption{$L^1$ norms (Eq.\ (\ref{eq: L1 norm})) of the flow variables ($1^{\text{st}}$ data column), of the 
discretisation errors $\epsilon_G^{\phi}$ on different grids $G$ (data columns 2 -- 5; $LR$ stands for Locally Refined 
grid, see Fig.\ \ref{fig: LR grids}), and of the difference $\delta_{M_2}^{M_1}$ of solutions obtained with different 
regularisation parameters $M_1$ and $M_2$ on the $512\times 512$ grid (last two columns). Data columns 6 and 7 ($q$ and 
$q^*$) display the order of grid convergence (Eq.\ (\ref{eq: order q})), calculated in two different ways (see text).
Data in columns 2 -- 5 and 8 -- 9 are expressed as a percentage of the data in column 1. Unless otherwise stated, 
$M=400$.}
\label{table: integral results}
\begin{center}
\begin{scriptsize}   
\renewcommand\arraystretch{1.25}   

\begin{tabular}{r | c c c c c c c c c }
\hline
 & $|\phi|_1$ & $|\epsilon^{\phi}_{128}|_1$\% & $|\epsilon^{\phi}_{256}|_1$\% & $|\epsilon^{\phi}_{512}|_1$\%
 & $|\epsilon^{\phi}_{LR}|_1$\% & $q$ & $q^*$ & $|\delta^{100}_{200}|_1$\% & $|\delta^{200}_{400}|_1$\% \\
 \hline
 $Bn=1$ & \# volumes $\rightarrow$ & 16,384 & 65,536 & 262,144 & 268,456 & & & & \\
 \hline
 $\phi \equiv u$ & $1.322\times10^{-1}$ & 1.762 & 0.466 & 0.119 & 0.045 & 1.90 & 2.00 & 0.020 & 0.012 \\
             $v$ & $1.317\times10^{-1}$ & 1.840 & 0.487 & 0.125 & 0.047 & 1.90 & 2.00 & 0.015 & 0.009 \\
             $p$ & $5.316\times10^{-2}$ & 3.142 & 0.823 & 0.213 & 0.089 & 1.93 & 2.04 & 0.019 & 0.012 \\
 \hline
 $Bn=10$ & \# volumes $\rightarrow$ & 16,384 & 65,536 & 262,144 & 284,896 & & & & \\
 \hline
 $\phi \equiv u$ & $5.555\times10^{-2}$ & 3.550 & 1.166 & 0.383 & 0.166 & 1.60 & 1.76 & 0.831 & 0.437 \\
             $v$ & $3.260\times10^{-2}$ & 4.982 & 1.639 & 0.534 & 0.243 & 1.60 & 1.71 & 1.091 & 0.577 \\
             $p$ & $1.052\times10^{-2}$ & 7.000 & 2.443 & 0.834 & 0.393 & 1.50 & 1.75 & 1.055 & 0.603 \\
 \hline
 $Bn=100$ & \# volumes $\rightarrow$ & 16,384 & 65,536 & 262,144 & 279,280 & & & & \\
 \hline
 $\phi \equiv u$ & $2.818\times10^{-2}$ & 9.618 & 5.721 & 2.498 & 0.673 & 0.27 & 1.53 & 1.856 & 1.006 \\
             $v$ & $1.106\times10^{-2}$ & 8.948 & 5.504 & 2.379 & 0.826 & 0.14 & 1.55 & 2.927 & 1.619 \\
             $p$ & $7.233\times10^{-2}$ & 7.831 & 2.912 & 1.078 & 0.448 & 1.42 & 1.63 & 1.294 & 0.784 \\
 \hline
\end{tabular}

\end{scriptsize}
\end{center}
\end{table}

\begin{figure}[!t]
\centering
\noindent\makebox[\textwidth]{
 \subfigure[{$Bn = 1$}] {\label{sfig: LR grid Bn=1}
  \includegraphics[scale=0.9]{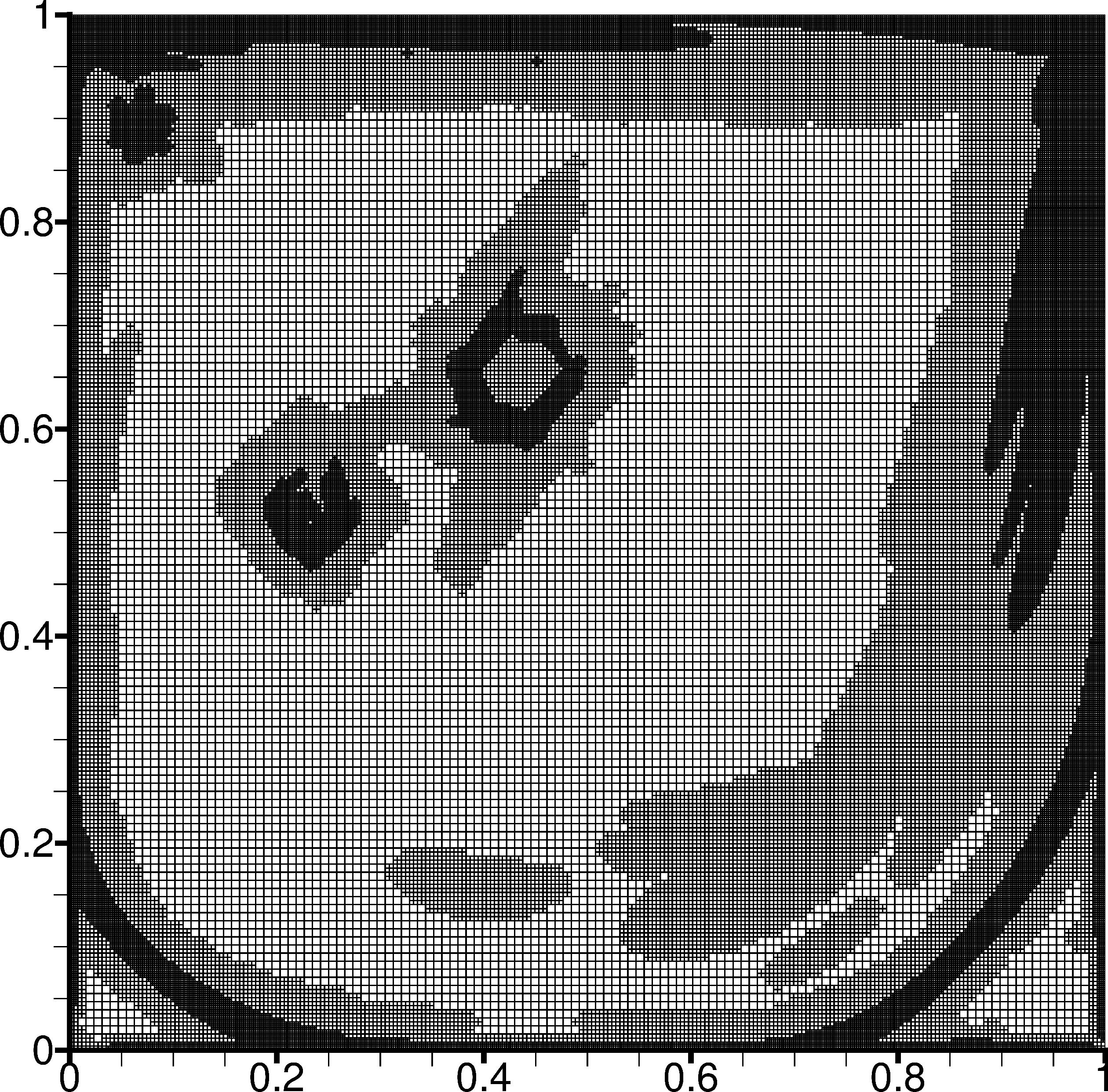}}
 \subfigure[{$Bn = 10$}] {\label{sfig: LR grid Bn=10}
  \includegraphics[scale=0.9]{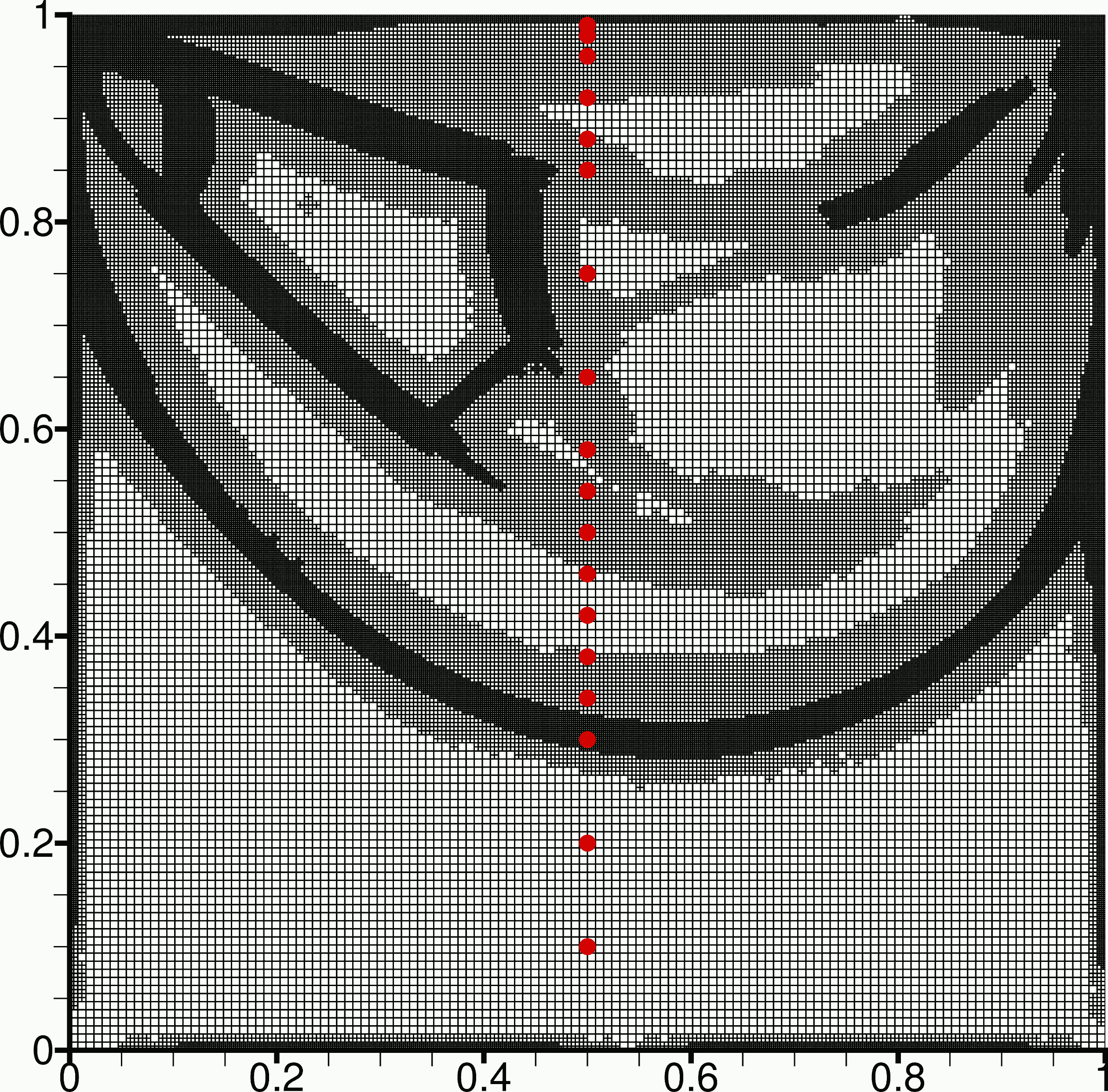}}
}
\noindent\makebox[\textwidth]{
 \subfigure[{$Bn = 100$}] {\label{sfig: LR grid Bn=100}
  \includegraphics[scale=0.9]{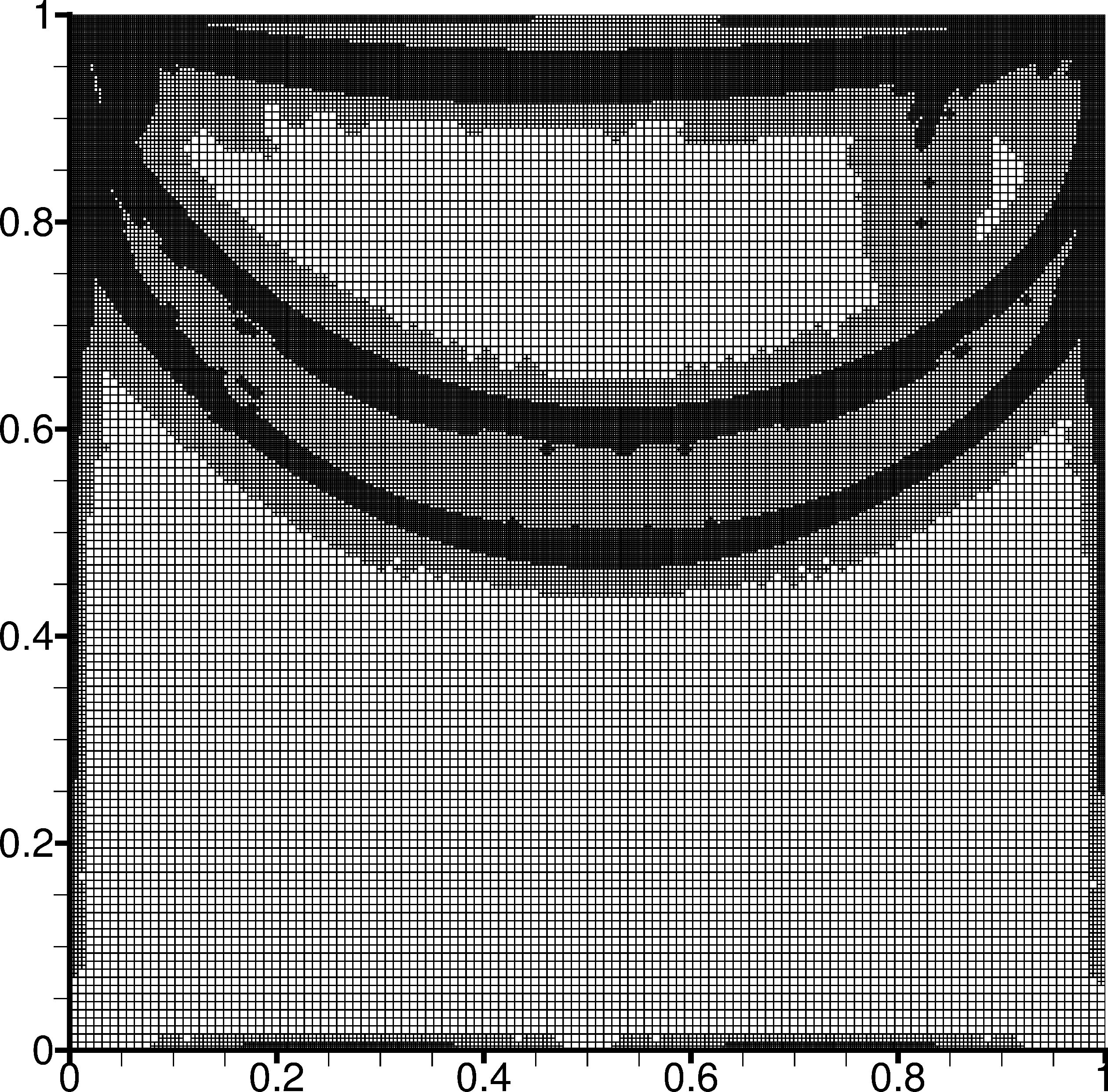}}
}
\caption{Locally refined grids, for $Re$ = 1000 and different Bingham numbers. Actually, for clarity, the grids shown 
are the \textit{underlying} grids (see \cite{Syrakos_06b}), which are twice as coarse as the actual ones used. For $Bn$ 
= 10, the dots along the centreline mark the points that are used in Table \ref{table: U error, Re=1000, Bn=10}.}
\label{fig: LR grids}
\end{figure}

Table \ref{table: integral results} summarises most of the results. Use is made of the following norm:

\begin{equation} \label{eq: L1 norm}
 \left| w \right|_1 \;=\; \frac{1}{\Omega} \cdot \sum_{P=1}^{N} \left| w_P \right| \cdot \Omega_P
\end{equation}
where $w$ is an arbitrary quantity, $w_P$ is the value of this quantity at the centre of cell $P$ of the grid, 
$\Omega_P$ is the volume of cell $P$, $N$ is the total number of cells of the grid, and $\Omega$ is the total volume of 
the domain. Table \ref{table: integral results} contains the following columns:

\begin{itemize}
 \item The first data column displays the norm $|\phi|_1$, on the 2048 grid (or the 1024 grid, for $Bn$ = 100), where 
$\phi$ stands for each of the three main flow variables, $u$, $v$ and $p$ (one per row).

 \item The next four columns display the $|\epsilon^{\phi}_G|_1$ norms of the discretisation errors of the variable 
$\phi$ on grid $G$, computed by comparison against the Richardson extrapolation solution. They are expressed as a 
percentage of the norm $|\phi|_1$. In the last column, $LR$ stands for the Locally Refined grid shown in Figure 
\ref{fig: LR grids}.

 \item The next column is the order of grid convergence $q$ defined as (see \cite{Ferziger_02})

\begin{equation} \label{eq: order q}
 q \;\equiv\; \frac{\log\left( \frac{|\epsilon^{\phi}_{4h}|_1 - |\epsilon^{\phi}_{2h}|_1}
                                    {|\epsilon^{\phi}_{2h}|_1 - |\epsilon^{\phi}_{h}|_1} \right)}{\log(2)}
\end{equation}
where the subscripts denote the grid where $\phi$ was calculated: $h$, $2h$ and $4h$ are the 512, 256 and 128 grids, 
respectively. Since the equations were discretised using 2$^{\text{nd}}$-order accurate central differences, $q$ should 
normally equal 2.

 \item The column $q^*$ provides an alternative calculation of the order of grid convergence. It uses the solutions on 
the three finest available grids (512, 1024 and 2018 for $Bn$ = 1 and 10; 256, 512 and 1024 for $Bn$ = 100) instead of 
those used by $q$, and applies a pointwise version of equation (\ref{eq: order q}) (i.e. without the norm) at each cell 
centre of the coarsest of these grids, to calculate the local order of convergence there. Then, $q^*$ is the average 
over all grid cells. In this calculation, cells where Eq.\ (\ref{eq: order q}) returns an undefined or negative result 
are excluded from the averaging.

 \item Finally, the last two columns are the norm (\ref{eq: L1 norm}) of the difference $\delta_{M_2}^{M_1} = \phi(M_2) 
- \phi(M_1)$ of the solutions obtained with different regularisation parameters $M_1$ and $M_2$ on the 512 grid, 
expressed again as a percentage of $|\phi|_1$. As both results are obtained on the same grid, no interpolation is 
necessary.
\end{itemize}

The Table has three sections, for $Bn$ = 1, 10 and 100 respectively. The header rows of these sections display also the 
number of volumes of each grid. Of course, the number of volumes of the uniform grids is independent of the Bingham 
number. The Locally Refined grids slightly vary in volume number (268456, 284896 and 279280 volumes for $Bn$ = 1, 
10 and 100, respectively), but in every case the number of volumes is very close to that of the 512$\times$512 grid 
(262144 vols.), and therefore a direct comparison can be made between the accuracies on the Locally Refined and 
512$\times$512 grids.

The results of Table \ref{table: integral results} show that for $Bn$ = 1 the finite volume method exhibits its nominal 
2$^{\text{nd}}$-order convergence, but for $Bn$ = 10 the order of convergence drops to 1.60 - 1.75. For $Bn$ = 100, $q$ 
is very small because the differences $(|\epsilon_{4h}^{\phi}|_1 - |\epsilon_{2h}^{\phi}|_1)$ and 
$(|\epsilon_{2h}^{\phi}|_1 - |\epsilon_{h}^{\phi}|_1)$ are nearly equal. However, the fact that $\epsilon_{512}^{\phi}$ 
is already quite small relative to $\epsilon_{256}^{\phi}$ and $\epsilon_{128}^{\phi}$ means that convergence is 
accelerating as the grid is refined, and the result $q \approx 0.2$ is too pesimistic. A much more optimistic picture 
is given by the index $q^* \approx 1.55$ which is calculated on finer grids. It is noticeable that increasing the 
Bingham number causes also a significant increase of the discretisation error, as a percentage of the solution; it 
increases by a factor of 3-4 if $Bn$ is raised by an order of magnitude, and this factor appears to increase as the 
grid is refined.

\begin{figure}[!t]
\centering 
\noindent\makebox[\textwidth]{
 \subfigure[{Newtonian}] {\label{sfig: ABS(taux) Newtonian}
  \includegraphics[scale=1.0]{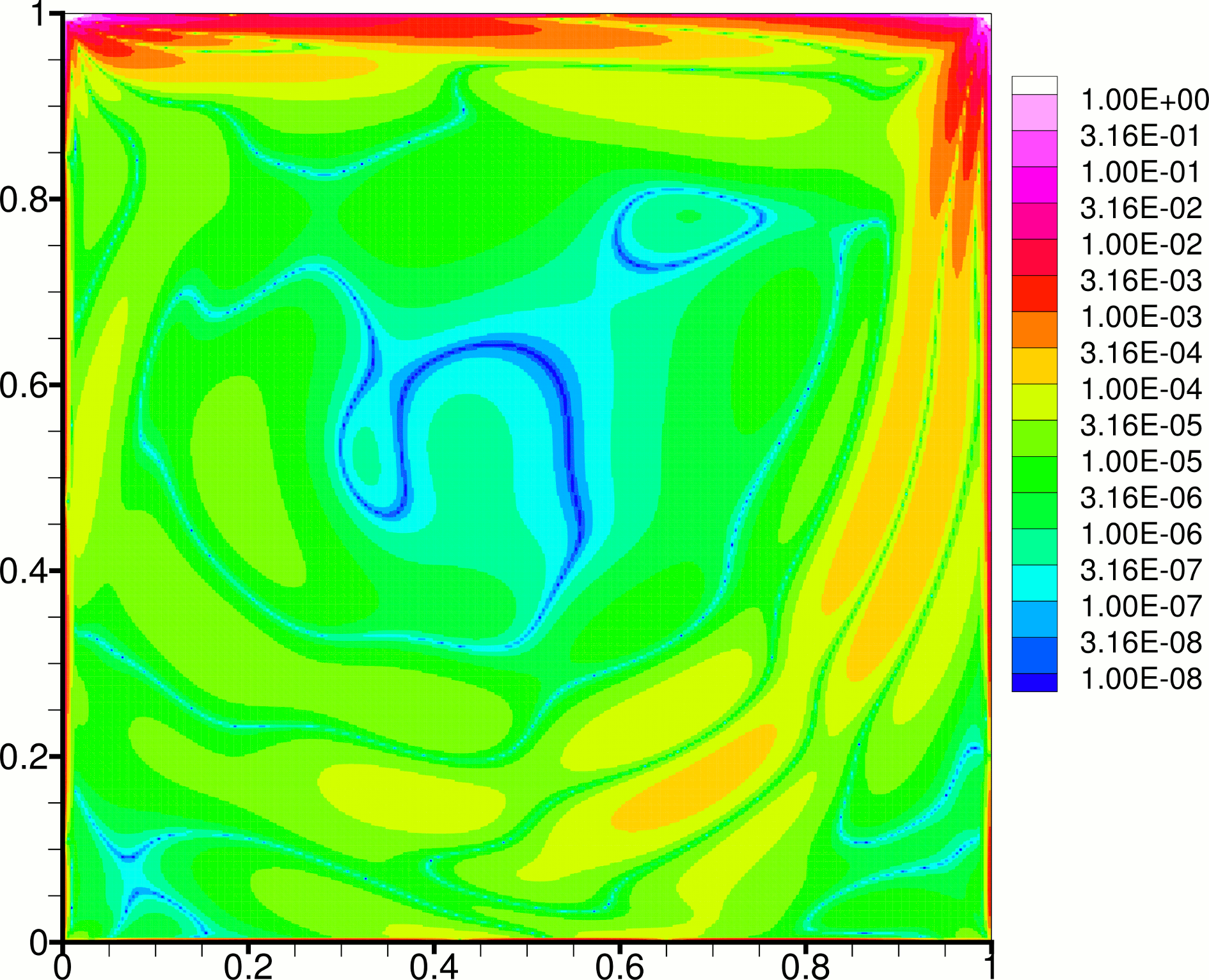}}
 \subfigure[{$Bn = 1$}] {\label{sfig: ABS(taux) Bn=1}
  \includegraphics[scale=1.0]{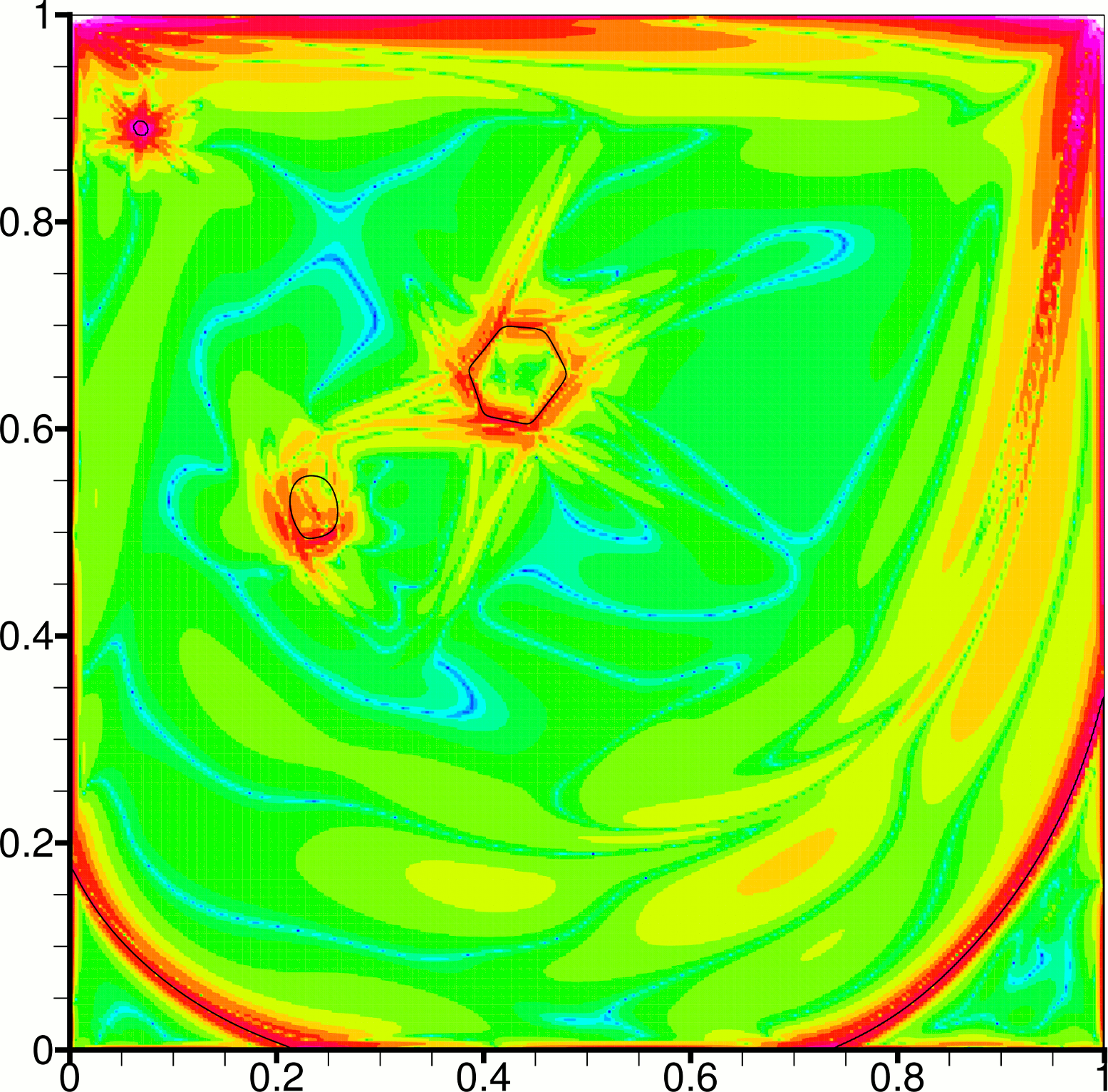}}
}
\noindent\makebox[\textwidth]{
 \subfigure[{$Bn = 10$}] {\label{sfig: ABS(taux) Bn=10}
  \includegraphics[scale=1.0]{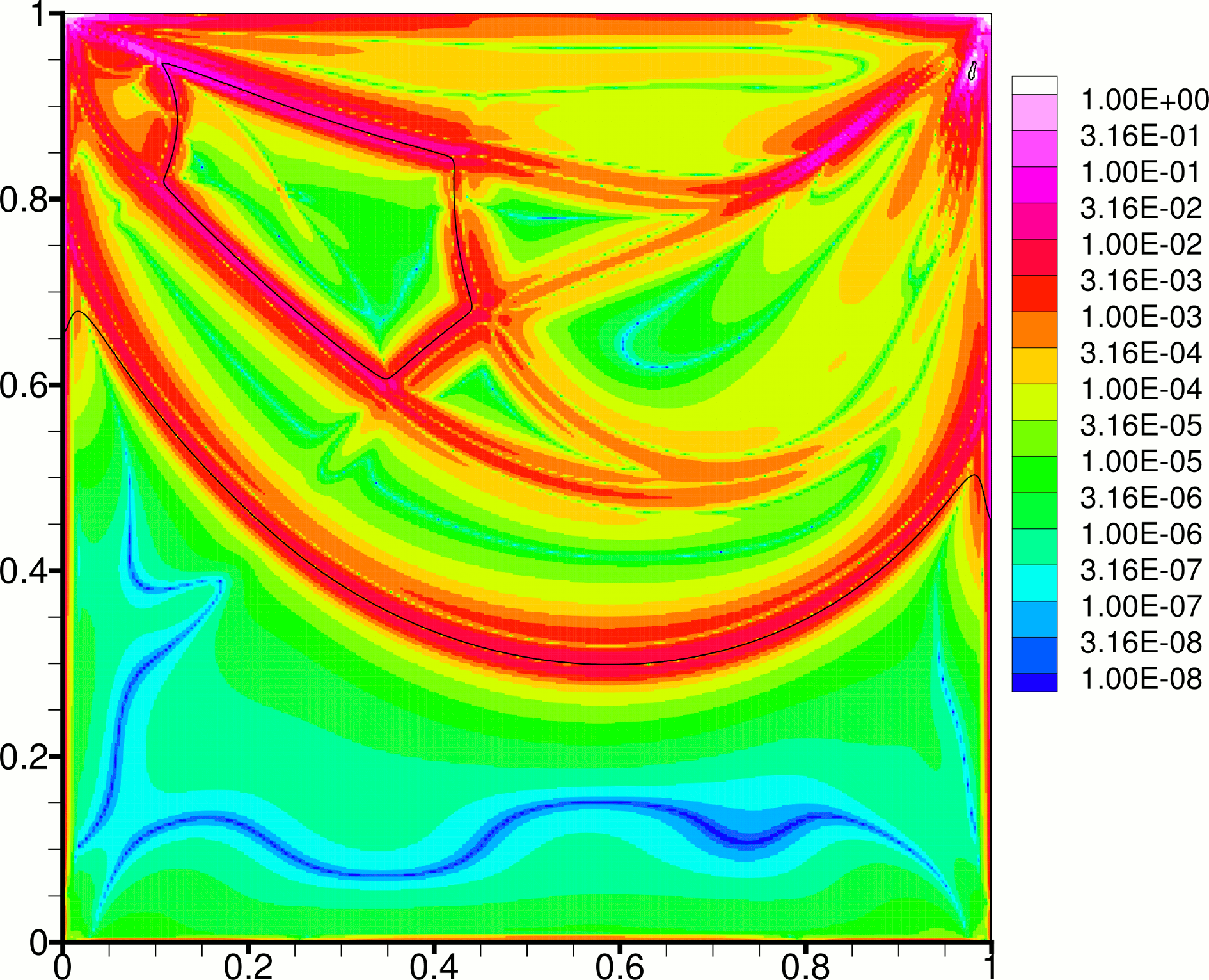}}
 \subfigure[{$Bn = 100$}] {\label{sfig: ABS(taux) Bn=100}
  \includegraphics[scale=1.0]{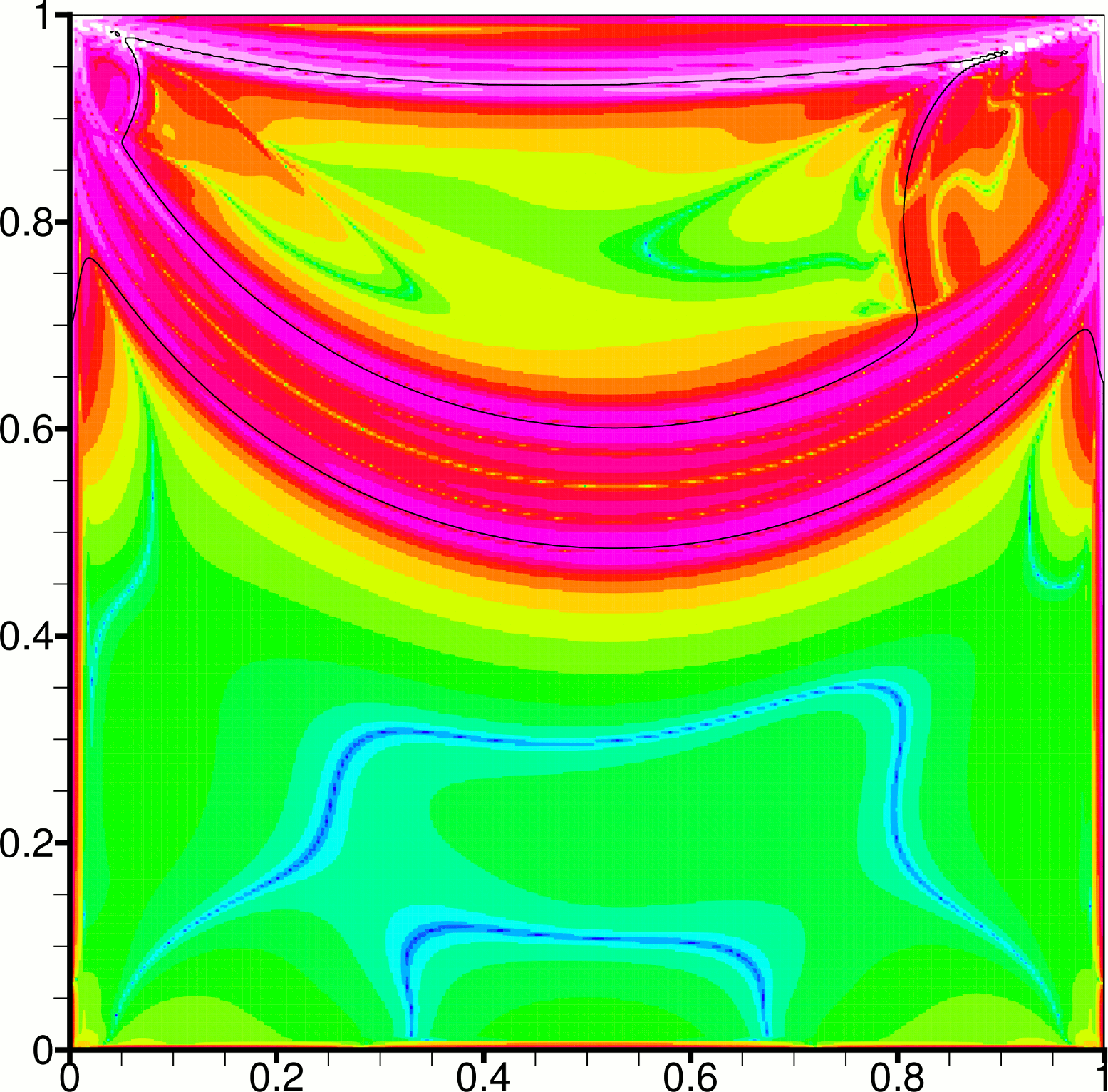}}
}
\caption{Contours of the absolute value of the truncation error of the $x-$momentum equation, on the 512 $\times$ 512 
grid, according to the estimate (\ref{eq: truncation error estimate}). The yield lines are shown in black. In all cases, 
$Re$ = 1000.}
\label{fig: truncation errors}
\end{figure}

The fact that increasing the Bingham number causes a deterioration of the convergence rate and an increase in the 
relative discretisation error can be investigated by examining the truncation error, which is the source of the 
discretisation error. Figure \ref{fig: truncation errors} shows plots of the absolute value of the truncation error of 
the $x-$momentum equation, calculated according to estimate (\ref{eq: truncation error estimate}), for our three 
selected cases plus the Newtonian case. The truncation error can be seen to increase by several orders of magnitude as 
the Bingham number increases, which results in the loss of accuracy observed in Table \ref{table: integral results}. 
High truncation errors occur mostly in the vicinity of the yield surfaces, where the flow field looses its regularity, 
but they are also observed elsewhere in the domain. Very high high-order derivatives develop at these locations, giving 
rise to the high truncation errors, which in turn generate high discretisation errors that are convected and diffused 
everywhere in the domain.

These results suggest that it would be more efficient to use locally refined instead of uniform grids. Therefore, the 
selected cases were solved again, starting on the 256 $\times$ 256 uniform grid and allowing two grid refinements, 
according to the scheme described in Section \ref{ssec: method with refinement}. So, for each case the problem had to be 
solved three times: once on the 256 $\times$ 256 grid, and once after each of the two grid refinements. The final grids 
obtained are shown on Figure \ref{fig: LR grids}. They consist of three levels: the coarsest level has the same density 
as the 256 $\times$ 256 uniform grid; the intermediate level has the same density as the 512 $\times$ 512 grid; and the 
finest level has the same density as the 1024 $\times$ 1024 grid. One would then expect on these composite grids an 
accuracy greater than that of the 256 $\times$ 256 grid, but less than that of the 1024 $\times$ 1024 grid. The value of 
local refinement is demonstrated in Table \ref{table: integral results}, where one can see that the error on the 
locally refined grids is nearly 2.5 times smaller than on the equally-sized (in terms of number of volumes) 512 
$\times$ 512 uniform grid. Table \ref{table: U error, Re=1000, Bn=10}, which displays pointwise data for the $Bn$ = 10 
case, shows that this improvement of accuracy occurs everywhere in the domain, including in regions where the composite 
grid is more coarse than the 512$\times$512 grid (see the points marked in Fig.\ \ref{sfig: LR grid Bn=10}).

\begin{table}[t]
\caption{\{$Re$ = 1000, $Bn$ = 10\} case: Values of $u$ (the $x-$ velocity component) and related discretisation errors 
$\epsilon^u_G$, obtained on various grids $G$ and expressed as a percentage of $u$, at selected points of the vertical 
centreline whose vertical coordinates are shown in the first column. The values of $u$ shown were obtained from the 
2048$\times$2048 grid solution with linear interpolation. $LR$ stands for the Locally Refined grid shown in Fig.\ 
\ref{sfig: LR grid Bn=10}, where the selected points are indicated.}
\label{table: U error, Re=1000, Bn=10}
\begin{center}
\begin{scriptsize}   
\renewcommand\arraystretch{1.1}   
\begin{tabular}{ c | c r r r r }
 \hline
  $y$ & $u$ & $\epsilon_{256}^u$\% & $\epsilon_{512}^u$\% & $\epsilon_{1024}^u$\% & $\epsilon_{LR}^u$\% \\
 \hline
  1.000  &  1.00000  &  0.00\%  &  0.00\%  &  0.00\%  &  0.00\%  \\
  0.990  &  0.78310  &  0.09\%  &  0.04\%  &  0.01\%  &  0.02\%  \\
  0.980  &  0.58701  &  0.25\%  &  0.07\%  &  0.01\%  &  0.02\%  \\
  0.960  &  0.29412  &  0.60\%  &  0.18\%  &  0.06\%  &  0.04\%  \\
  0.920  &  0.04561  &  5.07\%  &  1.79\%  &  0.47\%  &  0.71\%  \\
  0.880  & -0.02002  &  8.29\%  &  2.88\%  &  0.79\%  &  1.37\%  \\
  0.850  & -0.03879  &  0.63\%  &  0.32\%  &  0.12\%  &  0.19\%  \\
  0.750  & -0.05521  &  0.02\%  &  0.02\%  &  0.01\%  &  0.05\%  \\
  0.650  & -0.06820  &  0.84\%  &  0.27\%  &  0.07\%  &  0.09\%  \\
  0.580  & -0.07597  &  0.28\%  &  0.10\%  &  0.03\%  &  0.00\%  \\
  0.540  & -0.07941  &  0.19\%  &  0.08\%  &  0.03\%  &  0.01\%  \\
  0.500  & -0.07966  &  0.69\%  &  0.21\%  &  0.05\%  &  0.17\%  \\
  0.460  & -0.06587  &  1.35\%  &  0.39\%  &  0.09\%  &  0.32\%  \\
  0.420  & -0.04296  &  3.02\%  &  0.92\%  &  0.22\%  &  0.49\%  \\
  0.380  & -0.02024  &  7.25\%  &  2.32\%  &  0.57\%  &  0.95\%  \\
  0.340  & -0.00467  & 16.51\%  &  8.26\%  &  2.42\%  &  4.97\%  \\
  0.300  & -0.00083  &  4.72\%  &  1.21\%  &  0.27\%  &  0.68\%  \\
  0.200  & -0.00042  &  0.97\%  &  0.32\%  &  0.07\%  &  0.19\%  \\
  0.100  & -0.00020  &  0.56\%  &  0.17\%  &  0.03\%  &  0.10\%  \\
  0.000  &  0.00000  &  0.00\%  &  0.00\%  &  0.00\%  &  0.00\%  \\
 \hline                                                                                            
\end{tabular}
\end{scriptsize}
\end{center}
\end{table}

\begin{figure}[t]
\centering
\noindent\makebox[\textwidth]{
 \subfigure[{$Bn=2$, $M=400$: $\dot{\gamma}_y=0.01266$}] {\label{sfig: gamma Re=0, Bn=2, M=400, unyielded only}
  \includegraphics[scale=1.00]{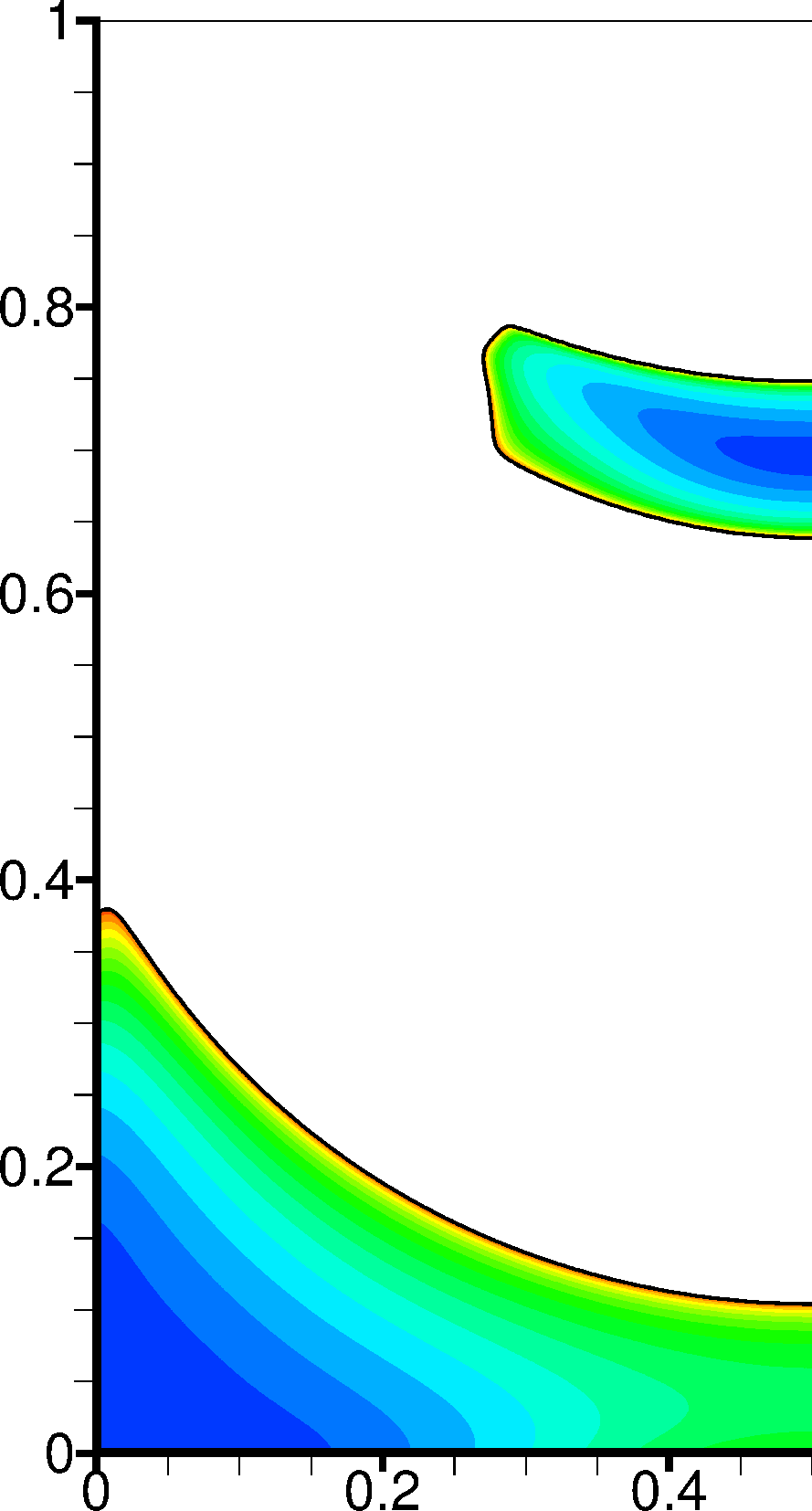}}
\qquad
 \subfigure[{$Bn=50$, $M=400$: $\dot{\gamma}_y=0.01961$}] {\label{sfig: gamma Re=0, Bn=50, M=400, unyielded only}
  \includegraphics[scale=1.00]{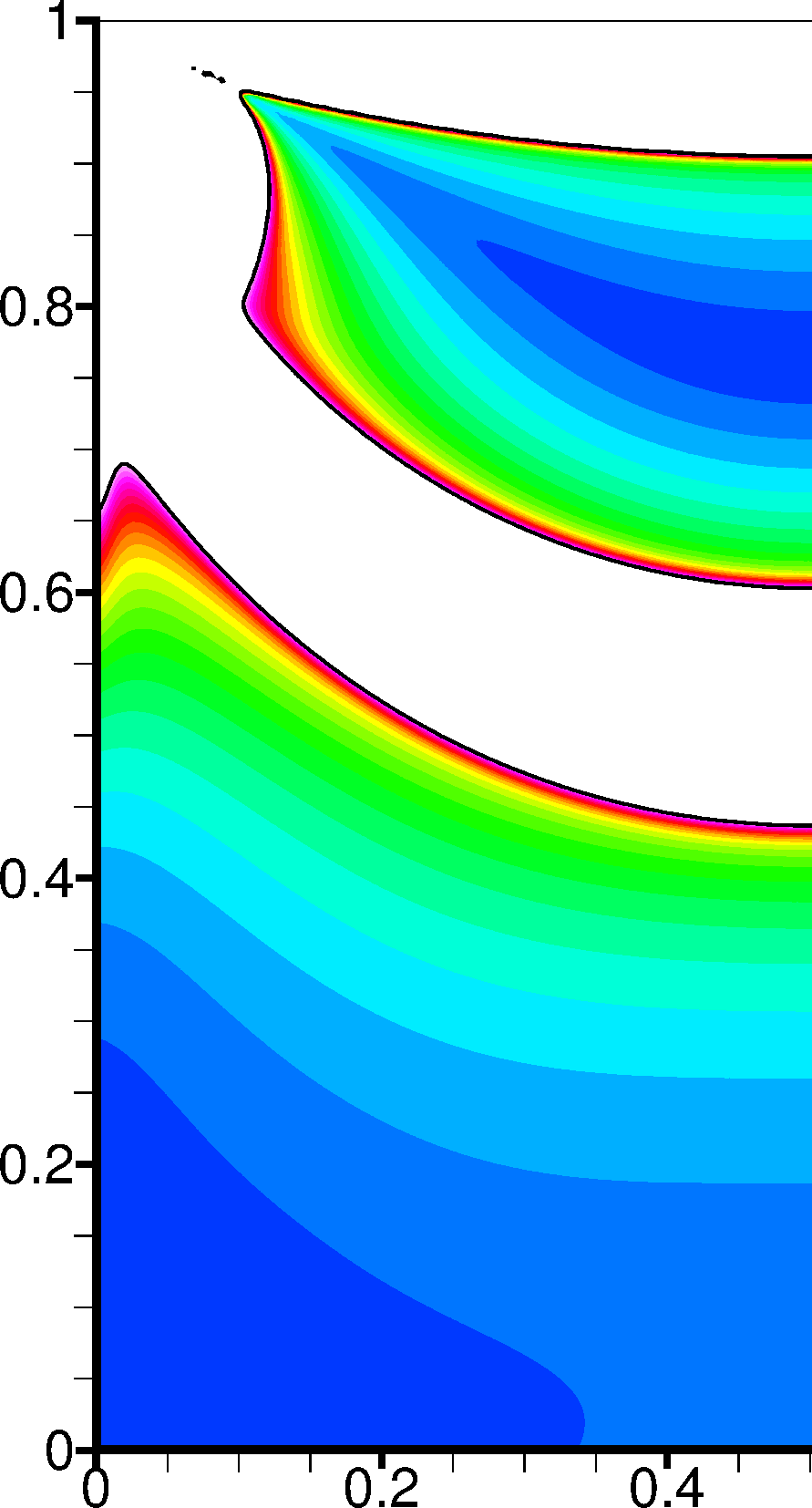}}
\qquad
 \subfigure {\label{sfig: gamma Re=0, M=400, unyielded only legend}
  \includegraphics[scale=1.00]{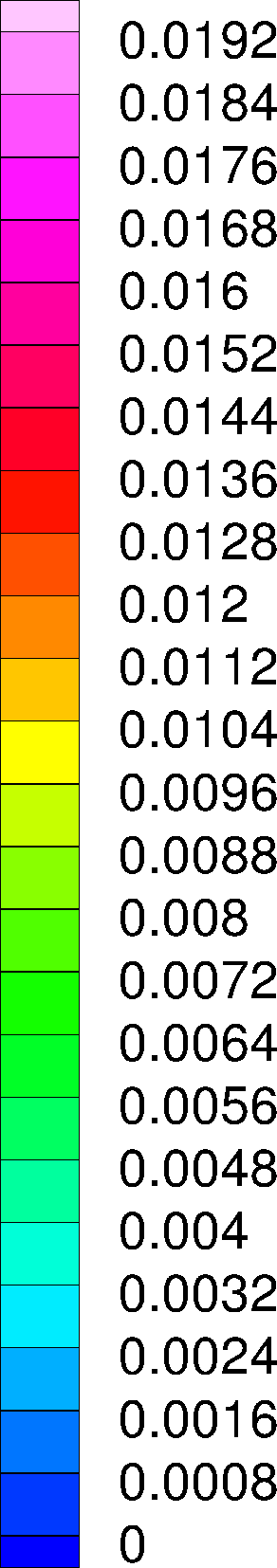}}
  \addtocounter{subfigure}{-1}
}
\caption{Contours of the magnitude of the rate of strain $\dot{\gamma}$ inside the unyielded zones (yielded regions are 
shown in white). The Reynolds number is zero, so the flow field is symmetric and only half the domain is drawn. The 
boundary of the unyielded zones, calculated as $|\dot{\gamma}|=\dot{\gamma}_y$ from equation (\ref{eq: gamma_y}), is 
marked with a black line.}
\label{fig: gamma Re=0 unyielded only}
\end{figure}

The last two columns of Table \ref{table: integral results} show the effect of the regularisation parameter $M$. The 
change inflicted on the flowfield by chaning $M$ from 200 to 400, $\delta^{200}_{400}$, is in every case 0.5-0.6 times 
that caused by changing $M$ from 100 to 200 ($\delta^{100}_{200}$). The difference $\delta^{200}_{400}$ can be viewed 
as a crude approximation to the error caused by regularisation. For $Bn$ = 1, this difference is of the order 
$\delta^{200}_{400} \approx$ 0.01\% and is much smaller than the discretisation error on the 512$\times$512 grid. For 
$Bn$ = 10 and 100, $\delta^{200}_{400}$ are significantly higher, of the order of 0.5\% and 1\% respectively, and they 
are comparable to the discretisation error on the 512$\times$512 grid. These results indicate the necessity of using 
larger $M$ parameters for simulating flows of higher $Bn$. In contrast to this result, it has been suggested by 
researchers who used the Papanastasiou regularisation, e.g.\ in \cite{Tsamopoulos_96, Burgos_99}, that smaller values of 
$M$ can be used with higher values of $Bn$, on the basis that the limit of the value of the viscosity $\eta$ (\ref{eq: 
viscosity}) as $\dot{\gamma}$ tends to zero is $M\!\cdot\!Bn + 1$. This means that if $M$ is kept constant then in the 
core of the unyielded regions the viscosity becomes higher as $Bn$ increases, thus providing a better approximation for 
the unyielded material. However, away from the core, near the yield surface, the approximation of the unyielded material 
in fact becomes worse as $Bn$ increases, if $M$ is kept constant. This can be seen if one rearranges equation (\ref{eq: 
gamma_y}) as  

\begin{equation} \label{eq: gamma_y alternative)}
 M \;=\; \frac{1}{\dot{\gamma}_y} \ln \left( \frac{Bn}{\dot{\gamma}_y} \right)
\end{equation}
Therefore, if $M$ is constant, then $\dot{\gamma}_y$ increases with $Bn$. Figure \ref{fig: gamma Re=0 unyielded only} 
shows an example where $\dot{\gamma}$ is plotted for two distinct $Bn$ numbers, $Bn=2$ and $Bn=50$. It can be seen that 
for the higher $Bn$ number, $Bn=50$, $\dot{\gamma}$ is smaller deep into the unyielded zones and larger near the yield 
lines than for the smaller $Bn=2$.

\subsection{Algebraic convergence of the SIMPLE/multigrid algorithm}
\label{ssec: algebraic convergence}

\begin{figure}[!b]
\centering
\noindent\makebox[\textwidth]{
\subfigure[{$Re$ = 1000; $Bn$ = 1; $M$ = 400}] {\label{sfig: convergence Bn=1}
 \includegraphics[scale=1.00]{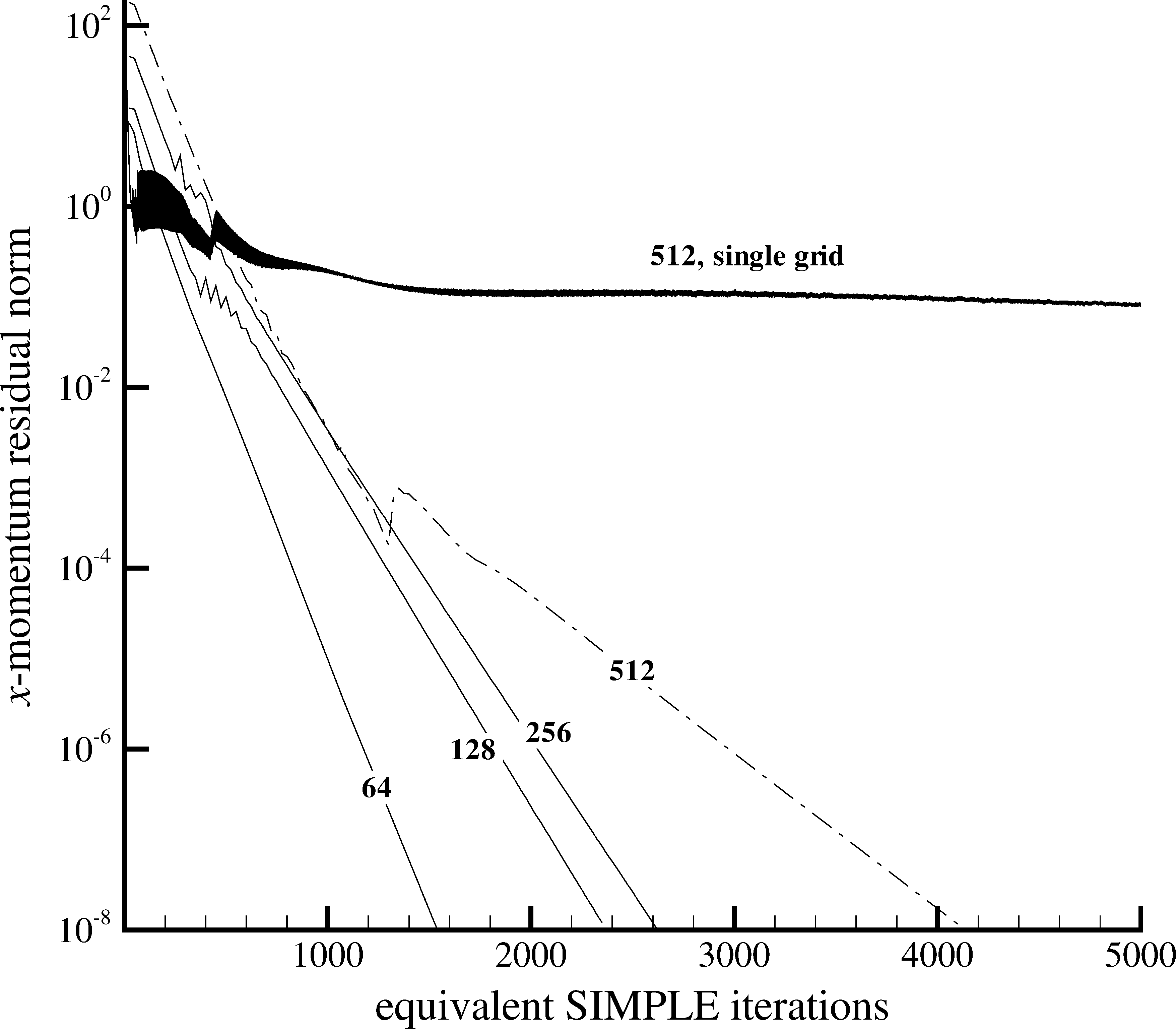}}
\subfigure[{$Re$ = 1000; $Bn$ = 10; $M$ = 400}] {\label{sfig: convergence Bn=10}
 \includegraphics[scale=1.00]{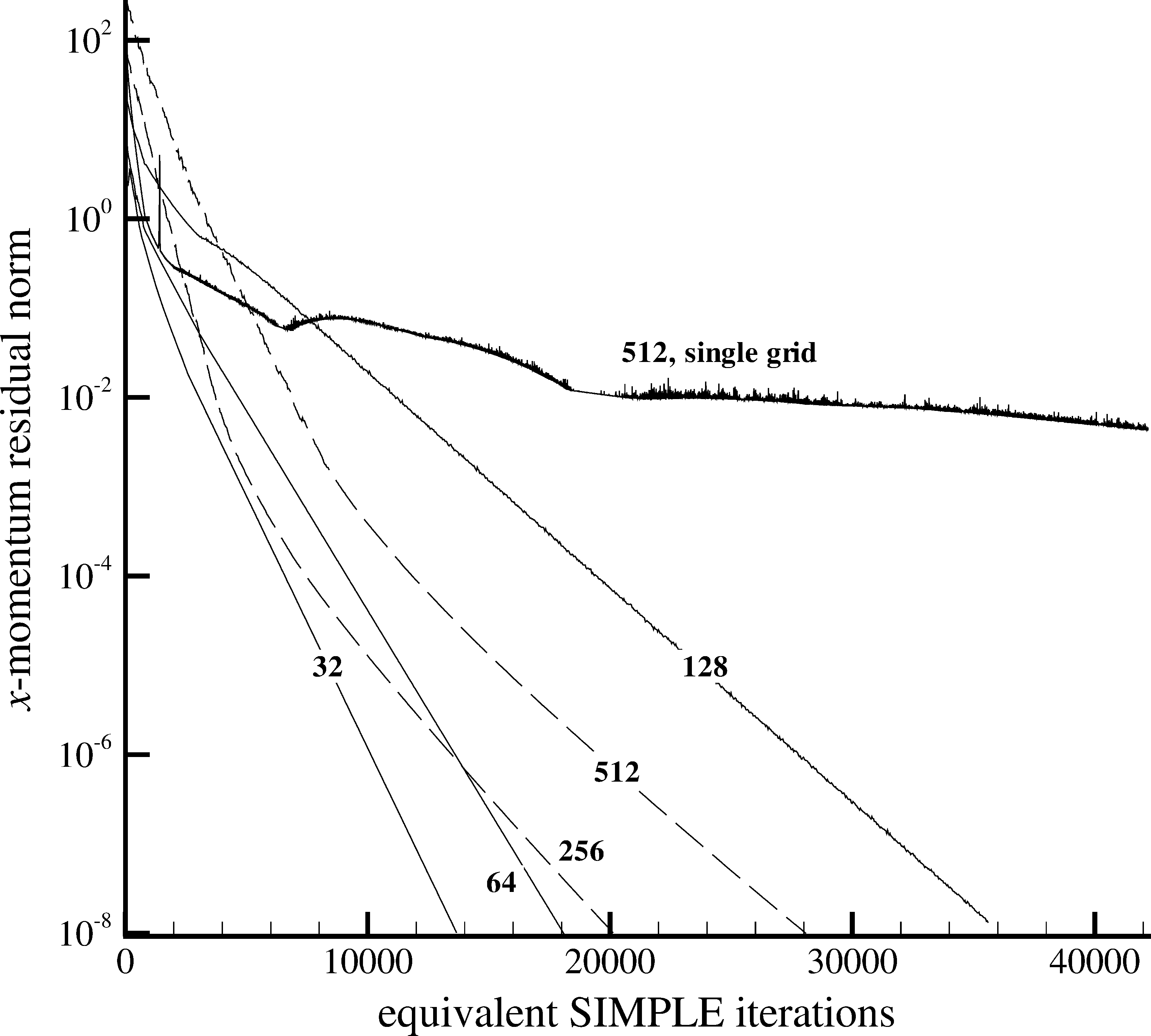}}
}
\noindent\makebox[\textwidth]{
\subfigure[{$Re$ = 1000; $Bn$ = 100; $M$ = 400}] {\label{sfig: convergence Bn=100}
 \includegraphics[scale=1.00]{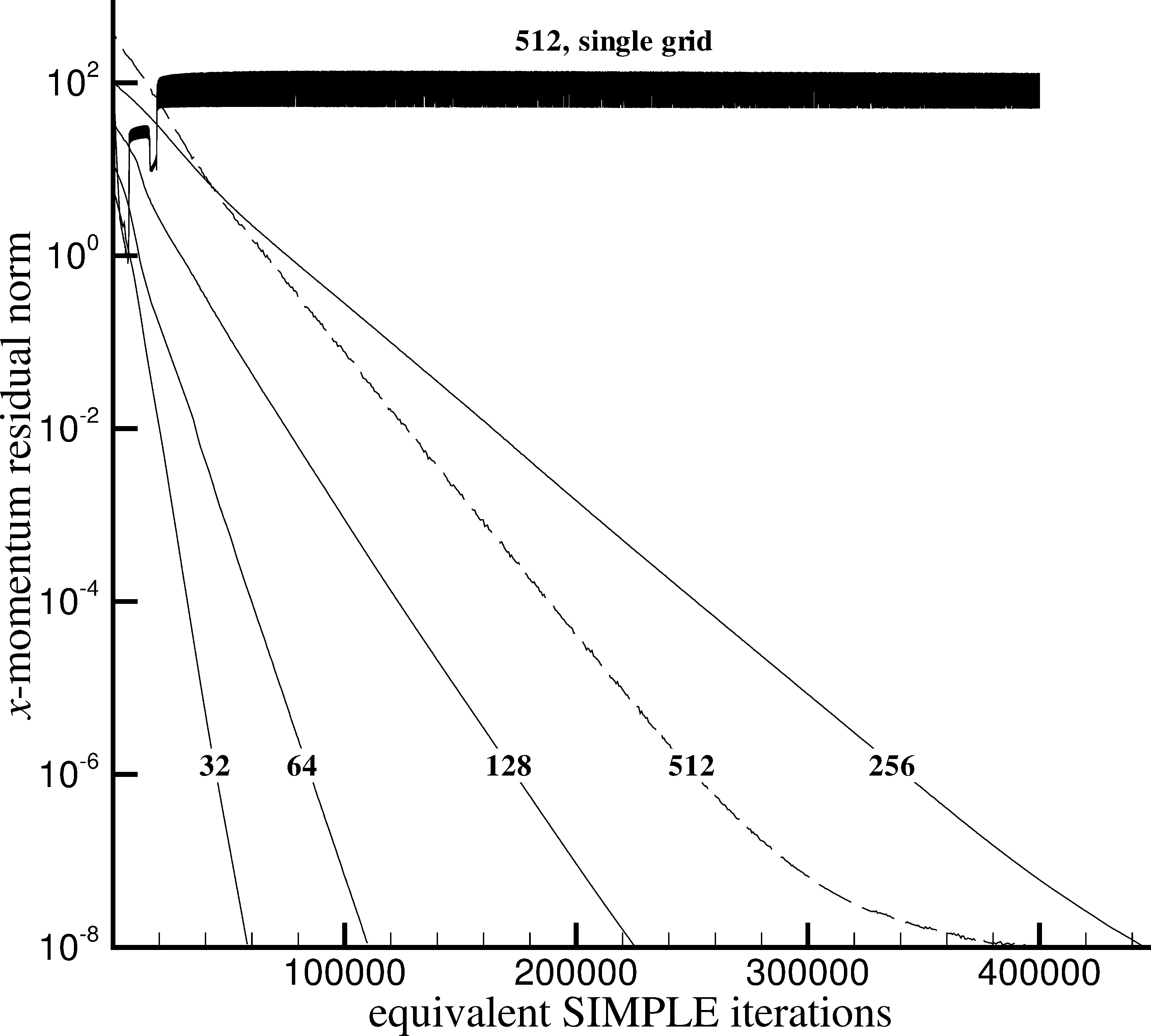}}
}
\caption{Maximum $x$-momentum residual per unit volume (\ref{eq: residual norm}) versus computational effort, for $Re$~=
1000, $M$~= 400, and $Bn$~= 1, 10, and 100. The results refer to the SIMPLE/multigrid algebraic solver, using the 
$8\times 8$ as the coarsest grid, except for a single-grid case which is indicated on each figure. The number on each 
curve indicates the finest grid ($n$ means that the finest grid has $n \times n$ control volumes). The algebraic solver 
parameters are the following (see Section \ref{sec: numerical method} for definitions): For $Bn$~= 1, W(5,5)-5 cycles, 
$\alpha_{MG}$~= 1.0, $a_u$~= 0.7, and $a_p$~= 0.02 (multigrid) or 0.2 (single grid). For $Bn$~= 10, W(6,6)-10 cycles, 
$\alpha_{MG}$~= 0.9, and \{$a_u$, $a_p$\}~= \{0.5, 0.02\} (multigrid) or \{0.7, 0.2\} (single grid). For $Bn$~= 100, 
W(9,9)-25 cycles, $\alpha_{MG}$~= 0.9, and \{$a_u$, $a_p$\}~= \{0.4, 0.002\} (multigrid) or \{0.6, 0.1\} (single grid). 
On each grid, the solution of the immediately coarser grid was used as the initial guess.}
\label{fig: algebraic convergence}
\end{figure}

In this subsection some results on the algebraic convergence of the SIMPLE/multigrid algorithm are reported. Figure 
\ref{fig: algebraic convergence} shows the reduction of the algebraic residuals as a function of the computational 
effort for $Re=1000$ and $Bn$ = 1, 10, and 100, with $M$ = 400, as typical examples. The ordinate is the the 
$L^{\infty}$-norm of the residual vector of the $x-$momentum equations,

\begin{equation} \label{eq: residual norm}
 \lVert r \rVert_{\infty} \;=\; \max_{P=1,\ldots,N} \left\{|r_P|\right\} \;,
\end{equation}
where $r_P$ is the residual, expressed per unit volume, of the $x-$momentum equation of volume $P$ and $N$ is the total 
number of volumes in the grid. The computational effort is measured in equivalent fine-grid SIMPLE iterations. For the 
multigrid cases, the number of equivalent fine-grid SIMPLE iterations is obtained by multiplying the number of cycles by 
the number of fine-grid SIMPLE iterations that cost computationally the same as a single cycle. In particular, $n_C$ 
cycles of type W($\nu_1,\nu_2$)--$\nu_3$ cost approximately the same as $n_S = n_C\cdot[2(\nu_1+\nu_2)+\nu_3]$ SIMPLE 
iterations on the finest grid (see e.g.\ \cite{Trottenberg_01} on how to calculate the cost of W cycles). For example, 
one W(6,6)-10 cycle costs the same as 34 fine-grid SIMPLE iterations. It should be noted that the cost of restriction 
and prolongation is omitted in this calculation, since it is very small compared to the cost of the SIMPLE iterations, 
especially if one considers that the numbers of pre- and post- smoothing iterations are large, and fine-grid iterations 
are also carried out between cycles. Therefore, multigrid and single-grid convergence rates are directly comparable in 
the Figure. The SIMPLE underrelaxation factors were chosen differently in the multigrid and single-grid cases, in order 
to make the solvers more efficient in each case. For $Bn$ = 100, we were unable to make the single-grid algorithm 
converge on the 512 $\times$ 512 grid, for any choice of underrelaxation parameters.

Figure \ref{fig: algebraic convergence} also shows that the multigrid procedure greatly accelerates the convergence of 
SIMPLE. One can notice that as the grid becomes finer, the multigrid convergence slows down in general. This non-typical 
multigrid behaviour is explained by the fact that the present multigrid method contains single-grid features, as 
described in Section \ref{sec: numerical method}. For $Bn$ = 10, on the $256\times 256$ and $512\times 512$ grids the 
procedure converges fast at the initial stages, due to a good initial guess, but slows down at later stages of 
iteration. For $Bn$ = 100 it is noticeable that convergence is faster on the $512 \times 512$ grid than on the $256 
\times 256$ grid; a possible explanation is that the solution on the $256 \times 256$ grid provides a good initial guess 
for the $512 \times 512$ grid, whereas this does not occur on coarser grids. The convergence rates decrease as $Bn$ 
increases, and are significantly worse than those typically exhibited in Newtonian flows. In fact one may notice that 
for every ten-fold increase in Bingham number ($Bn$ = 1 to $Bn$ = 10 to $Bn$ = 100) there is roughly also a ten-fold 
increase in the number of equivalent SIMPLE iterations required.

\begin{figure}[t]
\centering
 \includegraphics[scale=1.00]{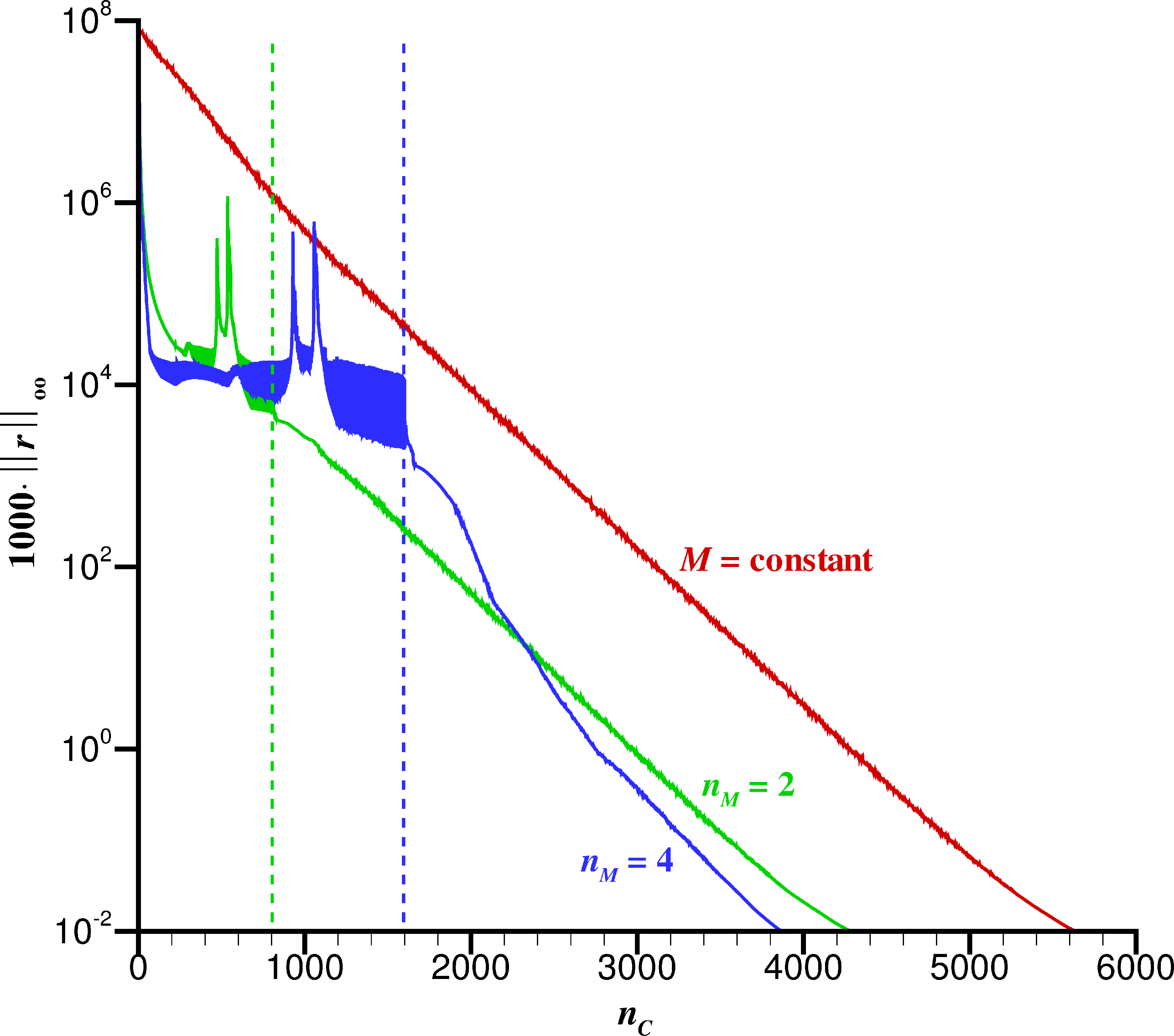}
 \caption{The $L^{\infty}$ norm of the $x$-momentum residual plotted against the number $n_C$ of W(6,6)-20 cycles 
($a_u=0.5$, $a_p=0.01$) on the $256\times 256$ grid, for $Re=0$, $Bn=20$, $M=400$. The solution of $128\times 128$ was 
used as the initial guess in each case. The red line corresponds to the case that $M=400$ is used throughout. The other 
lines depict convergence when $M$ is progressively increased by $1$ every $n_M$ cycles, until the maximum of $M=400$ 
(marked by dashed lines).}
 \label{fig: MG with nM}
\end{figure}

As mentioned in Section \ref{sec: numerical method}, it was observed that it is sometimes advantageous to gradually 
increase the value of $M$ as multigrid cycles progress, up to the maximum selected value, instead of keeping it at this 
value from the start of the calculations. As an example, Figure \ref{fig: MG with nM} shows convergence results for 
$Re=0$ and $Bn=20$ with $M=400$. The ``$M$=constant'' curve depicts convergence when $M=400$ throughout. The other two 
curves depict the convergence of a procedure where, starting with $M=1$, after every $n_M$ cycles the exponent $M$ is 
increased by one. The two curves correspond to $n_M=2$ and $n_M=4$. The point where $M=400$ is reached is marked with 
vertical dashed lines of the same colour, and from that point onwards the value of $M$ is held fixed at $400$. Actually, 
the residuals shown in Figure \ref{fig: MG with nM} prior to the dashed lines ($M<400$) are not the actual residuals of 
the exact, $M=400$, momentum equations, but of the temporary momentum equations with the current value of $M$ (the 
oscillations are due to the fact that every time $M$ is incremented there is a sharp increase in the residual, 
since the equations change). But from the dashed lines onwards the residuals can be directly compared among the three 
curves. It is evident that this technique can bring significant performance gains at no extra cost.

\section{Conclusions}
\label{sec: conclusions}

A popular method for solving fluid flow problems consists of a combination of a finite volume discretisation and the 
SIMPLE algebraic solver. Existing codes which use this method are easily extended to solve also viscoplastic flows by 
using a regularised version of the constitutive equation. Essentially, all that is required is to write a function to 
calculate the viscosity from the current estimate of the velocity field. The advantages of this approach are that 
minimum modifications to the code are required, and that all the other features of the code that may have been developed 
over time are available also for the viscoplastic flows; finite volume / SIMPLE solvers have a long tradition, and many 
existing codes have a rich set of features including meshing capabilities, numerical schemes, choices of models for 
different physical phenomena, graphical user interfaces, etc. all of which will be also available for the simulation of 
viscoplastic flows. On the other hand, regularisation introduces a deviation from the exact equations, which is 
controlled by a parameter, which therefore plays a role similar to the grid spacing (and the time step if the flow is 
transient) in determining the accuracy with which the original problem is approximated. Depending on the application, 
this deviation may not be very important as the exact constitutive equation is also an approximation to the behaviour of 
true viscoplastic fluids, and for some materials the regularised constitutive equation may actually be a better 
approximation of the true behaviour.

In the present work, the capability of the finite volume / SIMPLE method for solving viscoplastic flows was tested by 
applying it to the very popular lid-driven cavity flow problem, for a range of Bingham and Reynolds numbers, in 
combination with the regularisation scheme of Papanastasiou \cite{Papanastasiou_87}. The results showed that both the 
discretisation errors and the errors due to regularisation increase with the Bingham number. The discretisation error 
increase is due to the truncation error increase in the vicinity of the yield surfaces, because the flow field is nearly 
discontinuous there and the high-order derivatives of the flow variables attain very large values. Since the high 
truncation errors are localised, local grid refinement is the most efficient way to reduce them, as the present results 
verify. On the other hand, the increase of the regularisation error requires that the regularisation parameter is 
assigned larger values as $Bn$ increases. Unfortunately, with the present method the SIMPLE solver was found unable to 
cope with regularisation parameters larger than about 400, which is an important weakness of SIMPLE as a viscoplastic 
flow solver, since this value is rather low according to the literature. Nevertheless, it appears sufficient to produce 
satisfactory results in the range of Bingham numbers considered here, for the lid-driven cavity problem. This problem 
is well-behaved in the sense that, due to the confinement of the flow domain, the stress variation is rather rapid and 
extended regions where the magnitude of the stress is close to the yield stress are not present. Otherwise, very high 
values of $M$ might be required, as Frigaard and Nouar note \cite{Frigaard_05}. In that case, a stronger solver may be 
used instead, with all the additional complexity and modifications to the code. Finite Element methods usually use 
Newton solvers, but the calculation of the Jacobian matrix would be a very difficult task for a finite volume method 
which uses non-Cartesian grids, as is the case if local grid refinement is applied. However, one could use a 
Newton-Krylov method thus avoiding explicit calculation of the Jacobian matrix. Such a solver is used for for example by 
Evans et al. \cite{Evans_2006} for a phase change problem which also involves fluid and solid regions. In fact they use 
SIMPLE as a preconditioner, so that the existing SIMPLE routines can be exploited. This is planned to be the subject of 
a future study.

Another disadvantage of regularisation is that the yield surfaces are not clearly defined. Usually they are identified 
using the criterion $\tau = \tau_y$, but it is important to place the results under scrutiny, by using different values 
of $M$, comparing against $\tau = (1+\epsilon)\tau_y$ contours, and / or using an extrapolation technique such as that 
proposed by Liu et al. \cite{Liu_2002}. These techniques, which are all easily implemented in the post-processing stage 
without any modifications to the main code, were applied successfully and provided minor corrections to the yield 
surfaces predicted by the $\tau = \tau_y$ criterion.

Finally we note that regularisation errors can be avoided altogether by using a multipliers method, with all the 
additional complexity and programming effort involved. This has been implemented in a Finite Volume context by Vinay et 
al. \cite{Vinay_2005} and Glowinski and Wachs \cite{Glowinski_2011}.

\section*{Acknowledgements}
This work was co-funded by the European Regional Development fund and the Republic of Cyprus through the Research 
Promotion Foundation (research projects $\mathrm{AEI\Phi OPIA/\Phi Y\Sigma H}$/0609(BIE)/15 and $\mathrm{T\Pi 
E/\Pi\Lambda HPO}$/0609(BIE)/11).






\section*{REFERENCES}
\bibliographystyle{ieeetr}
\bibliography{syrakos_2013}








\end{document}